\documentclass[useAMS,usenatbib]{mn2e}
\usepackage{graphicx}
\usepackage{txfonts}
\usepackage[authoryear]{natbib}
\usepackage{subcaption}
\usepackage{array} % for defining a new column type
\usepackage{aas_macros}

\usepackage{multirow,array,varwidth}
\usepackage{float}
\title[Spin axes orbital alignment in binaries]{Tracking the spin axes orbital alignment in selected binary systems - Torun Rossiter-McLaughlin effect survey }
\author[P.~Sybilski et al.]
{\parbox{\textwidth}{P. Sybilski,$^{1,4}$\thanks{E-mail: \texttt{sybilski@ncac.torun.pl}} R. K. Paw{\l}aszek,$^{1}$ A. Sybilska,$^{2,3}$ M. Konacki,$^{1}$ K. G. He{\l}miniak,$^{1}$ \\ S. K. Koz{\l}owski,$^{1}$ M. Ratajczak,$^{5,1}$}\vspace{0.4cm}\\
$^{1}$Nicolaus Copernicus Astronomical Center, Polish Academy of Sciences, Rabia\'nska 8, 87-100 Toru\'n, Poland\\
$^{2}$Baltic Institute of Technology, al. Zwyciestwa 96/98, 81-451 Gdynia, Poland\\
$^{3}$European Southern Observatory, Karl-Schwarzschild-Strasse 2, 85748 Garching bei M\"{u}nchen, Germany\\
$^{4}$Sybilla Technologies Sp. z o. o., , Toru\'nska 59, 85-023 Bydgoszcz, Poland\\
$^{5}$Astronomical Institute, University of Wroc{\l}aw, Kopernika 11, 51-622 Wroc{\l}aw, Poland\\
}

\newcounter{daggerfootnote}

\begin{document}

\newcommand\NameEntry[1]{%
  \multirow{1}*{%
    \begin{minipage}{10em}% --- or minipage, if you prefer a fixed width
    #1%
    \end{minipage}}}

\date{Accepted ... Received ...; in original form ...}

\pagerange{\pageref{firstpage}--\pageref{lastpage}} \pubyear{2012}

\maketitle

\label{firstpage}

\begin{abstract}
We have obtained high-resolution spectra of four eclipsing binary systems (FM\,Leo, NN\,Del, V963\,Cen and AI\,Phe) with the view to gaining insight into the relative orientation of their stellar spin axes and orbital axes. The so called Rossiter-McLaughlin (RM) effect, i.e. the fact that the broadening and the amount of blue- or redshift in the spectra during an eclipse depend on the tilt of the spin axis of the background star, has the potential of reconciling observations and theoretical models if such a tilt is found. We analyse the RM effect by disentangling the spectra, removing the front component and measuring the remaining, distorted lines with a broadening function (BF) obtained from single value decomposition (SVD), weighting by the intensity centre of the BF in the eclipse. All but one of our objects show no significant misalignment, suggesting that aligned systems are dominant.  We provide stellar as well as orbital parameters for our systems. With five measured spin-orbit angles we significantly increase (from 9 to 14) the number of stars for which it has been measured. The spin-orbit angle $\beta$ calculated for AI\,Phe's secondary component shows a misalignment of 87$\pm$17 degrees. NN\,Del, with a large separation of components and a long dynamical timescale for circularisation and synchronisation, is an example of a close to primordial spin-orbit angle measurement.
\end{abstract}

\begin{keywords}
binaries: eclipsing -- methods: observational -- techniques: radial velocities -- methods: numerical -- stars: fundamental parameters.
\end{keywords}

\section{Introduction}
\label{sec1}
Eclipsing binaries are excellent laboratories for studying stellar formation and test beds for stellar evolution models. This is because even with the most advanced currently available instrumentation we are not able to resolve surface details of many stars. However, observing an eclipsing binary system gives us a rare opportunity to circumvent this limitation. During an eclipse, part of the light from one of the stars in a binary is blocked from our view. Measuring this light loss at different phases of the eclipse can provide clues as to the relative sizes of the stars, their temperatures or even structure of their atmospheres and estimates of the orbital parameters. Additionally, when this light loss can also be assessed as a function of wavelength, one is also able to determine the orientation of the spin axes of the stars projected to the sky plane, which is described by the so called Rossiter-McLaughlin (RM) effect (\citealt{rossiter:1924}, \citealt{mclaughlin:1924})\footnote{Spin axis and rotational axis of the star are used interchangeably in the text, as well as the orbital pole and orbital axis, with directions of these axes following the "right-hand rule" convention.}.

The most up-to-date and thorough study of the spin-orbit alignment effect has been conducted within the framework of the BANANA Project \citep{albrecht:2011,albrecht:2012b,albrecht:2013,albrecht:2014} and the initial publications which lead to it \citep{albrecht:2007,albrecht:2009}. Table~1 in \cite{albrecht:2011} summarizes the available measurements and estimates of spin-orbit angles in eclipsing binaries. Altogether 32 binaries have been examined up to now and obliquity has been measured for nine stars. With the addition of our sample of four binaries and obliquity measurements of five stars therein, the sample is significantly enlarged, providing new data for the subject of formation, evolution and timescales shaping binary systems.

The RM effect takes advantage of the fact that during an eclipse different portions of a binary system's component are  covered by a companion star or a planet in front of it, which translates into a shift in the radial velocity measurements and the change in the broadening function of the observed spectrum. If the spin axes of the components are aligned with the orbital axis of the system, then the amount of blueshift observed in the recovered velocity during the first part of the eclipse will be equal to the amount of redshift during the second part of the eclipse. The RM effect will have a point symmetry against the midpoint of the eclipse. If, however, the spin axis of the background star is tilted with respect to the system, there will appear a red- or blueshift bias visible in the velocity, line broadening function and shape, dependent on the amount of the tilt. Other factors involved are the relative radii of the stars, the line-of-sight inclination of the orbital plane which determines the percentage of coverage during the eclipse, limb and gravity darkening, turbulence, ellipticity of the orbit and the position of the periastron relative to the observer. 

Chances of encountering a spin-orbit misalignment (SOM) are higher in systems with highly eccentric orbits, given the fact that the spin-orbit alignment timescale is shorter than the circularisation timescale, and as such the stars in the system have not yet had their orbits circularized and their axes aligned. The latter is expected to happen unless a third body on a highly inclined orbit is present in the system (e.g. a planet or an unseen star). This is due to the fact that a third-body companion may pump the eccentricity via the Lidov-Kozai effect (\citealt{lidov:1962}, \citealt{kozai:1962}) and may also be the source of disturbed spin axes of the binary's components. The misalignment has been proposed by e.g. \cite{albrecht:2009} as a possible explanation for the apparent discrepancy between the sometimes observed precession of the orbit (i.e. the so called apsidal motion) and the value for it predicted from theoretical models. Moreover, the eccentricity of the orbit increases the chances of good separation of components in the spectrum of two-line binary star in the eclipse, which will make the task of the subtraction of the foreground component easier and more precise.

%Aims:
The aim of our study is to characterize the stellar and orbital parameters in four binary systems: FM\,Leo, NN\,Del, V963\,Cen and AI\,Phe, with particular attention paid to a robust determination of the orientation of the rotation axes of the stars. The first system, FM\,Leo, has a circular orbit and was selected as a target of opportunity. The remaining three systems have been chosen from the catalogue of \cite{bulut:2007} from among those with highly eccentric orbits to maximize the chances of finding spin-orbit misalignment (see Table~\ref{AggrgatedTab} for a compilation of their published stellar and orbital parameters, as available). Using single value decomposition (SVD) to obtain the broadening function (BF) with the method presented in \cite{rucinski:1999} we have developed a set of tools for the determination of spin-orbit alignment parameters of the studied systems and their components. Bringing four more systems to the still small group of systems for which the $\beta$ parameter in one or both components has so far been determined, we can provide more insight on the frequency of occurrence of the spin-orbit misalignment. 

The paper is structured as follows. In Section~\ref{sec2} we describe the observations and the data reduction process, in Section~\ref{sec3} we describe the method for measuring the obliquity, in Section~\ref{sec4} we summarize measured values for different binaries and in Section~\ref{sec5} we provide comments and conclusions.

\section{Observations and data reduction}
\label{sec2}
\subsection{Observations}
The observations were carried out over 20 nights during the period of 2014-2016 at the Cerro Tololo Inter-American observatory with the use of the 1.5m CTIO telescope equipped with the CHIRON\footnote{Operated by the SMARTS Consortium} echelle spectrograph \citep{schwab:2012, tokovinin:2013}. The instrument has a spectral resolution of $\sim$\,80000 or $\sim$\,27000 (with or without image slicer, respectively), a spectral range of 410-870\,nm and a total efficiency of 6\% \footnote{http://www.ctio.noao.edu/noao/content/CHIRON}. These parameters make CHIRON ideal for precise radial velocity measurements. After a few test observations we decided to use the instrument in the image slicer mode due to the high resolution obtained from the instrument (resolution of 80000) with still high enough signal-to-noise ratio of 50-100 for objects from our survey list. Their brightness is higher than 9 magnitudes and with exposure times between 10 and 15 minutes the desired SNR was reached. This choice gives us good radial velocity resolution for small effects and enough photons even during an eclipse to measure the RM effect.

\subsection{Sample}
We selected eclipsing double-lined binaries from the Southern Hemisphere, with non-zero eccentricity (except for FM\,Leo), brightness higher than 9\,mag and a spectral type of A or later, with preference given to short-period binaries from the \cite{bulut:2007} catalogue and from the objects already studied by ourselves. Earlier spectral type objects were excluded due to the lack of at least a~few narrow spectral lines in most of the echelle spectra orders. We observed our selected targets during the primary and/or secondary eclipse to obtain the RM measurements, as well as between the eclipses to calculate accurate orbital parameters to model the orbits and disentangle the stars' spectra. Below we provide a short description of each of the four studied systems.

\subsubsection{FM\,Leo}
FM\,Leo (HD 97422, HIP 54766), of spectral type F7V\,C \citep{houk:1978} was first classified as an eclipsing binary in \cite{kazarovets:1999} and then used among other double-lined spectroscopic (SB2) eclipsing binaries for color-surface calibration in \cite{kruszewski:1999}. The eclipsing ephemeris of the system was first derived in \cite{solonovich:2003} from their photometric data, however, the period found was $3.3643328$ days that is approximately half of the currently derived period (e.g. the 6.728606 days of \citealt{ratajczak:2010}). The absolute properties of FM\,Leo were studied in \cite{ratajczak:2010} where the combination of photometric and spectroscopic measurements allowed for the derivation of orbital and physical parameters. The analysis included the determination of effective temperatures, radii, masses and surface gravities of the components and the estimation of the evolutionary stage of the system. The distance to the system is given in \cite{eker:2014} as 130.4(25.5)\,pc, later confirmed by \cite{stassun:2016} as 137(11.7)\,pc.

\subsubsection{NN\,Del}
NN\,Del (HD 197952, HIP 102545) is a long-period system of F8\,D spectral type, also first described as an eclipsing binary in \cite{kazarovets:1999}. \cite{gomez:2003} performed a photometric-only study and found the period of the binary to be $99.268$ days and from the secondary eclipses taking place at 0.188 phase derived the eccentricity value of 0.5176, as well as provided $T_{eff}$ estimates. The study by \cite{griffin:2014}, relying on photometric and spectroscopic data, provided the first derivation of the orbital parameters of NN\,Del and masses of the components.

\subsubsection{V963\,Cen}
V963\,Cen (HD 115496, HIP 64941) is a G2V\,C, highly eccentric (0.42), eclipsing binary, located at a distance of ca. 93\,pc \citep{holmberg:2009} that was described by \cite{clausen:1999} (based on {\it uvby} light curves) as having a $15.27$-day period and estimated masses of the primary and secondary component of around $1.1$ M$_{\odot}$. \cite{clausen:2001} subsequently presented a more precise period estimate of $15.269389$ days, as well as modeled absolute dimensions of V963\,Cen. The study of \cite{casagrande:2011} provides the latest estimates of effective temperature, surface gravity, and chemical enrichment parameters of the system

\subsubsection{AI\,Phe}
AI\,Phe (HD 6980, HIP 5438), of the G3V\,C type, is the most extensively investigated system of the four selected for this survey. The first photometric study of the system to provide an accurate period estimate, established to be $24.5923(1)$ days, was carried out by \cite{reipurth:1978}. \cite{imbert:1979} combined spectroscopic and photometric observations to provide a full orbital and physical model assuming an inclination of $90^{o}$ (with the estimates of the radii for a range of inclinations between $88^{o}$ and $90^{o}$. Combining reanalysed literature results with new spectroscopic measurements, \cite{andersen:1988} obtained estimates for masses, radii, and chemical composition of the system. Later, \cite{milone:1992} calculated the parameters of the system by applying an improved light curve modelling technique incorporating Kurucz model atmospheres, with the results in agreement with those of the previous authors.

More recently, the system was studied by \cite{helminiak:2009} who determined its parameters relying on a novel implementation of the iodine cell technique of \cite{konacki:2009} for spectroscopic binaries. In another recent study, the WASP team \citep{kirkbykent:2016} showed improved precision of the radii estimates of the system components, also showing that the times of primary minimum of the system are not constant, with further studies needed to determine the cause of the variation.

\subsection{Data reduction}
Basic data reduction, CCD reduction, spectrum extraction and wavelength calibration, was provided by the on-site facility with the pipeline developed at Yale University \citep{tokovinin:2013}. For the wavelength calibration, we used exposures of Thorium-Argon lamp taken before each science exposure. We used \textsc{iraf}\footnote{\textsc{iraf} is written and supported by the \textsc{iraf} programming group at the National Optical Astronomy Observatories (NOAO) in Tucson, AZ. NOAO is operated by the Association of Universities for Research in Astronomy (AURA), Inc. under cooperative agreement with the National Science Foundation. http://iraf.noao.edu/} \textit{rvsao.bvcorr} task for barycentric velocity and time corrections. For the radial velocity measurements, we used our own implementation of the {\sc todcor} technique \citep{zucker:1994}, which finds velocities of two stars $V_1$ and $V_2$, simultaneously. Templates for the initial solution were taken from \textsc{atlas9} and \textsc{atlas12} codes \citep{kurucz:1992} and for the final fit the disentangled spectra were used. Single measurement errors were calculated with a bootstrap approach as in \cite{helminiak:2012}, and used for weighting the measurements during the orbital fit. The photometric observations for FM\,Leo were taken from \cite{ratajczak:2010}, for V963\,Cen and AI\,Phe from the All Sky Automated Survey (ASAS-3; \citealt{pojmanski:2002}), and for NN\,Del from \cite{gomez:2003}. All photometric observations were conducted in the Johnsons' V-band. The physical parameters of observed binaries were calculated with the help of \textsc{JKTEBOP}\footnote{\textsc{JKTEBOP} is written and supported by John Southworth. http://www.astro.keele.ac.uk/jkt/codes/jktebop.html } (\citealt{southworth:2004a}, \citeyear{southworth:2004b}b), based on EBOP (\textit{Eclipsing Binary Orbit Program}; \citealt{popper:1981,etzel:1981}).

\section{Methods}
\label{sec3}
To obtain radial velocity measurements outside of the eclipse for {\sc v2fit} modelling of the binary's orbit \citep{konacki:2010}, we used {\sc todcor}. For the points in the eclipses and RM effect measurements we used a broadening function. As an additional test, we used BF outside of the eclipses to compare the results with {\sc todcor}. The results were in very good agreement, well within the measurements uncertainties. We then disentangled the spectra to produce single-line, high signal-to-noise templates for both components. The foreground component was subtracted from the observed spectra and the BF was calculated for the remaining spectra with the background's template. From the gravitational centre of the obtained function a displacement of the RV was calculated. The set of such displacements was used to model the RM effect and to obtain the $\beta$ parameter for the eclipsed component. Errors were calculated with the bootstrap method.

\subsection{Disentangling of the spectra}

Our modelling of the RM effect is based on single spectra of each component, deconvolved from the observed spectra. To disentangle the component's spectra we used a tomographic approach presented by \cite{konacki:2010}, which is based on the idea introduced by \citet{bagnuolo:1991}, and uses the maximum entropy method to solve the tomographic problem. In our approach, several observations made in various orbital phases are required, and spectra need to be continuum-corrected. The recovered spectra were then re-scaled to have them normalized to 1. They were later used as templates in {\sc todcor} when RVs outside of the eclipses were calculated. These RVs were finally used (combined with available light curves) to obtain absolute parameters of the component, which were used as an input for the RM fitting procedure.

\subsection{Orbital and physical parameters}
Before modelling the RM effect, we derived orbital and stellar parameters of all the studied systems. Not all systems have their parameters available in the literature (Table~\ref{AggrgatedTab}), so we decided to analyse all of them in a uniform way, using the same methods and sources of data. Radial velocities were fitted with our own procedure {\sc v2fit}, and light curves with the {\sc JKTEBOP} code. Uncertainties in {\sc v2fit} were calculated with a bootstrap procedure, which allows to correctly account for various systematics. In the {\sc jktabsdim} we used a Monte-Carlo approach (task 8) to calculate the errors. Absolute values of parameters (such as masses and radii) were calculated with the {\sc jktabsdim}  code (available with {\sc JKTEBOP}), with the results of the two previous codes used as input (namely: velocity amplitudes $K_1$, $K_2$ and eccentricity $e$ from {\sc v2fit}, inclination $i$, period $P$ and fractional radii $r_1$, $r_2$, brightness of the system in a given band $L$, also known as light scale factor, from {\sc JKTEBOP}). More detailed description of the use of {\sc V2FIT}, {\sc JKTEBOP}, and {\sc jktabsdim} can be found for example in \citet{ratajczak:2010}, \citet{helminiak:2014}, or \citet{helminiak:2015}.
	
The results are shown in Table~\ref{tab:JKTEBOP_Parameters}. They should be treated as preliminary, as there were some simplifications used during the fitting, and those models were not the main goal of this study. For example, we obtain an overall better fitting result for V963\,Cen when the eccentricity and period are fixed to the values obtained from {\sc V2FIT} rather than fitting all of them simultaneously. A very good coverage of the radial velocity curve and limited number of light curve points in the eclipse seem to be the cause. More photometric data should improve the overall solution. Publications that focus on specific targets, with models derived from more observational data (spectra from various instruments, dedicated high-precision photometry) are in preparation.

\subsection{Broadening function}
To recover line profiles and stellar radial velocities from the disentangled spectra, we apply the broadening function (BF) method of \cite{rucinski:1992}, a fully linear approach where a transformation between a typically very narrow model spectrum and the observed spectrum, broadened and shifted in velocity space, is sought (see Figure~\ref{sfig:fmleo-bfGrid-b} for an example of BFs in eclipse). The method's advantage over a cross-correlation function is that it establishes a good baseline, close to zero, which is important for line strengths determination and is not influenced by "peak-pulling" which can affect radial velocity measurements in the case of asymmetries and the presence of more than one peak in the signal \citep{rucinski:1999} The Fourier transform could also be used but due to the likely amplification of high frequency noise and necessity of frequency filtering we decided that a single value decomposition would be the most stable and robust approach. In our RM effect analysis we used echelle spectra orders which contained clearly visible lines of both components close to the order's centre. This was based mainly on the automatic filtering of outside-of-the-eclipse measurements with two clearly visible peaks for BFs. When in one of the peaks the SNR was smaller than 3, the row was rejected.  Noise was calculated as the standard deviation of the data away from the peak (2 FWHMs). 25\% of each order was ignored on each side during this evaluation due to the uncertainties of normalization of the spectra at the order's edges influencing the fit of templates to the observed spectra. A second filtering was done for the measurements in the eclipse. The Kappa-sigma clipping algorithm was used for the series of measurements with $\sigma$ equal to 2 or 3. If the row was rejected in more than three measurements, it was rejected in all measurements in the final modelling of the RM effect. The manual approval and rejection was used to set the clipping value and to visually inspect the distribution of points. Even in the most difficult case of AI\,Phe more than ten rows were accepted. Each approved order was used in measuring the radial velocity in the eclipse and from the set of measurements the average and its standard deviation were obtained. The standard deviation of the average was used as an estimate of the error of the measured radial velocity.

The method relies on finding a Doppler broadening kernel in the following convolution equation:
\begin{equation}
P(x) = \int B(x')T(x-x')dx'
\end{equation} 
where P denotes the observed spectrum and T -- a template, each having a length of the adopted spectral window \citep{rucinski:2015}. 

\subsection{Subtraction}

Outside of the eclipse, the disentangled spectrum of the foreground star can be subtracted and the velocity calculated for the remaining component (see Figures~\ref{sfig:fmleo-bfGrid-e}, \ref{sfig:fmleo-bfGrid-f} and~\ref{sfig:fmleo-bfGrid-g} for an example). During the eclipse, however, a changing part of the light from the background star is hidden from the observer's view, i.e. the light ratio of the components is a function of the phase of the eclipse, so a photometric model needs to be applied when a BF is being recovered. The synthetic light curve from {\sc JKTEBOP} modelling is used with parameters provided in Table~\ref{tab:JKTEBOP_Parameters}. This is an initial guess of the light ratio of the components during the eclipse. This fraction is a free parameter which is then fitted with 10\% freedom when two spectra are fitted to form the observed one. The boundary was never crossed when finding the best solution. The subtraction is done on the spectra level.

The radial velocity, as well as its anomaly related to the RM effect, is subsequently obtained by determining the so called centre of gravity of the broadening function, i.e. not the maximum of the fitted function but rather a dividing point where the area under the curve on each side is identical. This is robust against, e.g. double-peaked functions \citep{albrecht:2007}. 

\subsection{Modelling RM effect}
We developed our own code for modelling obliquity based on \cite{gimenez:2006a} with erratum ~(\citeyear{gimenez:2006c}) and ~(\citeyear{gimenez:2007}),  where the author provides analytical formulae for the computation of RM effect on the radial velocity. The formulae are based on the typical parameters of a binary system and the angles for star's spin and orbit. The foundation for the method was laid by \cite{kopal:1979} with the approach of evaluating a fractional loss of light as a cross-correlation of two apertures. We use two angles to describe the position of a star's rotational axis: the inclination of the star's rotational axis with respect to the line of sight $i_{rot}$ and the angle between the position of the spin axis projection on the plane of the sky with that of the pole of the orbit $\beta$. Other parameters which we use for the description of the RM effect model are the equatorial rotational velocity of the star $V_{rot}$, argument of periastron $\omega$, time of the minimum of the primary eclipse $T_0p$ and first order limb darkening coefficient $u$. Final values and their errors were calculated with the bootstrap approach, where parameters were allowed to vary within the $2\sigma$ limit of values from the starting values from our modelling (see Table~\ref{tab:JKTEBOP_Parameters}). Increasing the allowed variation to three and four $\sigma$ did not change the result but did increase the computation time. Fixed values slightly increased the root mean square (RMS) of the solution but without any significant change to the model. Initial and resulting values are within three $\sigma$ error bar limits for all systems. The set of fitting parameters for global minimum and parameters average from bootstrap results are very similar. Their difference when rounded to the obtained errors is negligible with the small exception for AI\,Phe modelling. During the fitting procedure we tested linear and quadratic limb darkening models, finding the first one to be more stable and sufficient for the available data and obtained precision. The RM effect is described in our code with Equation (10) from \cite{gimenez:2006c}:

\begin{equation}
\delta V = \frac{V^*}{\delta}\frac{\sum_{n=0}^{N}C_n\alpha_n^1}{1-\sum_{n=0}^{N}C_n\alpha_n^0}
\end{equation} 

with parameters following the naming convention of the referred paper: $V^*$ -- rotational parameter, $\delta$ -- projected relative separation of the centres of the stars, coefficients $C_n$ and $\alpha_n^m$ -- functions calculated in the code with Henkel transform and integrated in terms of Jacobi polynomials with an expansion of alpha function estimation to twenty elements.

The equations are valid for any degree of limb darkening, any configuration of the components and orbital eccentricity, with the only condition being that both components of the binary system can be approximated as circular disks moving in front of each other \citep{gimenez:2006c}. The graphical illustration of the used parameters in geometrical projection with similar approach can be found in \cite{hosokawa:1953}. Our algorithm is very similar to the broadening function's centre measurement and RM effect modelling approach described in \cite{albrecht:2007} with the difference of using other algorithms for spectra disentanglement and RM effect modelling.

\begin{figure*}
	\includegraphics[width=1.75\columnwidth]{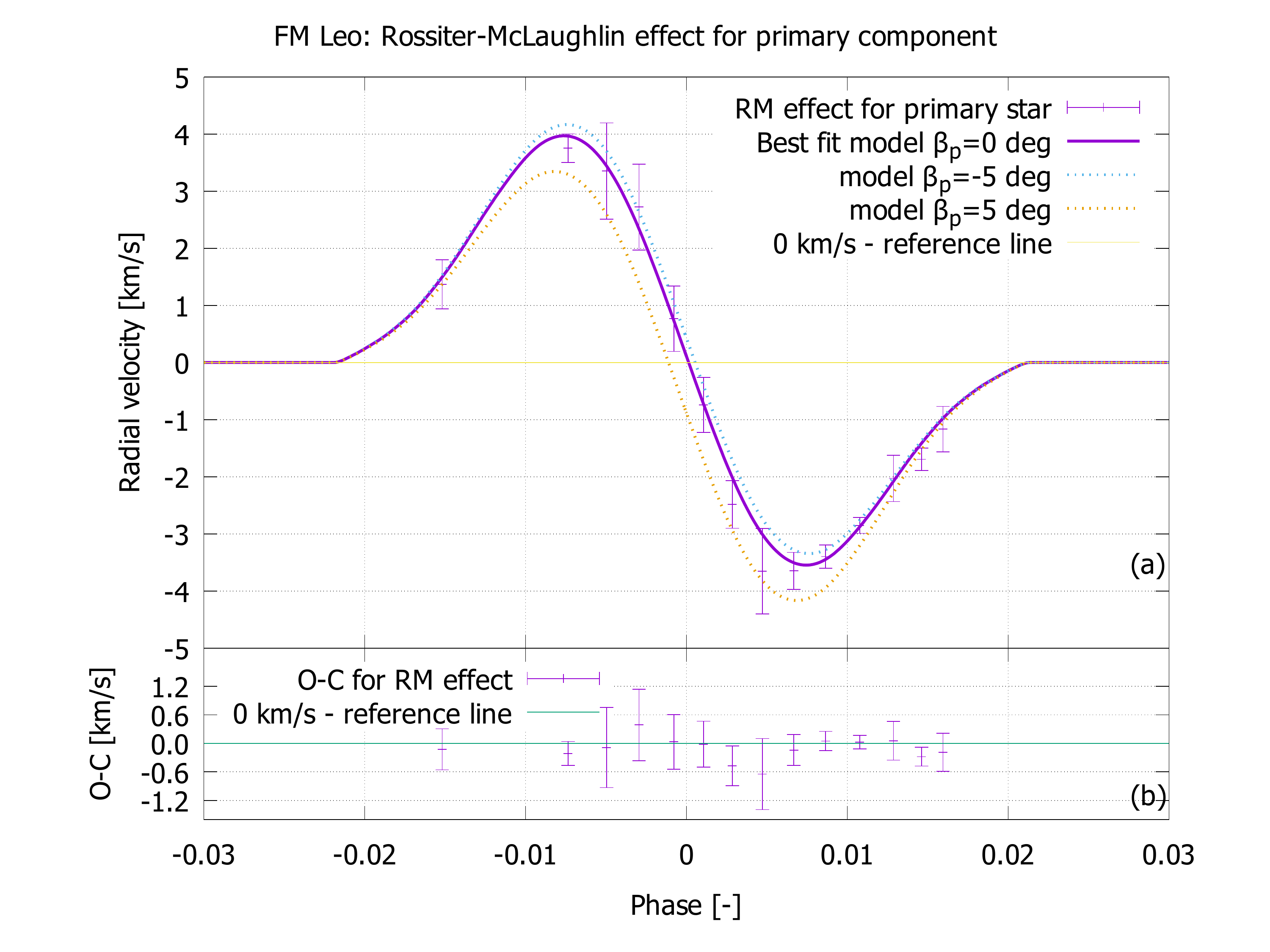}
	\includegraphics[width=1.75\columnwidth]{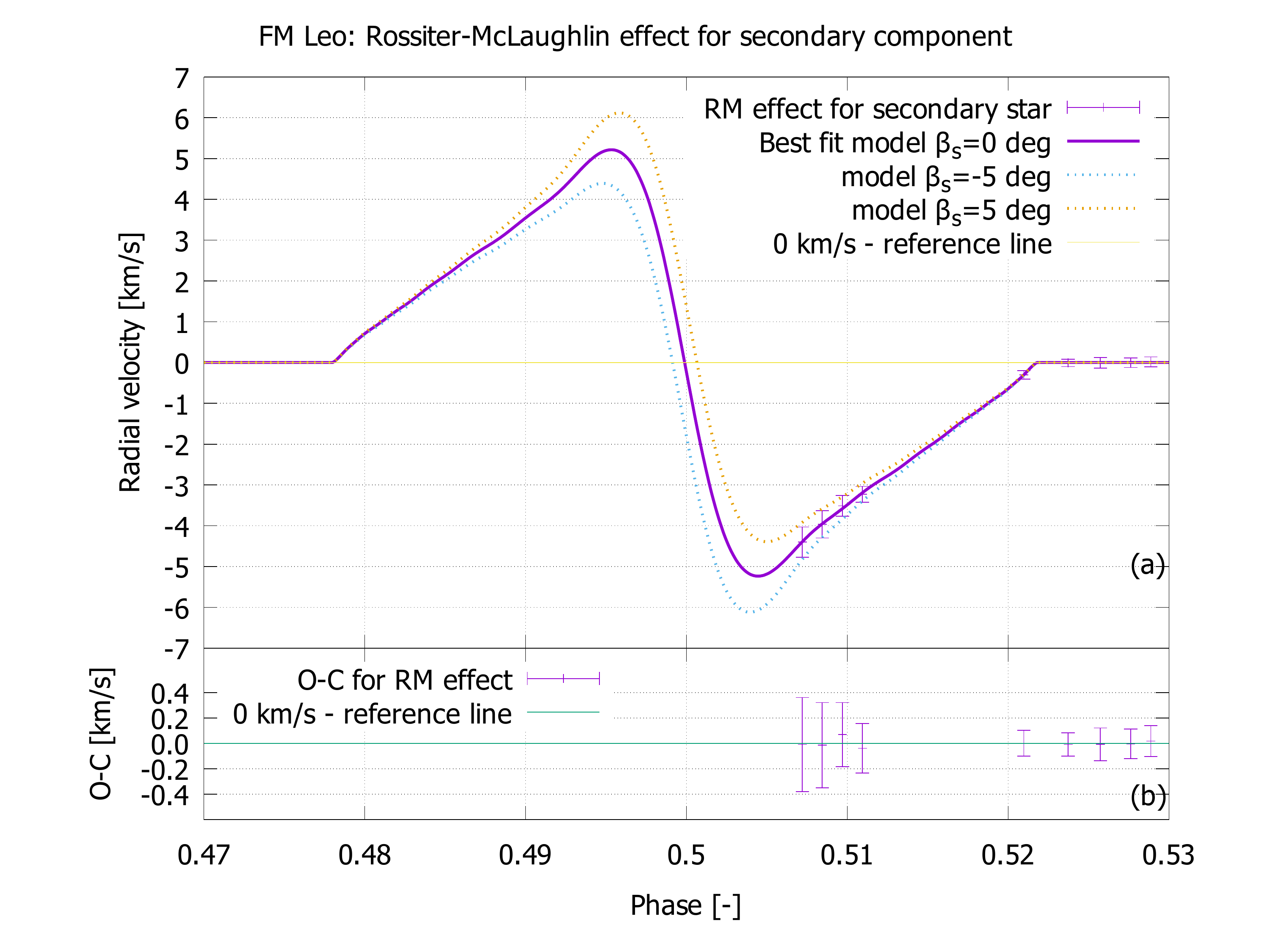}
	\caption{Rossiter-McLaughlin effect for the primary and secondary components of the FM\,Leo. In each plot the upper panel shows radial velocity as a function of phase and the lower panel shows the differences between the observed and calculated velocity for the best-fit model. Models for varying $\beta$ are also shown for comparison in the case of the primary component. In the case of the secondary component, due to the weak constrain and ambiguity caused by the correlated $\beta$ and $V_{rot}$ parameters in a system's configuration with $i$, $i_{rot}$ almost equal to 90 degrees, the models with different $V_{rot}$ are shown. In the case of the primary component the smaller size of the secondary star and inclinations deviation from the ideal edge-on case of 0.1\% is are enough to show a visible change with changing $\beta$.}
	\label{rm-FMLeo}
\end{figure*}

\begin{figure*}
	\includegraphics[width=1.7\columnwidth]{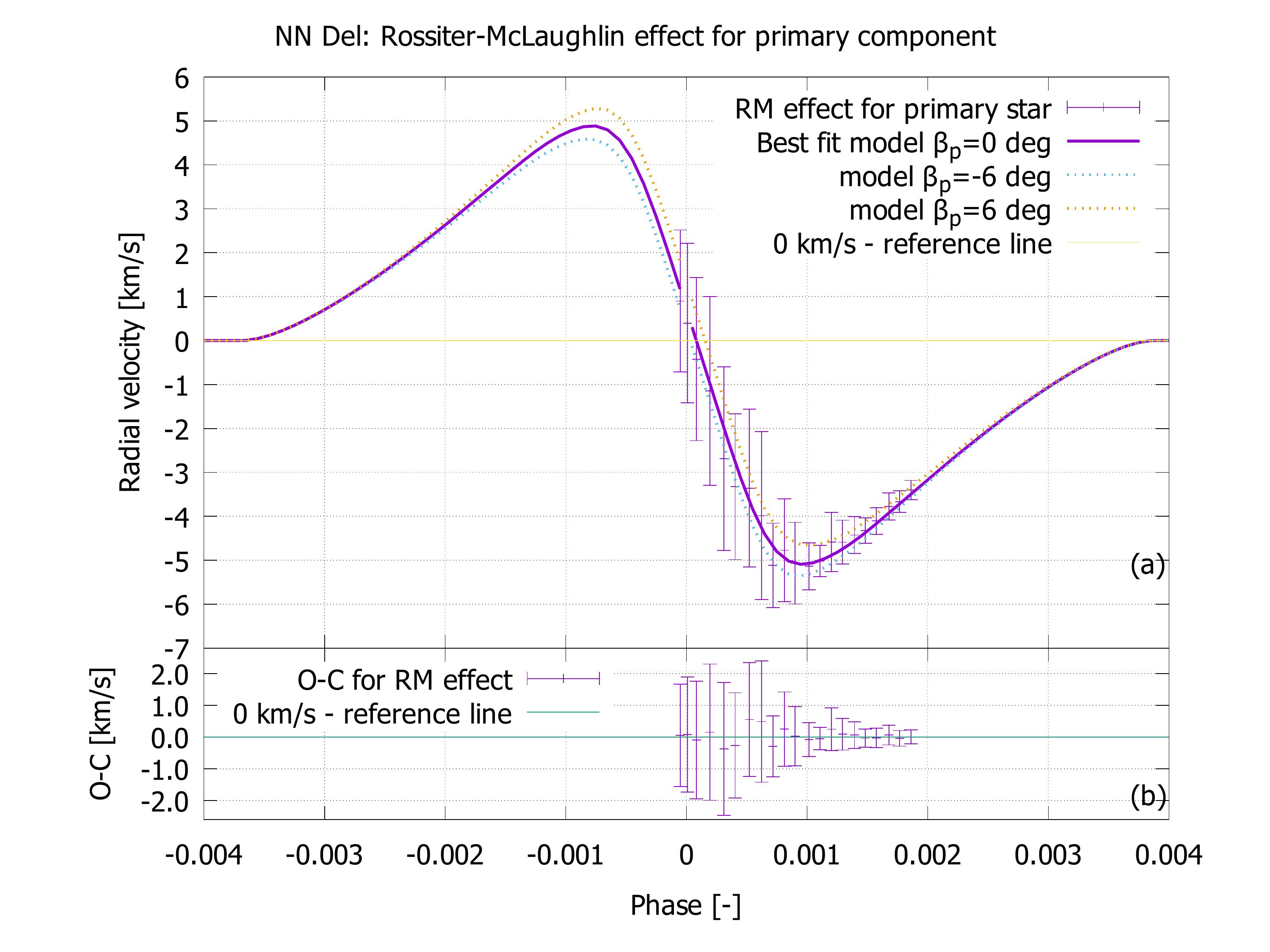}
	\caption{As in Figure \protect\ref{rm-FMLeo} but for primary component of NN\,Del binary.}
	\label{rm-NNDel}
\end{figure*}

\begin{figure*}
	\includegraphics[width=1.7\columnwidth]{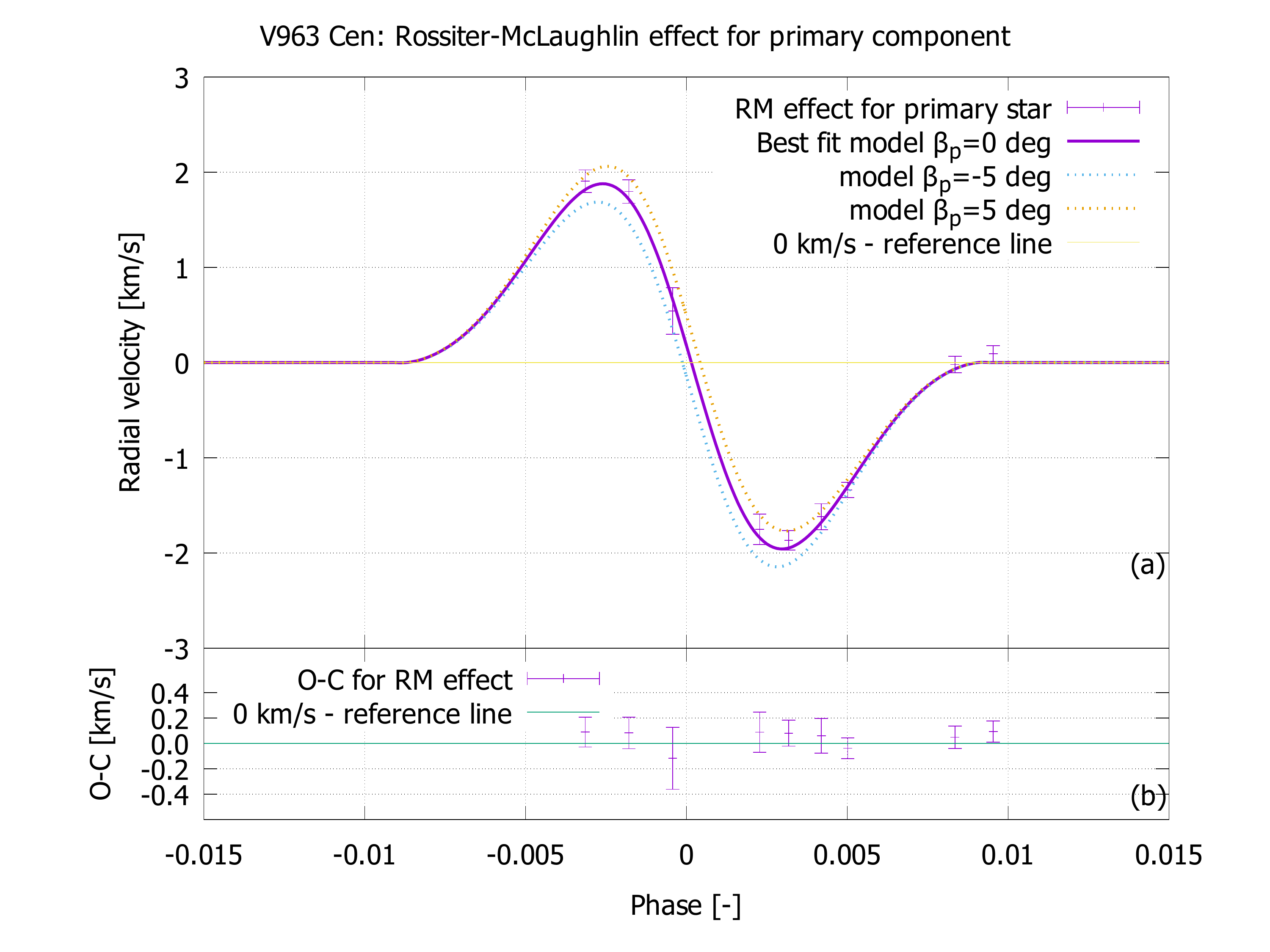}
	\caption{As in Figure \protect\ref{rm-FMLeo} but for primary component of V963\,Cen binary.}
	\label{rm-V963Cen}
\end{figure*}

\begin{figure*}
	\includegraphics[width=1.7\columnwidth]{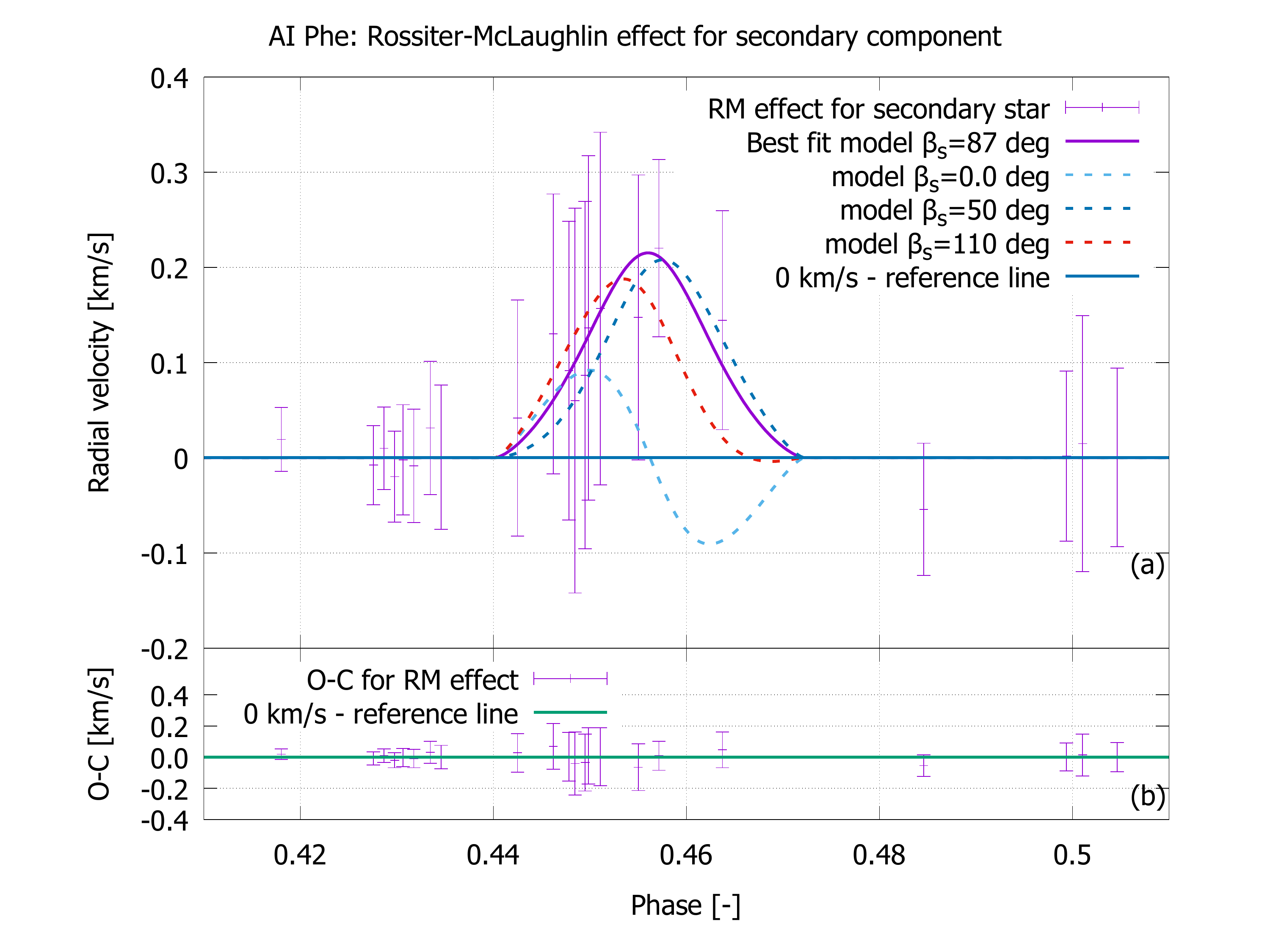}
	\caption{As in Figure \protect\ref{rm-FMLeo} but for secondary component of AI\,Phe binary.}
	\label{rm-AIPhe}
\end{figure*}

Errors for RM effect modelling were calculated with the bootstrap method based on 10\,000 iterations or more. The optimization of the model parameters was performed with HeuristicLab's genetic algorithm \citep{wagner:2014} based on Covariance Matrix Adaptation Evolution Strategy (CMA-ES) with the best single objective test function (also known as fitness function with the minimization of solution's RMS as a goal) being the weighted square root of the sum of squares of the observed minus computed values. The weights for the observed values were the estimated RV measurement errors. CMA-ES is an evolutionary algorithm which performs very well in difficult non-linear, non-convex problems in unconstrained or bounded optimization problems. However, it may also be used in simpler cases but will be much slower when compared to, for example, gradient methods. The heuristic algorithm was used due to its robustness and the ability to find a global minimum in the noisy data with multiple local minima with similar depths. Interested readers can consult \cite{goldberg:1989} for more details on genetic algorithms and \cite{hansen:2006} on CMA-ES and  HeuristicLab's approach. The algorithm also proved very successful in finding solutions in the case of AI\,Phe, where two-system orientations are almost equally possible. Figures \ref{rm-FMLeo}, \ref{rm-NNDel}, \ref{rm-V963Cen}, \ref{rm-AIPhe} depict the RM effect for the primary and/or secondary components (where available), showing models with similar $\beta$ or $V_{rot}$ for comparison. The rotation limit for the synchronous rotation of aligned components can be obtained from the binary and components' parameters. Summary of the estimates is presented in Table~\ref{rotation-parameters-table}. The estimation uses a simple formula where $V_{rot}=\frac{2\pi R}{P}$. We also used the JKTABSDIM code to conduct similar calculations and to obtain additionally the synchronization and circularisation timescales. Estimations come from \textsc{JKTABSDIM}\footnote{\textsc{JKTABSDIM} is written and supported by John Southworth. http://www.astro.keele.ac.uk/jkt/codes/jktabsdim.html }\citep{southworth:2005}, based on theoretical work of \cite{zahn:1975, zahn:1977}. The spin-orbit alignment timescale is based on the equation 41 from \cite{anderson:2017} with the constant tidal lag time $\Delta t_{tag}$ arbitrarily set to 0.1 s and star spin period $P_{\star}$ estimated on the basis of a synchronous rotation when no data was available from the RM modelling of $V_{rot}\sin{i_{rot}}$ for a given component. In the second case, the $\sin{i_{rot}}$ was assumed to be equal to one. The results are presented in Table~\ref{absdim-table}, with the source of rotation velocity estimatate provided in the last column thereof.

The $u$ linear limb darkening parameter was fitted for all components, however, no significant change in the obtained spin-orbit alignment was visible between the fixed parameters as listed in Table \ref{tab:JKTEBOP_Parameters} and the parameters from the best fit of the RM effect. The $u$ parameter is a weak constraint in our models.

\begin{table*}
\caption{Parameters of the studied binary systems collected from the literature. Values for both components are provided where available.}
\label{AggrgatedTab}
\begin{centering}
\begin{tabular}{r r r r r}
\hline
											& FM\,Leo									& NN\,Del					& V963\,Cen						& AI\,Phe										\\
\hline
Hipparcos catalogue (HIP) number			& 54766								& 102545					& 64941							& 5438											\\
Right ascension (FK5) [hh mm ss.sss]		& 11 12 45.095 $^{(1)}$				& 20 46 49.225 $^{(1)}$		& 13 18 44.357 $^{(1)}$ 		& 01 09 34.195 $^{(1)}$							\\
Declination (FK5) [dd mm ss.ss]				& +00 20 52.83 $^{(1)}$				& +07 33 10.44 $^{(1)}$		& -58 16 01.26 $^{(1)}$			& -46 15 56.09  $^{(1)}$						\\
Parallax [mas] 								& 5.88(1.01) $^{(1)}$				& 6.28(89) $^{(1)}$			& 10.72(1.10) $^{(1)}$			& 3.50(1.04) $^{(1)}$						\\
											& 7.29(64) $^{(2)}$					& 							& 								& 5.98(31) $^{(12)}$						\\
Distance [pc]								& 130.4(25.5) $^{(6)}$ 				& -							& 93 $^{(2)}$					& 68 $^{(2)}$; 162(6) $^{(6)}$       							\\
											& 137.2(11.7) $^{(12)}$				&							& 								& 167.3(8.8)	$^{(12)}$ 							\\
											& 									&							& 								& 173(11) $^{(13)}$						\\
											&	  								&							& 								& 170(9) $^{(22)}$ 						\\
Mean radial velocity [km $\textrm{s}^{-1}$]	& 11.87(13) $^{(9)}$				& -							& -30.7(1) $^{(3)}$		& -2.2(1) $^{(3)}$ 						\\
											& 									&  							& 								& -1.91(7) $^{(6)}$ 						\\
Spectral type								& F7V C $^{(8)}$					& F8 D $^{(16)}$ 			& G2V C $^{(4)}$				& G3V C  $^{(7)}$; G8  $^{(13)}$ 								\\
											& 									& 							& 								& K0IV, F7V  $^{(11)}$ 							\\
Period [days]								& 6.728606(6) $^{(9)}$				& 99.26840000  $^{(5)}$		& 15.26938900  $^{(5)}$			& 24.592325(8) $^{(13)}$ 						\\
											& 6.72863 $^{(18)}$					& 							& 15.2693 $^{(18)}$				& 24.59241(8) $^{(14)}$ 						\\
											& 									& 							& 								& 24.592483(17) $^{(10)}$ 						\\
											& 									& 							& 								& 24.5923(1) $^{(21)}$ 						\\
Apsidal motion period [years]				& -									& 						-   & 50000.000  $^{(5)}$			& - 											\\
Semimajor axis of apparent orbit [R$_{\odot}$]& 20.631(52) $^{(9)}$				& - 						& -								& 47.80(5)$^{(6)}$ 						\\
Orbit eccentricity							& 0 $^{(9)}$						& 0.51759(2) $^{(19)}$		& 0.42000 $^{(5)}$ 			& 0.1821(51) $^{(10)}$							\\
											& 									&	 					 	& 								& 0.186 $^{(6)}$								\\
											& 									&	 					 	& 								& 0.188(2) $^{(13)}$							\\ 
											& 									&	 					 	& 								& 0.187(4) $^{(14)}$							\\
Inclination of orbit [deg]					& 87.98(6) $^{(9)}$					& 89.488(3) $^{(19)}$		& - 							& 88.45 $^{(6)}$ 								\\
											& 									& 	 						& 	 							& 88.45(5) $^{(13)}$		 					\\
											& 									&  							& 	 							& 84.4(5) $^{(14)}$								\\
											& 									&  							&  	 							& 88.502(39) $^{(10)}$							\\		
Radius [R$_\odot$]							& 1.648(43), 1.511(49) $^{(9)}$		& - 						& -								& 1.816(24), 2.930(48)  $^{(6)}$	\\
											& 									& 	 						& 								& 1.818(24), 2.932(48)  $^{(11)}$	\\
											& 									& 							& 								& 1.816(24), 2.930(48) $^{(13)}$  				\\
											& 									& 							& 								& 1.82(5), 2.81(7) $^{(14)}$  					\\
											& 									& 							& 								& 1.835(14), 2.912(14) $^{(10)}$  				\\
											& 									& 							& 								& 1.762(7), 2.931(5) $^{(22)}$  				\\
log(Luminosity) [L$_{\odot}$] 				& 1.806(86), 1.617(102) $^{(6)}$	& - 						& -								& 0.687(44), 0.672(43) $^{(11)}$	\\
												& 									&  						&  								& 0.73(5), 0.64(4) $^{(22)}$	\\
log(Surface gravity) [cm/s2]				& 4.124(23), 4.189(28) $^{(9)}$		& - 						& 4.02 $^{(3)}$					& 3.24 $^{(3)}$ 								\\
											& 4.124(23), 4.189(28) $^{(17)}$	&  						& 								& 3.996(11), 3.595(14)$^{(11)}$ 	\\
											& 									&  						& 								& 3.985(8), 3.605(4)$^{(17)}$ 	\\
											& 									&  						& 								& 3.997(12), 3.596(14)$^{(13)}$ 	\\
											& 									&  						& 								& 4.021(4), 3.593(3)$^{(22)}$ 	\\
Effective temperature [K]					& 6316(240), 6190(211) $^{(9)}$		& 6500, 6437(5) $^{(19)}$	& 5597 $^{(2)}$					& 5559 $^{(2)}$ 								\\
											& 									&	 						& 								& 5588(127) $^{(3)}$      					\\
											& 									&							& 5815(80) $^{(3)}$ 			& 6310(150), 5010(120)$^{(13)}$  		\\
											& 									&							& 								& 6310(150), 5151(150)$^{(22)}$											\\
V magnitude [mag] 							& 8.45(1)	$^{(15)}$				& 8.40(1) $^{(15)}$			& 8.59(1)	$^{(15)}$			& 8.60(1)	$^{(15)}$						\\
Age [Gyr]									& $\sim$\,2.5-4.0 $^{(9)}$			& - 						& 9.9$_{8.2}^{12.0}$ $^{(2)}$	& 13.5$_{11.0}^{16.3}$ $^{(2)}$ 				\\
											& 	 								& 	 						& 								& 4.39(32) $^{(10)}$ 						\\
											& 	 								& 	 						& 								& 3.98 $^{(13)}$ 								\\
Mass [M$_{\odot}$]							& 1.318(7), 1.287(7) $^{(9)}$		& 1.328(21), 1.454(25)$^{(20})$& (est.) 1.1/1.1	& 1.197(5), 1.238(4) $^{(6)}$  	\\
											& 1.318(11), 1.287(10) $^{(17)}$	& 							& 								& 1.1973(37), 1.2473(39) $^{(10)}$\\
											& 									& -							& -								& 1.1934(41), 1.2336(45) $^{(11)}$\\
											& 									& 							& 								& 1.1954(41), 1.2357(45) $^{(13)}$  			\\
											& 									& 							& 								& 1.2095(44), 1.2600(46) $^{(14)}$  			\\
											& 									& 							& 								& 1.193(4), 1.242(4) $^{(17)}$  	\\		
											& 									& 							& 								& 1.190(6), 1.231(5) $^{(22)}$  	\\		
$[Fe/H]$ [dex]								& -									& -							& -0.25 $^{(2)}$ 				& -0.15 $^{(2)}$  								\\
											& 									& 							& -0.05 $^{(3)}$ 				& -0.01 $^{(3)}$  								\\
											& 									& 							& 								& -0.14(10) $^{(13)}$  					\\
											& 									& 							& 								& -0.14(10) $^{(22)}$ 					\\
$[M/H]$ [dex]								& -									& -							& -0.02 $^{(3)}$ 				& 0.02 $^{(3)}$  								\\
$[\alpha/H]$ [dex]							& - 								& - 						& 0.04 $^{(3)}$					& 0.09 $^{(3)}$ 								\\

\hline
\end{tabular}
\end{centering}\\
\vspace{2mm} %5mm vertical space
\footnotesize{Notes: (1): \protect\cite{vanleeuwen:2007}, (2): \protect\cite{holmberg:2009}, (3): \protect\cite{casagrande:2011}, (4) \protect\cite{houk:1975}: , (5): \protect\cite{bulut:2007cat}, (6): \protect\cite{eker:2014}, (7): \protect\cite{houk:1978}, (8): \protect\cite{houk:1999}, (9): \protect\cite{ratajczak:2010}, (10): \protect\cite{kirkbykent:2016}, (11): \protect\cite{torres:2010}, (12): \protect\cite{stassun:2016}, (13): \protect\cite{andersen:1988}, (14): \protect\cite{helminiak:2009} (uncertainties on masses from \protect\citealt{kirkbykent:2016}), (15): \protect\cite{hog:2000}, (16): SIMBAD astronomical database (\protect\cite{wenger:2000}), (17) \protect\cite{graczyk:2017}, (18): \protect\cite{otero:2003}, (19): \protect\cite{gomez:2003}, (20): \protect\cite{griffin:2014}, (21): \protect\cite{reipurth:1978}, (22): \protect\cite{milone:1992}.}
\end{table*}

\section{Results} 
\label{sec4}
\begin{figure*}
	\includegraphics[width=0.8\textwidth]{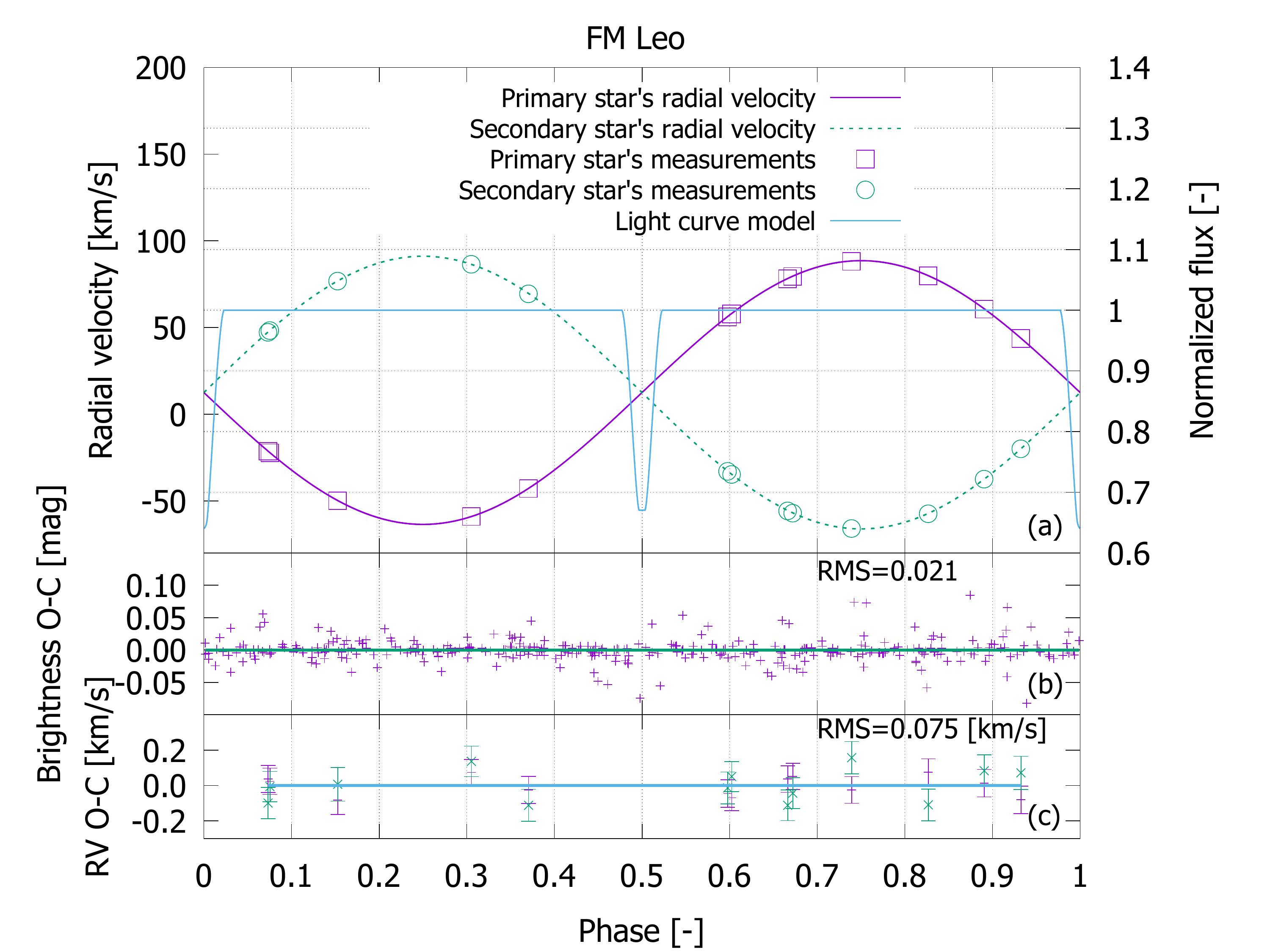}
	\includegraphics[width=0.8\textwidth]{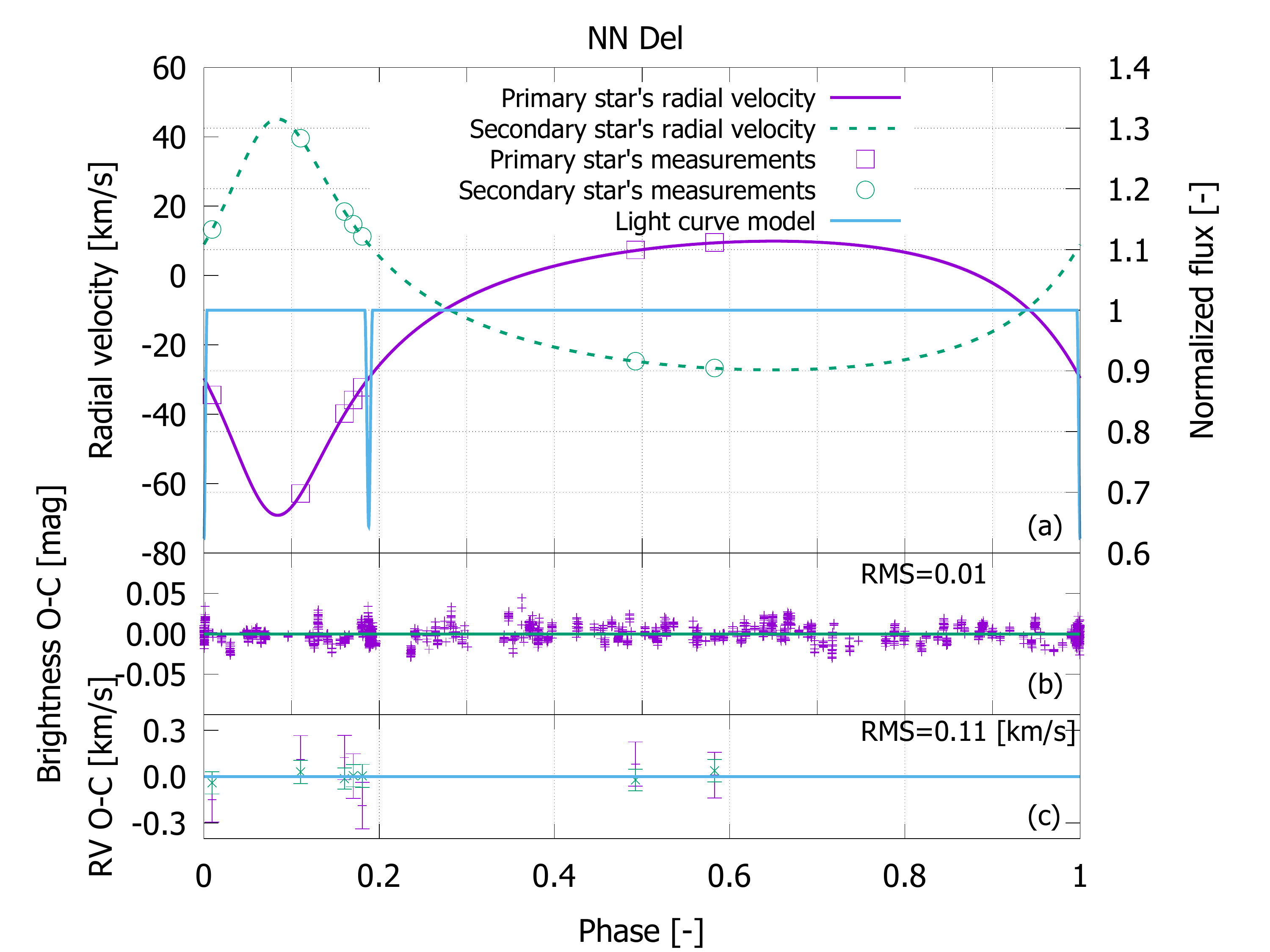}
	\caption{Radial velocity curves and ligthcurves for FM\,Leo and NN\,Del. In each of the two plots the top panels (a) show the model of the radial velocity curves of the primary (solid purple line) and secondary (dashed green line) components and the modelled lightcurve (solid light blue line), while the middle (b) and bottom panels (c) show differences between models and observations for the lightcurve and the radial velocities, respectively.}
	\label{grid13-a}
\end{figure*}

\begin{figure*}
	\includegraphics[width=0.8\textwidth]{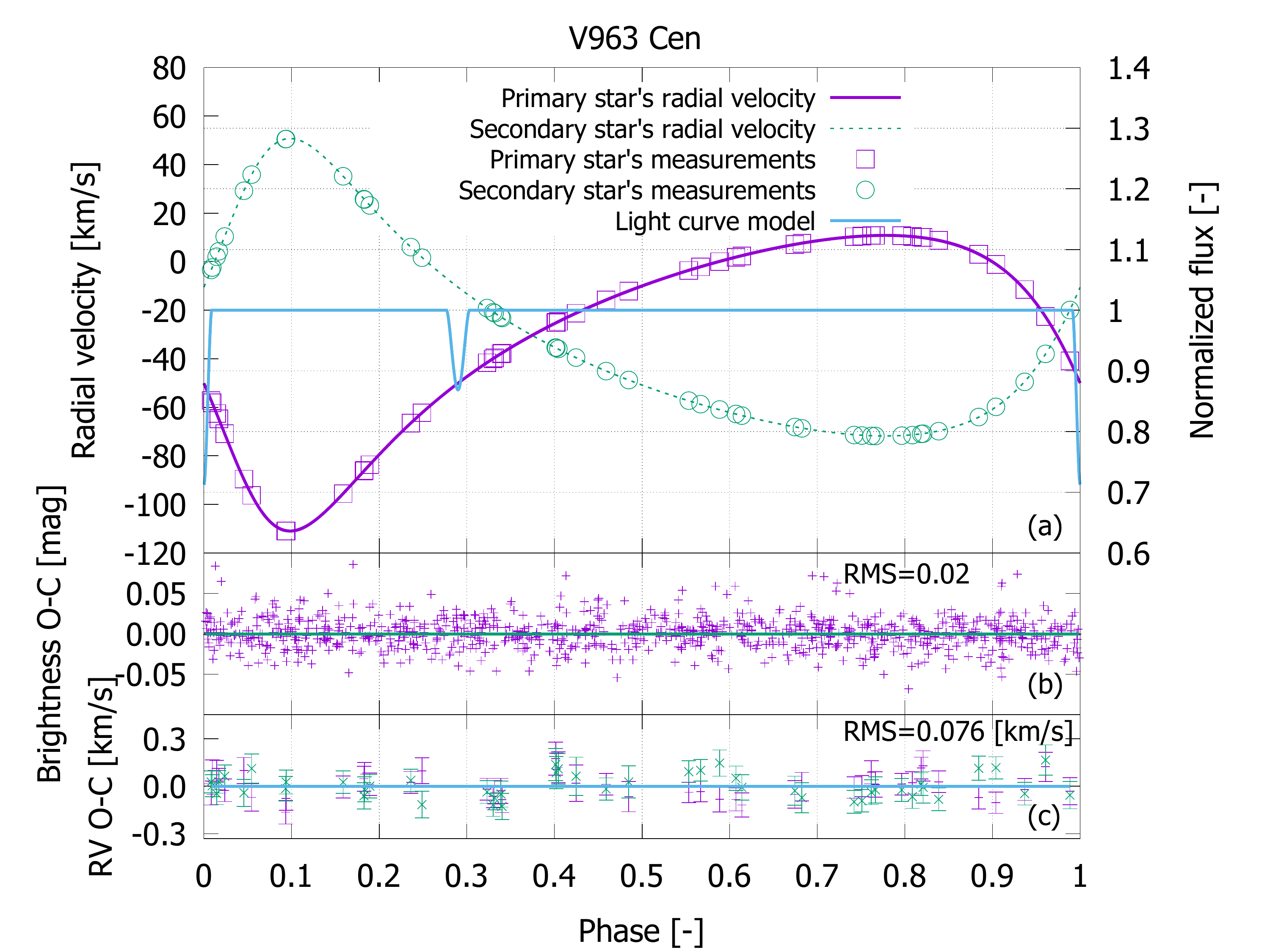}
	\includegraphics[width=0.8\textwidth]{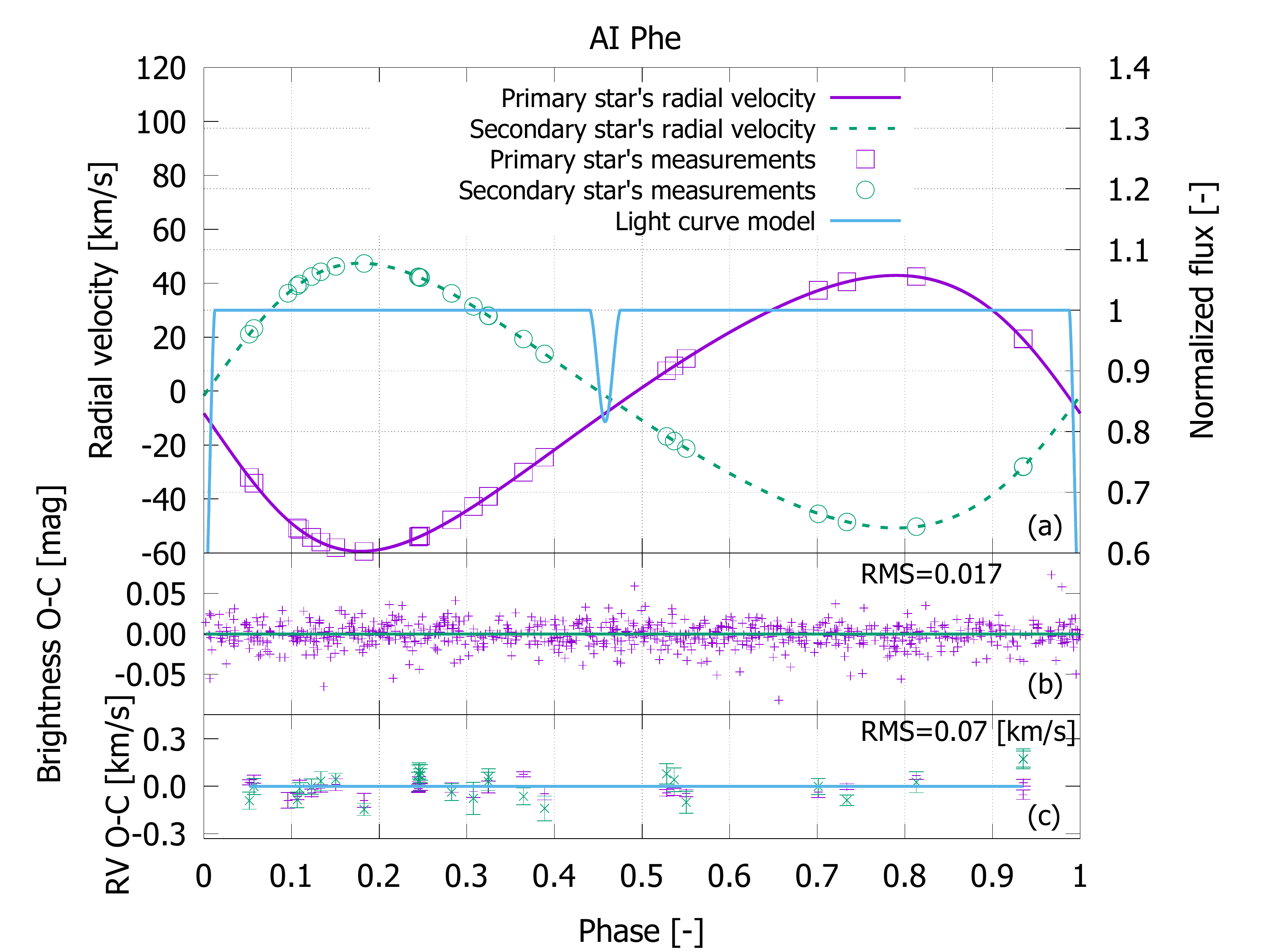}
	\caption{As in Figure \protect\ref{grid13-a} but for V963\,Cen and AI\,Phe.}
	\label{grid13-b}
\end{figure*}

Table~\ref{AggrgatedTab} lists the parameters of the four studied systems gathered from the literature. They are in agreement with the results of our modelling presented in Table~\ref{tab:JKTEBOP_Parameters}. These values, calculated from the source data with uniform tools, were used as input values for our RM effect modelling. The radial velocity and photometric measurements with their respective models are presented in Figures \ref{grid13-a} and \ref{grid13-b}. Figures \ref{fig:fmleo-bfGrid1}, \ref{fig:nndel-bfGrid}, \ref{fig:v963cen-bfGrid}, \ref{fig:aiphe-bfGrid} show the condensed view of the BFs in eclipse, RV measurements from the orbit's perspective and examples of spectra processing steps with the BF for the observed spectrum example for FM Leo, NN Del, V963 Cen and AI Phe, respectively. The results of the calculations are presented in Table~\ref{tab:JKTEBOP_Parameters} for typical binary systems parameters and for RM effect modelling in Table~\ref{rm-table}.

We were able to measure spin-orbit alignment for five stars from the binary systems presented in this study with Rossiter-McLaughlin effect. Four of them show clear alignment and the fifth one is an interesting example of non-alignment. 

\begin{figure*}
	\includegraphics[width=1.7\columnwidth]{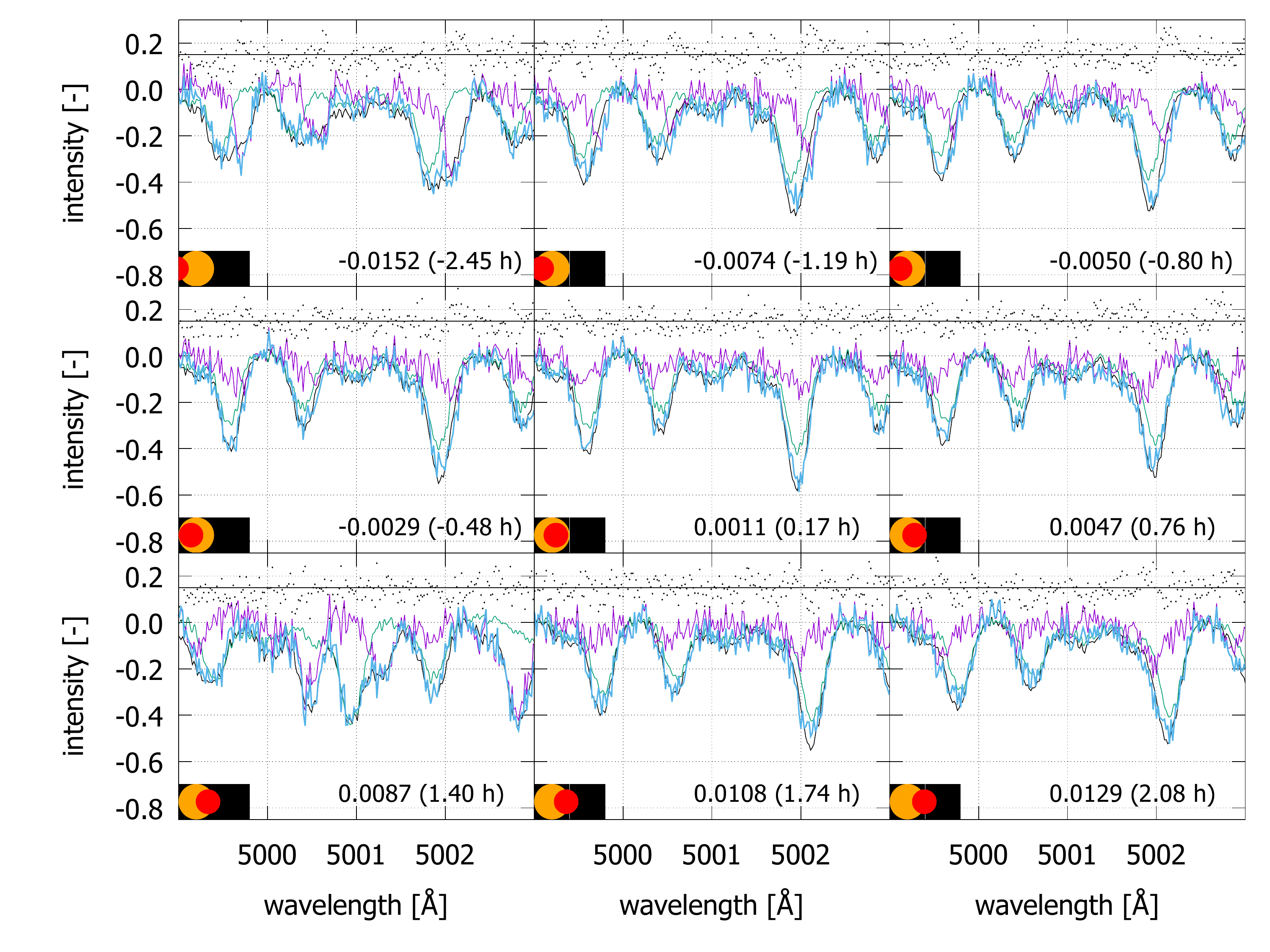}
	\caption{Selected part of the echelle spectrum during nine phases of a primary eclipse of FM\,Leo. The primary component is shown in orange and the secondary one in red. The colours are arbitrarily selected and are not based on the components' temperature. In the bottom part of each panel we quote the eclipse phase, with the time elapsed since the maximum eclipse given in parenthesis. The observed spectrum is marked with light blue, the front and background component in green and purple, respectively. The background component's spectrum is calculated from the difference of the scaled front component's spectrum and observed spectrum of the binary. Dotted black line shows the fit of the two disentangled spectra to the observations with scaled intensity. This line is used as an illustration on what the sum of spectra would look like if we ignored the fact that part of the background component is obscured and only a portion of light coming from it has been reduced. Outside the eclipse the sum of disentangled spectra is within the error limits equal to the observed spectra. Dots in the upper part of each panel illustrate the difference between the observed spectrum and the sum of the two disentangled spectra - the primary component's intensity is reduced in the summing, on the base of the light curve. The intensities shown are in arbitrary units.}
	\label{grid33-fmleoprimary}
\end{figure*}

\begin{figure*}
	\includegraphics[width=1.7\columnwidth]{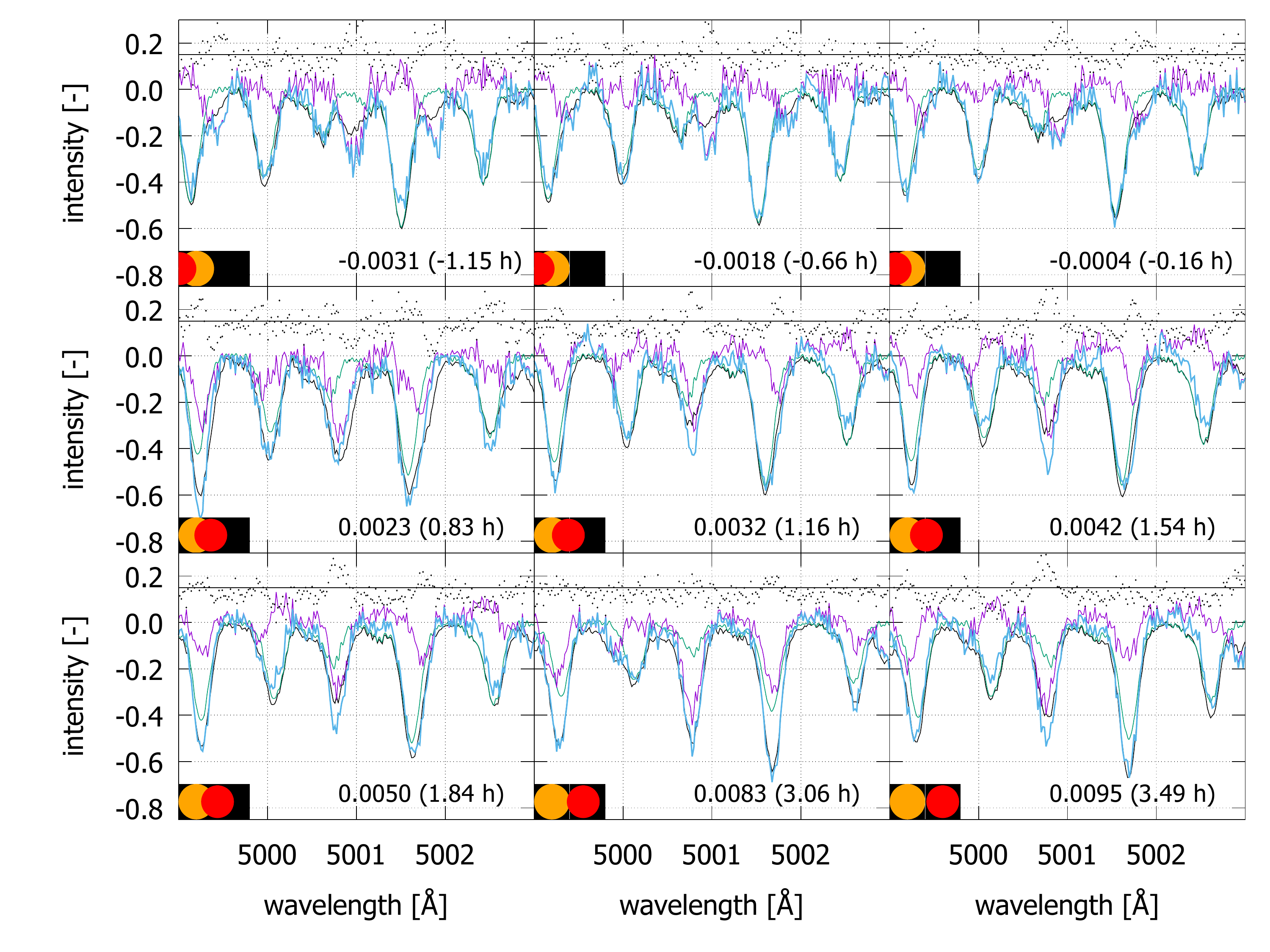}
	\caption{As in Figure \protect\ref{grid33-fmleoprimary} but for primary component of V963\,Cen.}
	\label{grid33-v963cen}
\end{figure*}

%FM\,Leo Primary and V963\,Cen
The very good coverage of the FM\,Leo binary primary eclipse leaves no doubt to the fact that spin and orbital axes are aligned. The primary eclipse of V963\,Cen is a similar case, even if the number of available measurements is only a half of the measurements available for the FM\,Leo primary. Examples of the measured spectra and their models in the eclipse for both systems are presented in Figures \ref{grid33-fmleoprimary} and \ref{grid33-v963cen}.

\begin{figure*}
	\includegraphics[width=1.7\columnwidth]{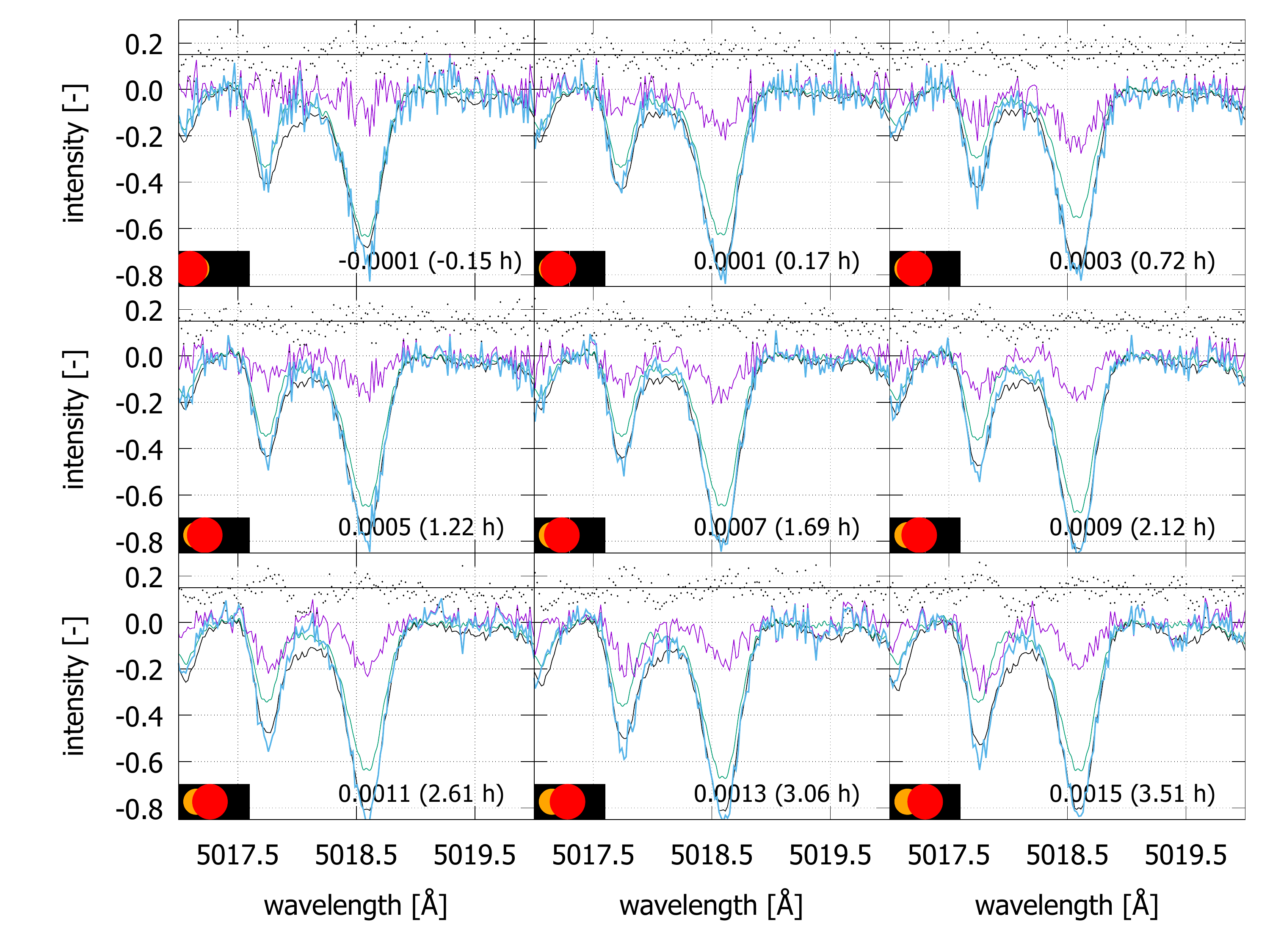}
	\caption{As in Figure \protect\ref{grid33-fmleoprimary} but for primary component of NN\,Del.}
	\label{grid33-nndel}
\end{figure*}

%NN\,Del Primary
NN\,Del is a system with a long period of almost 100 days and high cadence of measurements is possible for it. However, due to the weather conditions and using only one observatory, we obtained observations only for slightly over a half of the eclipse. Nevertheless, due to the high precision of the measurements and a good coverage of the mid point of the eclipse the alignment is also confirmed and the errors for $\beta_p$ and $V_{rot}\sin{i_{rot}}$ describing the spin-orbit alignment are 3.4 degrees and 3.8 km~$\textrm{s}^{-1}$ respectively. Example spectra measurements are presented in Figure \ref{grid33-nndel}. The synchronisation and circularisation timescales for this system presented in Table \ref{absdim-table} are close to 12 and 16 Gyr, respectively. Taking into account the system's spectral type of F8, calculated components' masses, and investigation conducted by \cite{gomez:2003} suggesting an age smaller than a few Gyr, we may initially assume that the binary is close to its primordial state in terms of spin axes alignment.

\begin{figure*}
	\includegraphics[width=1.7\columnwidth]{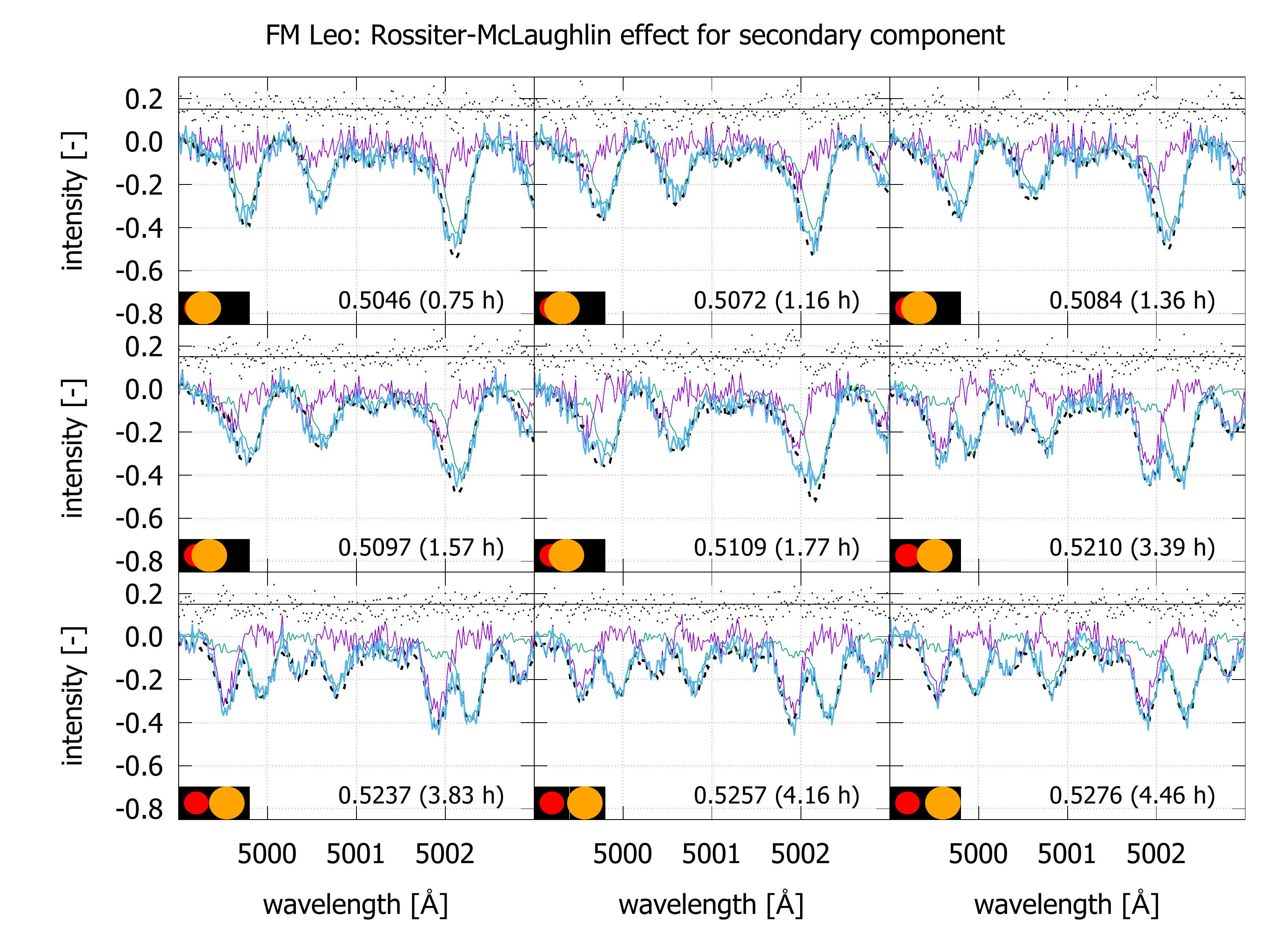}
	\caption{As in Figure \protect\ref{grid33-fmleoprimary} but for secondary component of FM\,Leo.}
	\label{grid33-fmleosecondary}
\end{figure*}

%FM\,Leo Secondary
The secondary component of FM\,Leo has the spin-orbit alignment degenerated due to the inclination of the orbit $i$ almost equal to 90 degrees. This is clearly visible in the equation (4) for the rotational parameter $V^{*}$ from \cite{gimenez:2006c}:

\begin{equation}
V^{*} = V_{rot}\sin{i_{rot}}(\sin{\beta}\cos{i}\cos{\theta}-\cos{\beta}\sin{\theta})
\end{equation}

\noindent where $V_{rot}$\footnote{Note that to denote equatorial rotation we have replaced $V$ of the original publication with $V_{rot}$ used in this work.} denotes equatorial rotation and $\theta$ denotes phase of the orbit. With first element $\cos{i}$ equal to zero, the dependency on $\beta$ is included via $\cos{\beta}$ element which is symmetrical in the context of distance from 90 degrees. In this case, the $\beta$ parameter is entangled with $V_{rot}$ parameter in the FM\,Leo secondary component. In fitting the parameters we decided to fix the $\beta$ parameter to 0 degrees and fit only the $V_{rot}$. With the limited number of observations we cannot completely rule out the misaligned case for the secondary component of FM\,Leo. However, with a small separation of components, short timescale for spin-orbit alignment (see Table \ref{absdim-table}) of 0.2 and 0.66 Myr when compared to the age of the system, the misalignment seems unlikely. The higher the misalignment, the higher the required equatorial rotational velocity of the star to explain the observations. With a synchronous rotation speed of 13.5 and 9.3 km $\textrm{s}^{-1}$ at the equator for the primary and secondary component, respectively (see Table \ref{rotation-parameters-table}), it seems that both components have measured values close to the these limits (see Table \ref{rm-table}). But the simplest explanation of the aligned spin-orbit is the most plausible solution with the lowest fit error. More measurements are required for a higher confidence measurement. In the case of a circular orbit and no trace of the third body in the system, the circularisation of the orbit and alignment of components' axes follows the expected theoretical evolution of the system. An example of the spectra order in the different phases of eclipse is presented in Figure \ref{grid33-fmleosecondary}.

\begin{figure*}
	\includegraphics[width=1.7\columnwidth]{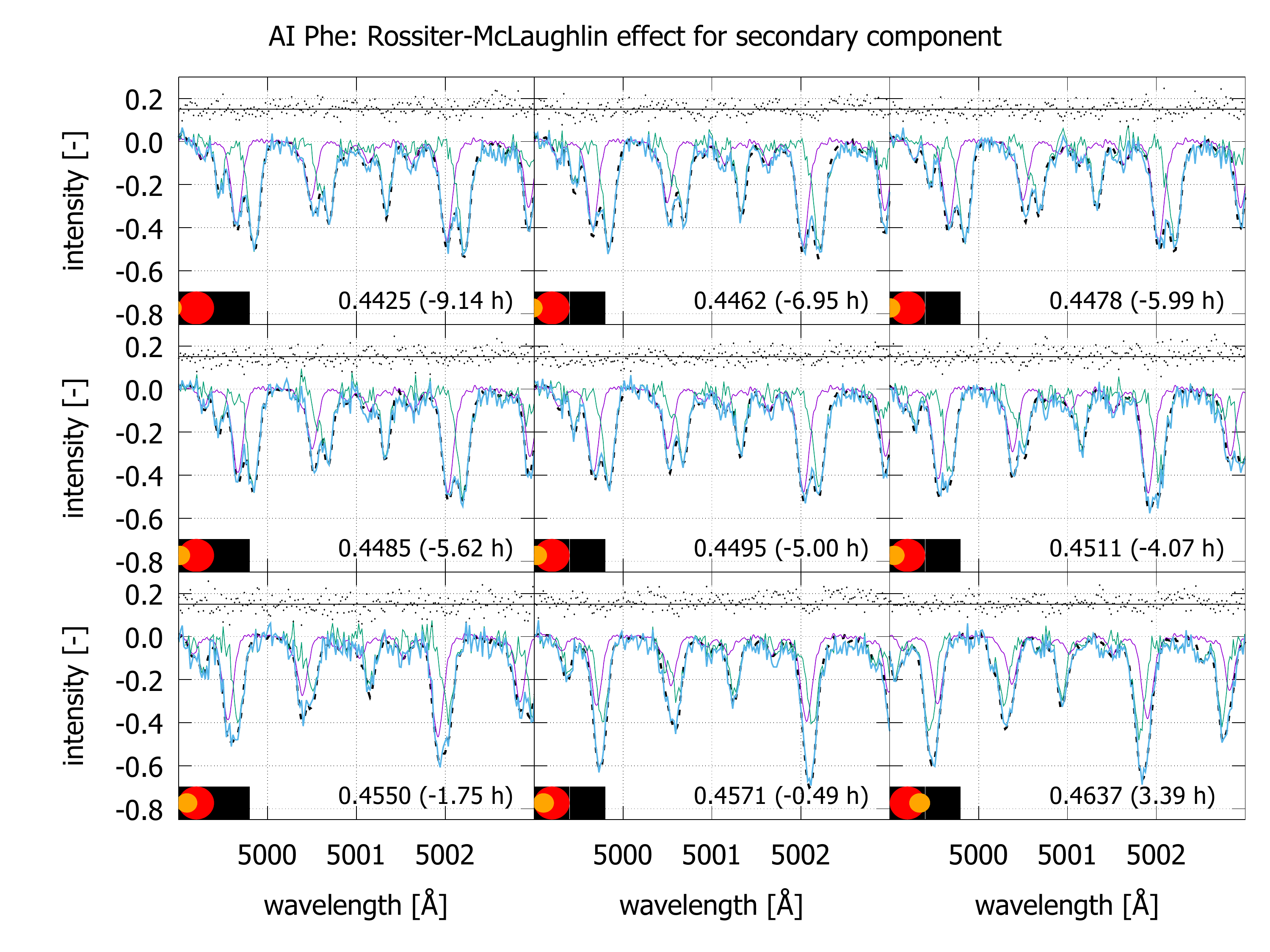}
	\caption{As in Figure \protect\ref{grid33-fmleoprimary} but for secondary component of AI\,Phe.}
	\label{grid33-aiphesecondary}
\end{figure*}

%AI\,Phe Secondary
Radial velocity data for the secondary component of AI\,Phe in eclipses shows a very low amplitude of the RM effect, definitely smaller than 1 km $\textrm{s}^{-1}$. This may be a sign of a misalignment, as for an aligned spin-orbit and synchronous rotation the velocity on the equator would reach 6 km $\textrm{s}^{-1}$. Moreover, all points are on the same side of the system's radial velocity curve, there is no point symmetry against the middle of eclipse, which is visible in other systems. These two properties suggest misaligned axes. Modelling of the RM effect was difficult due to the large errors, a small amplitude of the RM effect and the lack of a clear minimum in the $\chi^2$ values for the best fits. Multiple local optima exist and due to the small number of measurements and a low SNR there is no single global set of parameters. In our case the $\beta_s$ parameter can have two values - 87.5 or 87 degrees with a~similar RM effect. Complete freedom of the parameters during the fitting showed a tendency of the $e$, $\omega$, $T_0p$, $i$ parameters describing the orbit to slightly prefer higher $e$. However, the RMS decrease by 0.5\% for $e=0.2$ results in a model completely not fitting the radial velocity and light curves. The simultaneous fitting of all models and parameters would remove this trend but the necessary changes of the codes for LC, RV and RM fitting is beyond the scope of this work and paper. The ambiguity of solutions may be resolved with an increased SNR and more observations. At the moment we present two possible models in Table~\ref{rm-table}, with a slightly better fit in the case of the second model. The measured $\beta_s$ angle has a high uncertainty and so does not allow to draw definite conclusions. With a limited number of measurements and the fact that they come from three different nights separated by almost one hundred days, the result requires more observations and further investigation. Only one point is after the mid-eclipse. The misalignment hypothesis may be supported by the finding of a third body on a wide orbit around the binary (Konacki et al., in preparation) and explains the deviation in eclipse ephemeris noted by \citet{kirkbykent:2016}. The tertiary may be responsible for the so-called Kozai cycles, where the inner orbit's eccentricity is periodically modulated, and the orientation of rotation axes may be perturbed. Using the formalism from \cite{fabrycky:2007} we found that the Kozai cycles period is a few orders of magnitude shorter than the time scale of inner orbit's precession ($\sim$20~kyr vs. $\sim$20~Myr; assuming $P_{out}\sim20$~yr and that the tertiary is a late-M type star), which is the main condition for Kozai cycles to occur. Another one --  that the mutual inclination is large enough: $\cos^2(i_c)=3/5$ -- currently cannot be verified. A full orbital solution is not yet available (He\l miniak, private communication), but a long-term monitoring is being carried out. An example of one spectra order of the AI\,Phe in the different phases of an eclipse is presented in Figure~\ref{grid33-aiphesecondary}.

\section{Summary and conclusions}
\label{sec5}
We have observed four eclipsing binary systems: three (V963\,Cen, NN\,Del, AI\,Phe) characterized by high orbital eccentricity and one (FM\,Leo) with a circular orbit, with the view to determining their stellar and orbital parameters and, primarily, the relative orientation of their spin axes and the rotation axes of the systems. In the process we have developed a set of tools needed for the calculation of the RM effect for the eclipsed component from its spectra and deriving its accurate spin-orbit angle.

Based on our results, no significant spin-orbit misalignment is seen in three out of the four studied systems, the exception being the secondary component of AI\,Phe where $\beta_s = 87 \pm 17^o$. The misalignment which we have found likely results from the presence of a third body in the system (Konacki et al., in preparation). The stellar and orbital parameters obtained in the process increase -- from nine to fourteen -- the number of stars for which the $\beta$ parameter has been measured with errors. The theory of the spin-orbit alignment timescale seems to be satisfied in the case of FM\,Leo and V963\,Cen where the tidal dissipation impact on the spin-orbit alignment has a timescale of 0.2 - 0.66 Myr and 1.6 - 3.2 Myr, respectively, much shorter than the estimated ages of the systems. NN\,Del may represent a primordial orientation of the spin-orbit alignment. The system has a large separation of components relative to their radii, and all timescales presented in Table~\ref{absdim-table} for the primary component of NN\,Del are larger than 10 Gyr, definitely much larger than the age of the system. Even in the case of V963 Cen, where a measurement is only available for one binary component, we can expect the other component to be aligned due to tidal interactions. The primary component of AI Phe and the secondary component of NN Del may have $\beta$ angles significantly different from zero  due to the possible third body influence in the first case and a primordial difference in the second case. Further study of especially these two systems may provide important additional data for the subject of third-body influence and primordial spin-orbit orientations with minimal observing effort.

There is no agreement in the literature on the role of tidal dissipation in SOM removal in binary systems, or on whether the SOM effect is primordial or results from dynamical interactions with a perturber (see e.g. \citealt{anderson:2017} and references therein). A number of recent works have investigated the issue for stellar binary systems and/or systems hosting hot Jupiters.

\cite{fabrycky:2007} investigate the issue of orbital shrinking of stellar binary systems and star-planet systems, which is believed to affect the majority of them. From observations it is known that most or all short-period binaries have distant companions (tertiaries) and the authors show that long-period systems can evolve into short-period ones through a combined effect of the perturbations of a tertiary (Kozai oscillations) and tidal friction. In our case, only FM Leo satisfies the Fabrycky et al. criterion for a short-period binary (P $<$ 10 days). The other three studied systems have longer periods which could, in principle, mean that they may undergo further orbital shrinking if such a tertiary companion is indeed present in the system, which may be true for AI Phe. For the remaining systems, there is currently insufficient data available.

\cite{hale:1994} investigated co-planarity of equatorial and orbital planes in 73 systems and found out that solar-type binary systems with a separation less than 30-40 AU, a range where all our objects fall in, tend to be aligned. Our sample confirms this finding but is also an edge case when compared to \cite{hale:1994} sample with much larger separations in the range of 1-100 AUs.

For star and hot Jupiter interactions, \cite{albrecht:2012c} shows a correlation between the magnitude of the misalignment and tidal dissipation timescales, confirming that for short timescales the systems are aligned and for longer ones show a random distribution. \cite{anderson:2017} note it is possible to obtain obliquity and eccentricity in a binary system by including an effect of a third companion, assuming that a given system has a sufficiently large pericenter distance so as to make any possible tidal influence negligible.

\cite{lin:2017} also note that in binary systems, spin-orbit alignment should happen faster than orbit circularisation due to orbital angular momentum being much higher than spin angular momentum. Under this scenario, we would not expect to find many non-aligned systems with zero orbital eccentricity.  However, \cite{albrecht:2014} provide an example of just such a system, with a projected obliquity of -52 deg for the primary component, hypothesizing a third body's influence as the likely cause. Our candidate for a spin-orbit misaligned system, AI Phe, has an eccentricity of 0.188 and an estimated $\beta_s$ of -74 deg. Our upcoming work (Konacki et al., in preparation) confirms the presence of a third body in the system, providing a direct support to the argument that similar combinations of obliquity and eccentricity values in binary systems may be due to the influence of a third component, influencing the evolution of those parameters through e.g. the Lidov-Kozai effect.

Further observations of the remaining components to complement the current study would nevertheless be useful and could take advantage of the already existing analysis pipeline and the availability of disentangled spectra of both binary components. On the other hand, obtaining data for new systems would require at least a dozen measurement points outside eclipses, obtained with the same instrument and configuration. More observations for AI\,Phe primary and secondary components in eclipse are crucial for solving the puzzle of small rotational speed on the equator as well as the possible high spin-orbit inclination of the secondary and possibly also the primary component. Additional measurements of NN\,Del are necessary to measure the $\beta$ parameter for the secondary component, which will be important for the primordial alignment and formation analysis. Right now it remains an open question whether the orbital elements reflect the properties of the protostellar cloud, or result from post-formation dynamical evolution. Such new data for the sample studied here, combined with the existing data set, has the potential to significantly contribute to our knowledge on the source of misaligned spin-orbits in binary systems. We continue our photometric and spectroscopic study of these binaries and a separate publication will cover this aspect.

%FM Leo
\begin{figure*}
	\centering
	\begin{subfigure}[b]{.45\textwidth}
		\includegraphics[width=\linewidth]{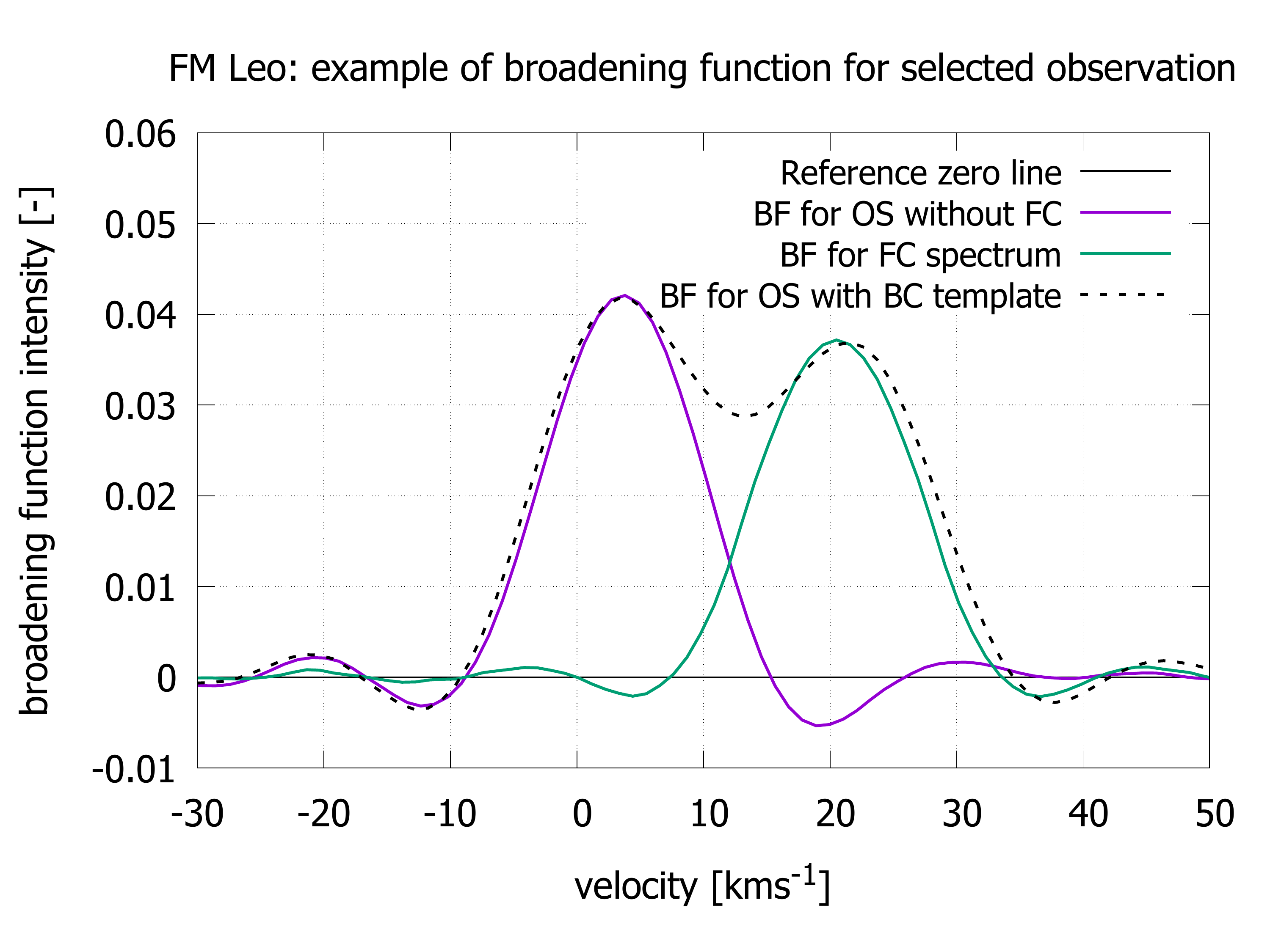}
		\caption{}\label{sfig:fmleo-bfGrid-a}
	\end{subfigure}
	\begin{subfigure}[b]{.45\textwidth}
		\includegraphics[width=\linewidth]{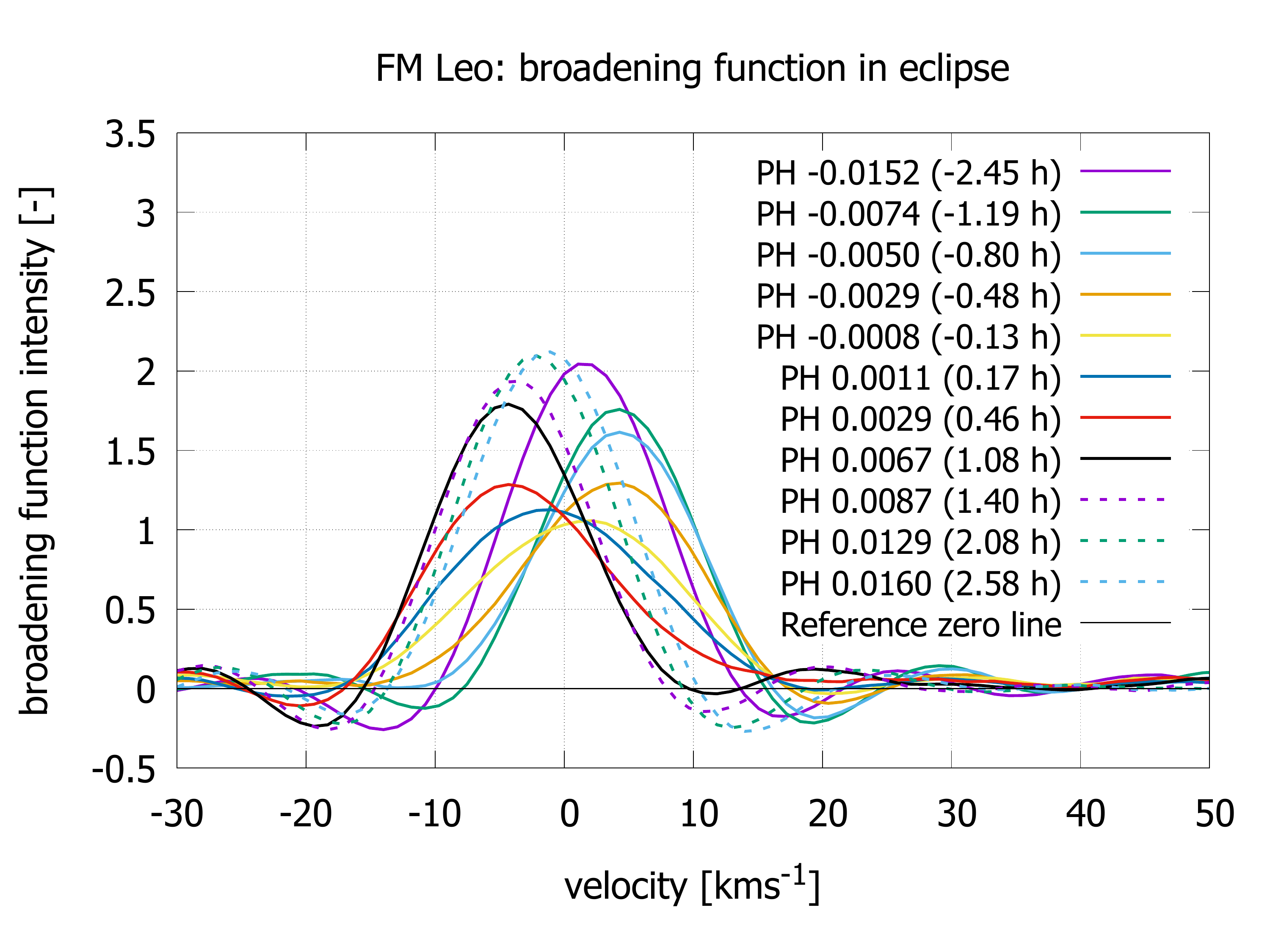}
		\caption{}\label{sfig:fmleo-bfGrid-b}
	\end{subfigure}
	
	\begin{subfigure}[b]{.45\textwidth}
		\includegraphics[width=\textwidth]{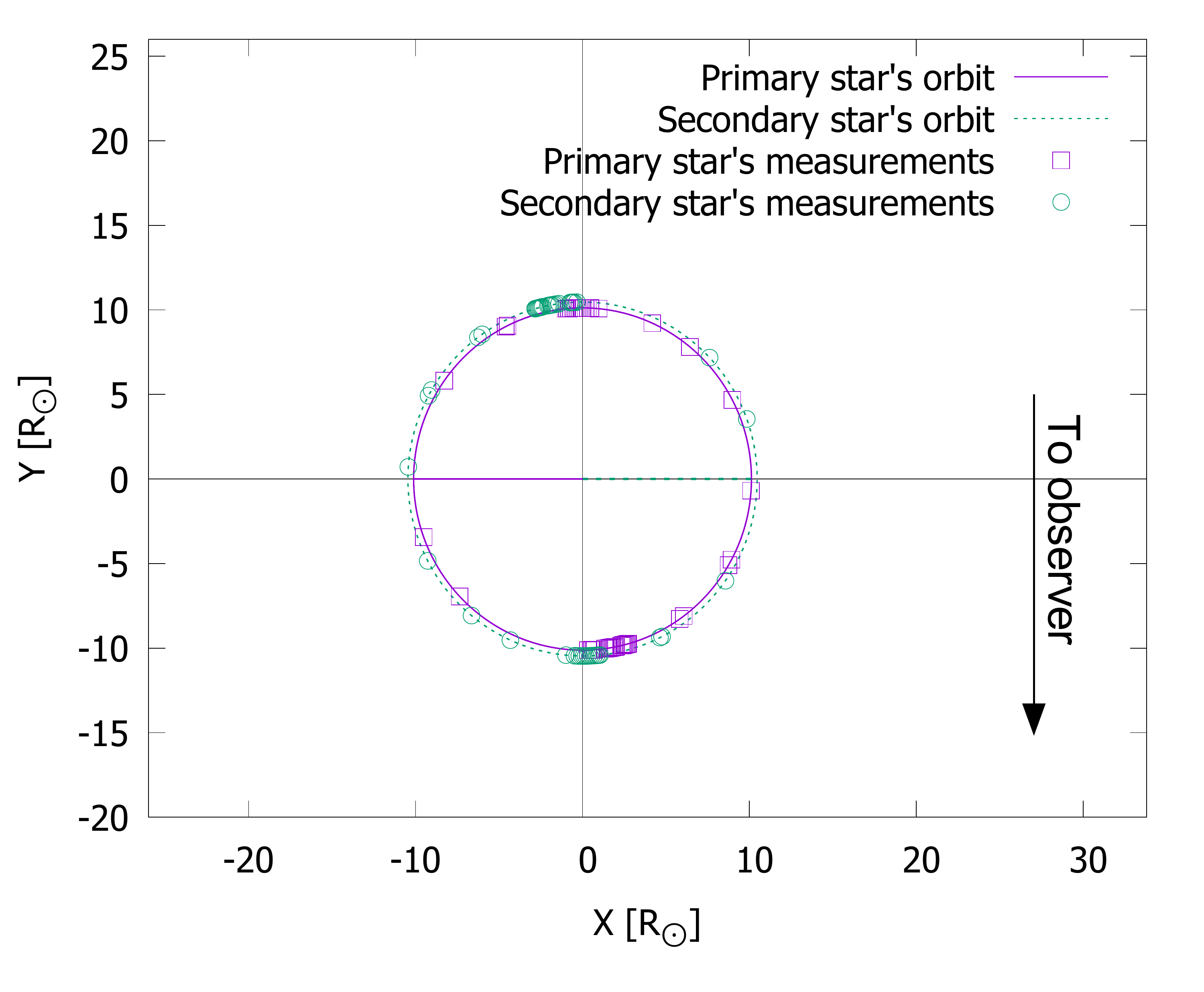}
		\caption{}\label{sfig:fmleo-bfGrid-c}
	\end{subfigure}
	\begin{subfigure}[b]{.45\textwidth}
		\includegraphics[width=\textwidth]{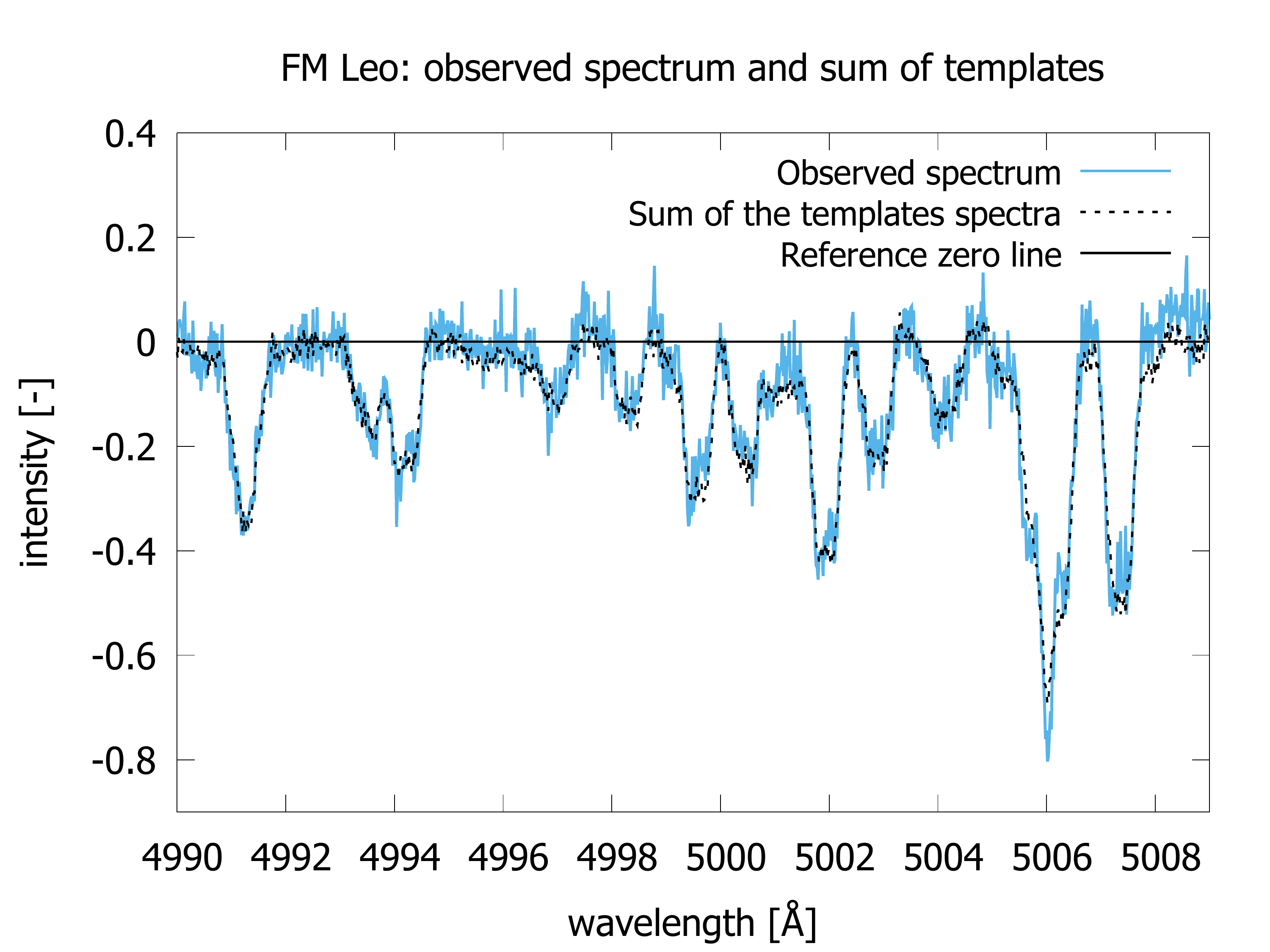}
		\caption{}\label{sfig:fmleo-bfGrid-d}
	\end{subfigure}
	\caption{
		a) Example of the broadening function calculated at the end of primary eclipse, at phase (PH) 0.01595. The BFs have been corrected for the Earth's velocity. The black dashed line denotes the BF for the observed spectrum with a background component (BC) template used for BF calculation. It is important to note that this is for illustration only, as the spectra of both components differ and for BF calculation of both components we used template of just one component. This has, for example, an impact on the shape and amplitude of the BF but also slightly changed position of the maximum of the function for the component whose spectrum does not fit the template. The green line represents the BF of the secondary star's spectra, foreground component (FC) which will be subtracted from the observed data. The purple line shows the BF for the primary star's spectra remaining after subtraction, which is in good agreement with the expected background component which is partially eclipsed. The amplitude of the components BFs was scaled to match the scale of observed spectra BF. The presented result is a sum of BFs from all used rows. A single row produces very a similar result but with a smaller amplitude. The BF is presented in the velocity domain.\\
		b) Broadening function of the background component, primary star, of the FM Leo binary in the velocity domain. The legend shows the phase of each curve as well as the difference between the observation time and the time of the eclipse. Clearly visible is the change of amplitude with the changing area of eclipsed star, and the shift of the peak to negative and positive values (the so called RM effect). The BF was shifted to the calculated velocity of primary component, so only the change due to the RM effect is visible. \\ 
		c) Orbits of the components of the binary system FM Leo, with observations marked according to the legend. The black arrow in the bottom right corner of the panel indicates the direction to the observer. \\ 
		d) Normalized observed spectrum, shown together with the sum of merged disentangled templates shifted to the appropriate reference frames according to the velocities of the components and scaled to the relative intensities based on the light curve. The observed spectrum is marked with light blue, and the sum of templates as a dotted black line. The BF from Figure \ref{sfig:fmleo-bfGrid-a} was calculated using this row of observed data as one of the inputs to the sum. \\ 
		e) Example of observed spectrum (OS) row marked with a light blue line and the observed spectrum with the foreground star spectrum subtracted marked with a purple line. Only the spectrum coming from the eclipsed star's visible fragment contributes to the purple spectrum. The green and purple lines order is switched in the case of the secondary eclipse to maintain the connection between the purple colour lines denoting everything related to primary star and green to the secondary star. \\
		f) Observed spectrum marked with a blue line and a template of the foreground component which will be subtracted for the BF measurement of the background component, marked with a green line. The row presented in this figure is the same row as the row in Figure~\ref{sfig:fmleo-bfGrid-d} and \ref{sfig:fmleo-bfGrid-f}. \\
		g) As in Figure~\ref{sfig:fmleo-bfGrid-b} but for the secondary eclipse. }
	\label{fig:fmleo-bfGrid1}
\end{figure*}
\begin{figure*}\ContinuedFloat
	\centering
	\begin{subfigure}[b]{.45\textwidth}
		\includegraphics[width=\textwidth]{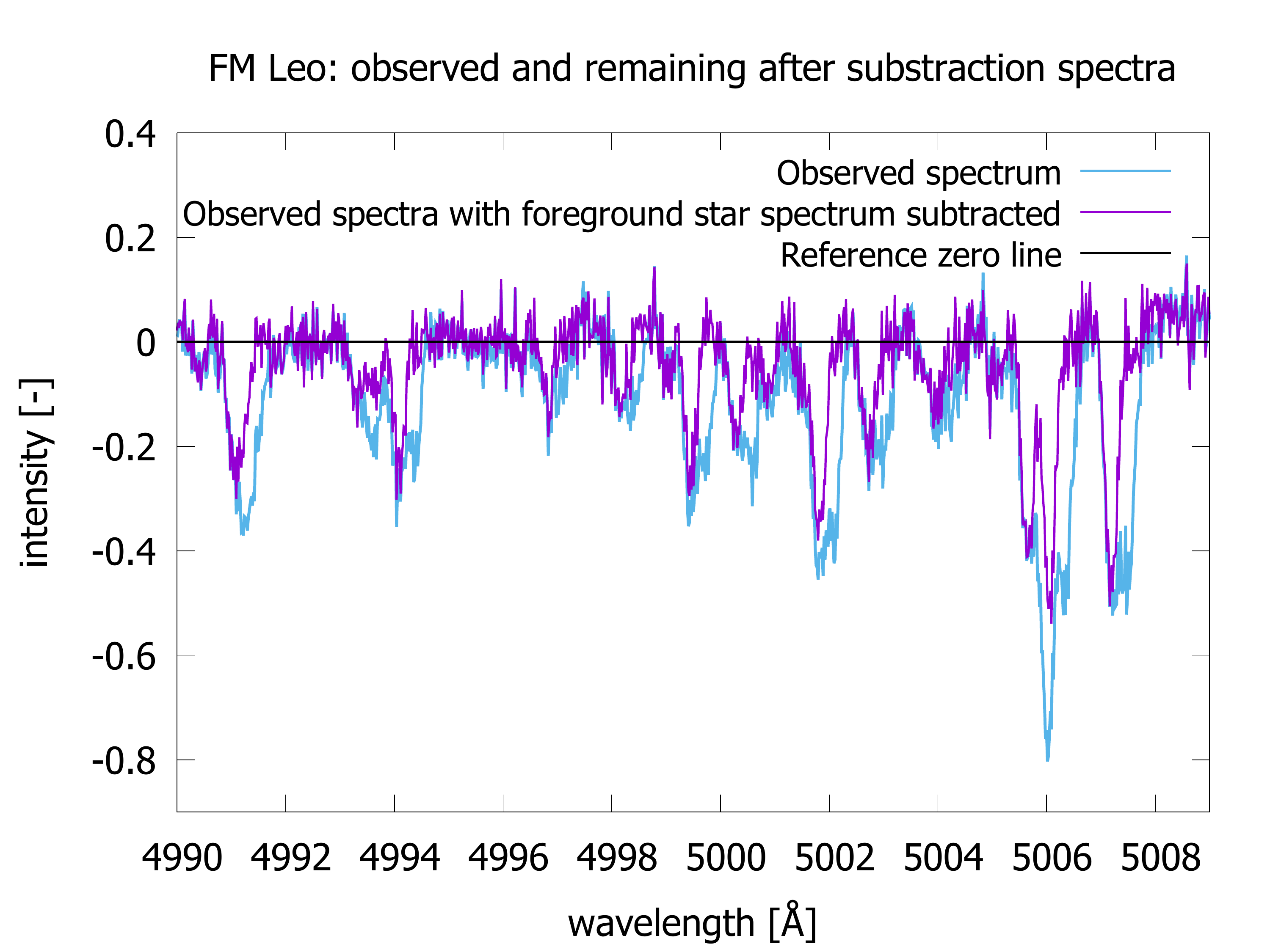}
		\caption{}\label{sfig:fmleo-bfGrid-e}
	\end{subfigure}
	\begin{subfigure}[b]{.45\textwidth}
		\includegraphics[width=\textwidth]{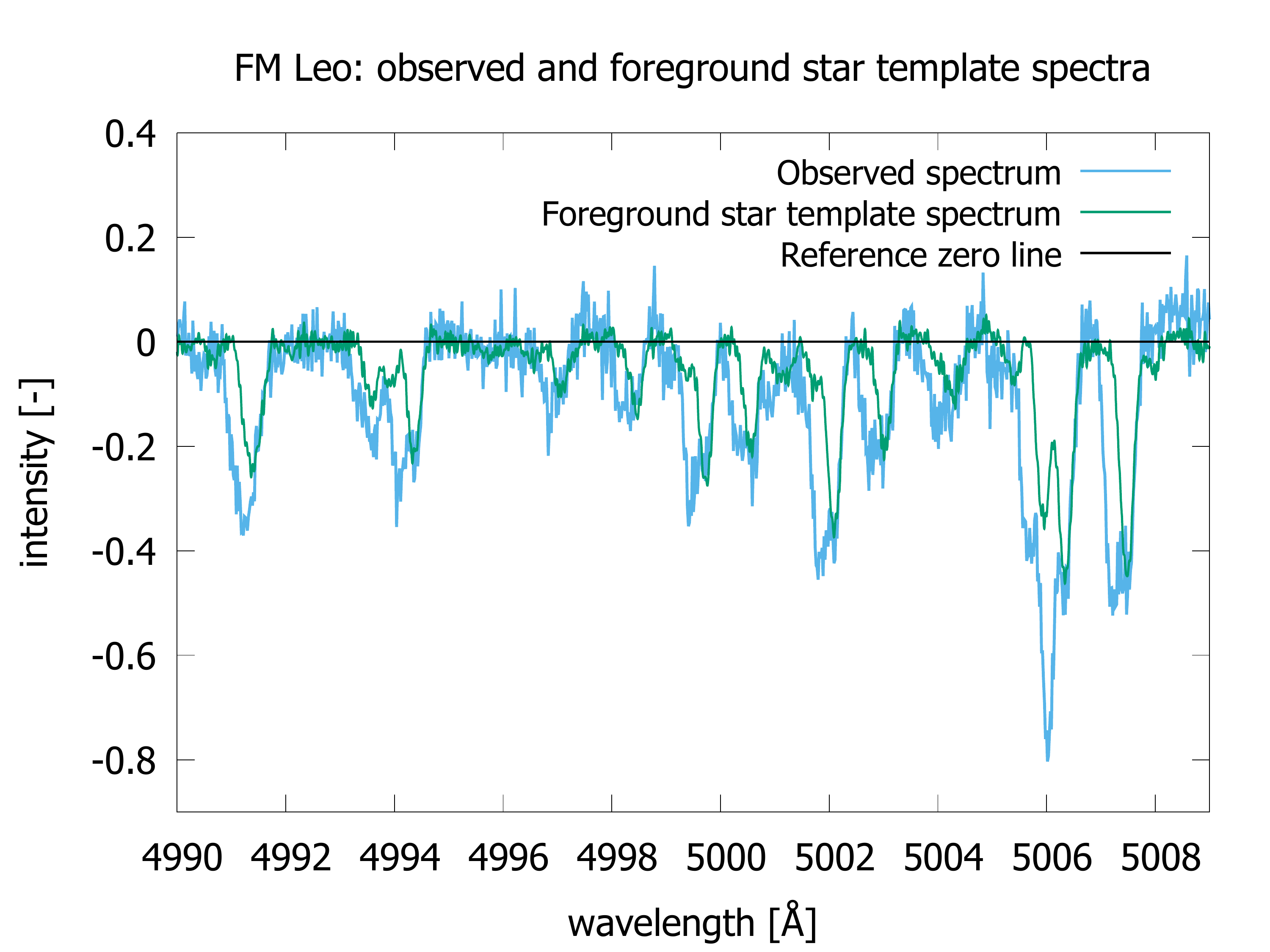}
		\caption{}\label{sfig:fmleo-bfGrid-f}
	\end{subfigure}
	
	\begin{subfigure}[b]{.45\textwidth}
		\includegraphics[width=\textwidth]{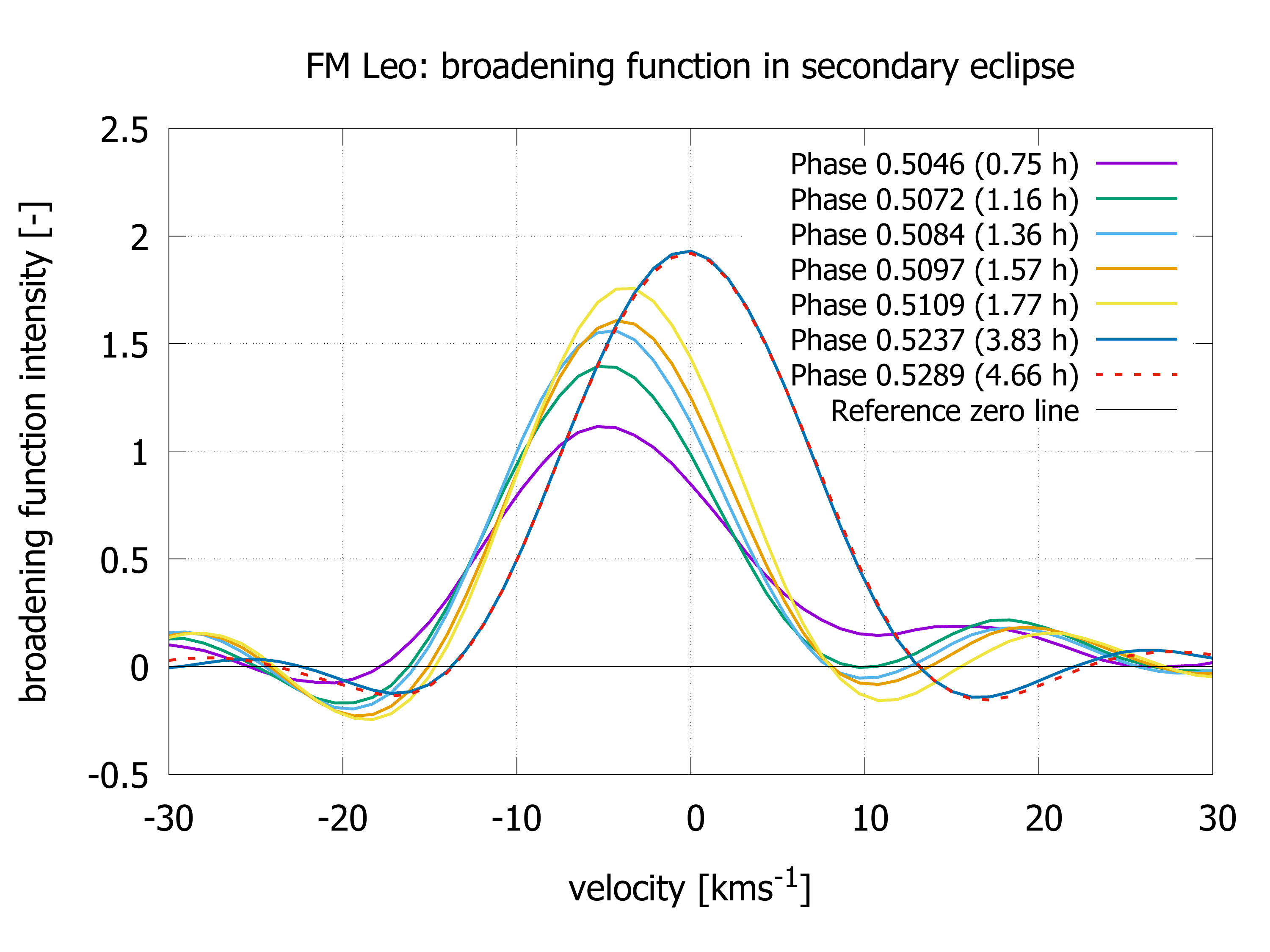}
		\caption{}\label{sfig:fmleo-bfGrid-g}
	\end{subfigure}
	\caption{Continued from previous page.}
	\label{fig:fmleo-bfGrid2}
\end{figure*}

%NN Del
\begin{figure*}
	\centering
	\begin{subfigure}[b]{.45\textwidth}
		\includegraphics[width=\linewidth]{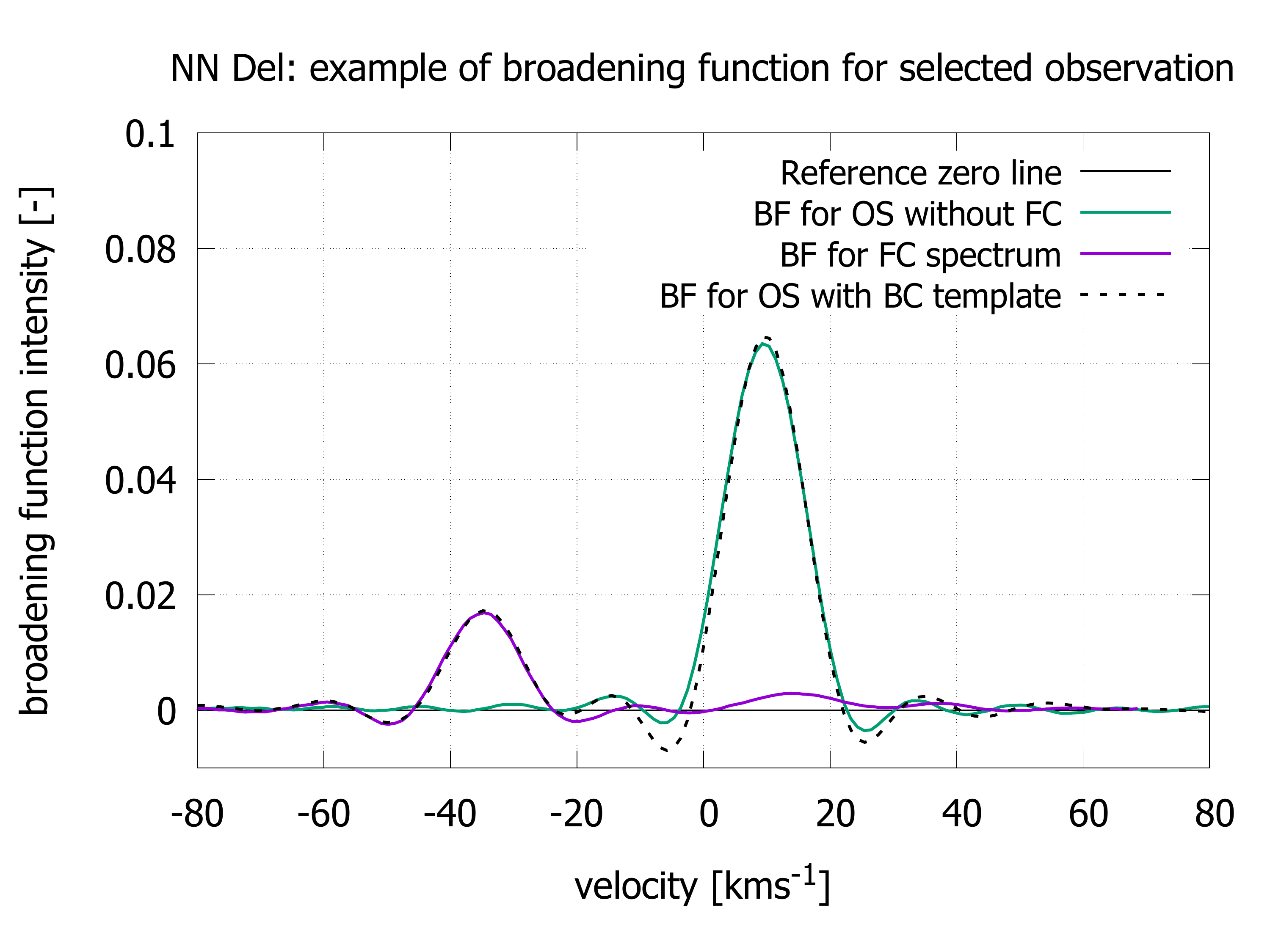}
		\caption{}\label{sfig:nndel-bfGrid-a}
	\end{subfigure}
	\begin{subfigure}[b]{.45\textwidth}
		\includegraphics[width=\linewidth]{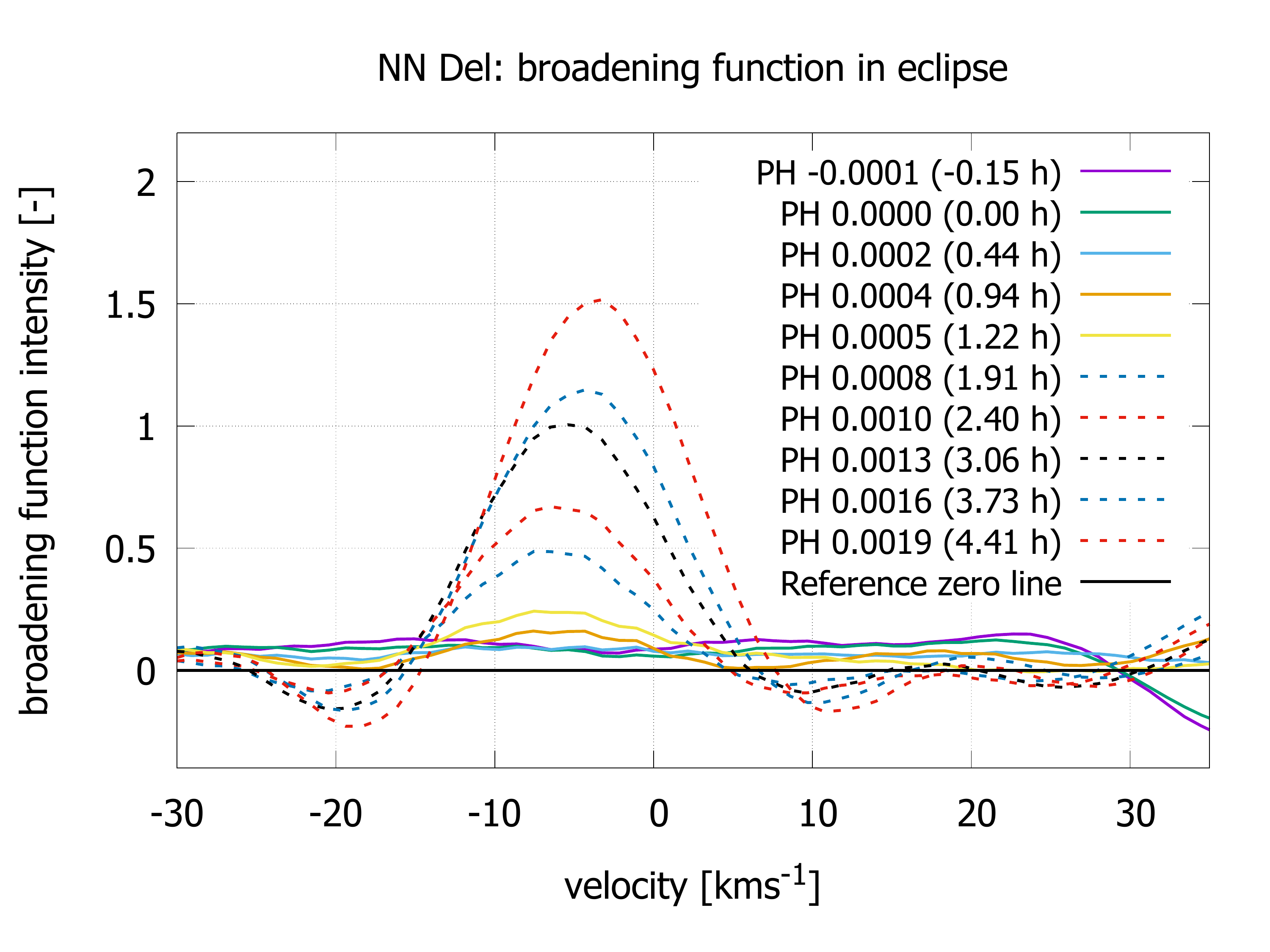}
		\caption{}\label{sfig:nndel-bfGrid-b}
	\end{subfigure}
	
	\begin{subfigure}[b]{.45\textwidth}
		\includegraphics[width=\textwidth]{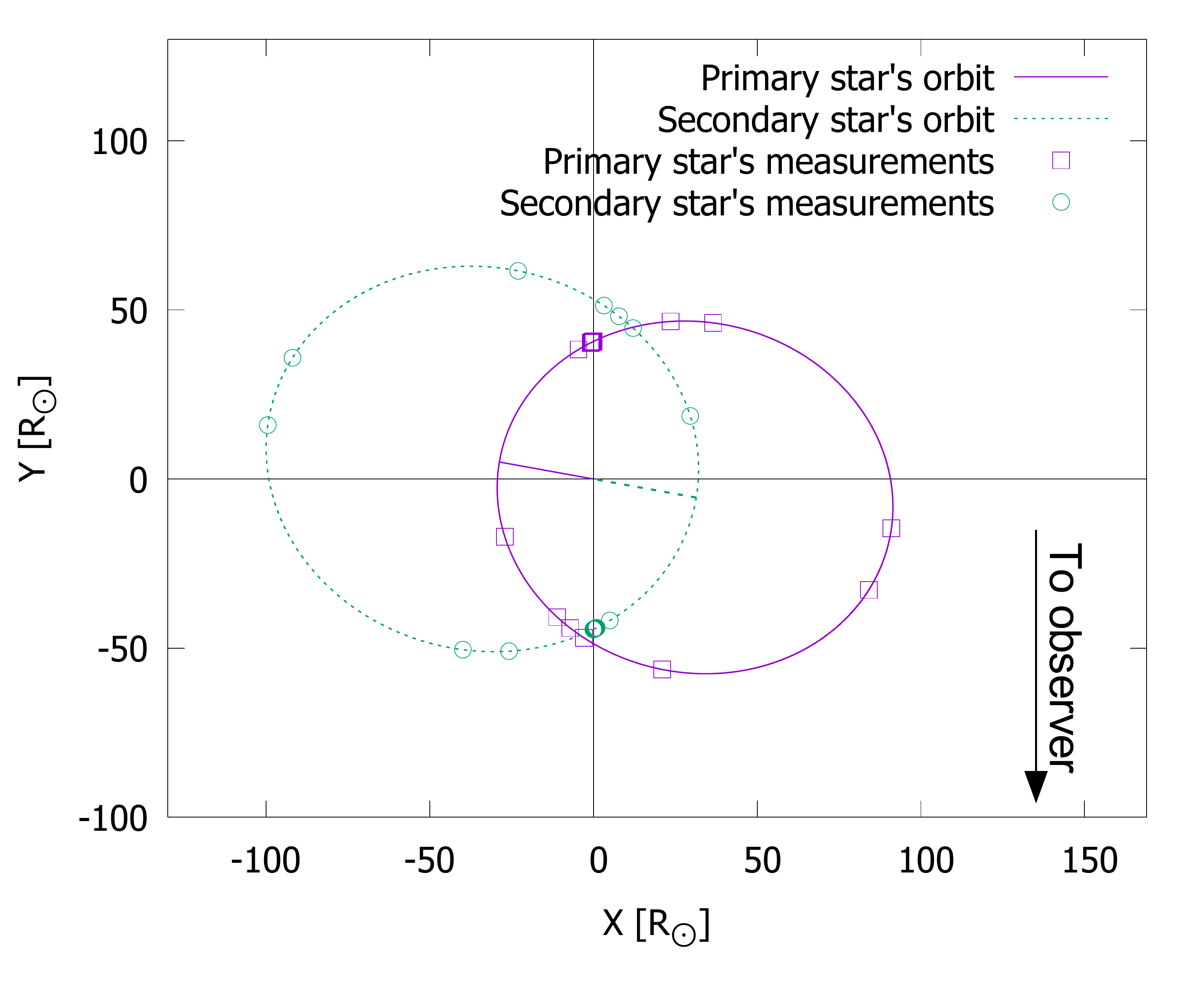}
		\caption{}\label{sfig:nndel-bfGrid-c}
	\end{subfigure}
	\begin{subfigure}[b]{.45\textwidth}
		\includegraphics[width=\textwidth]{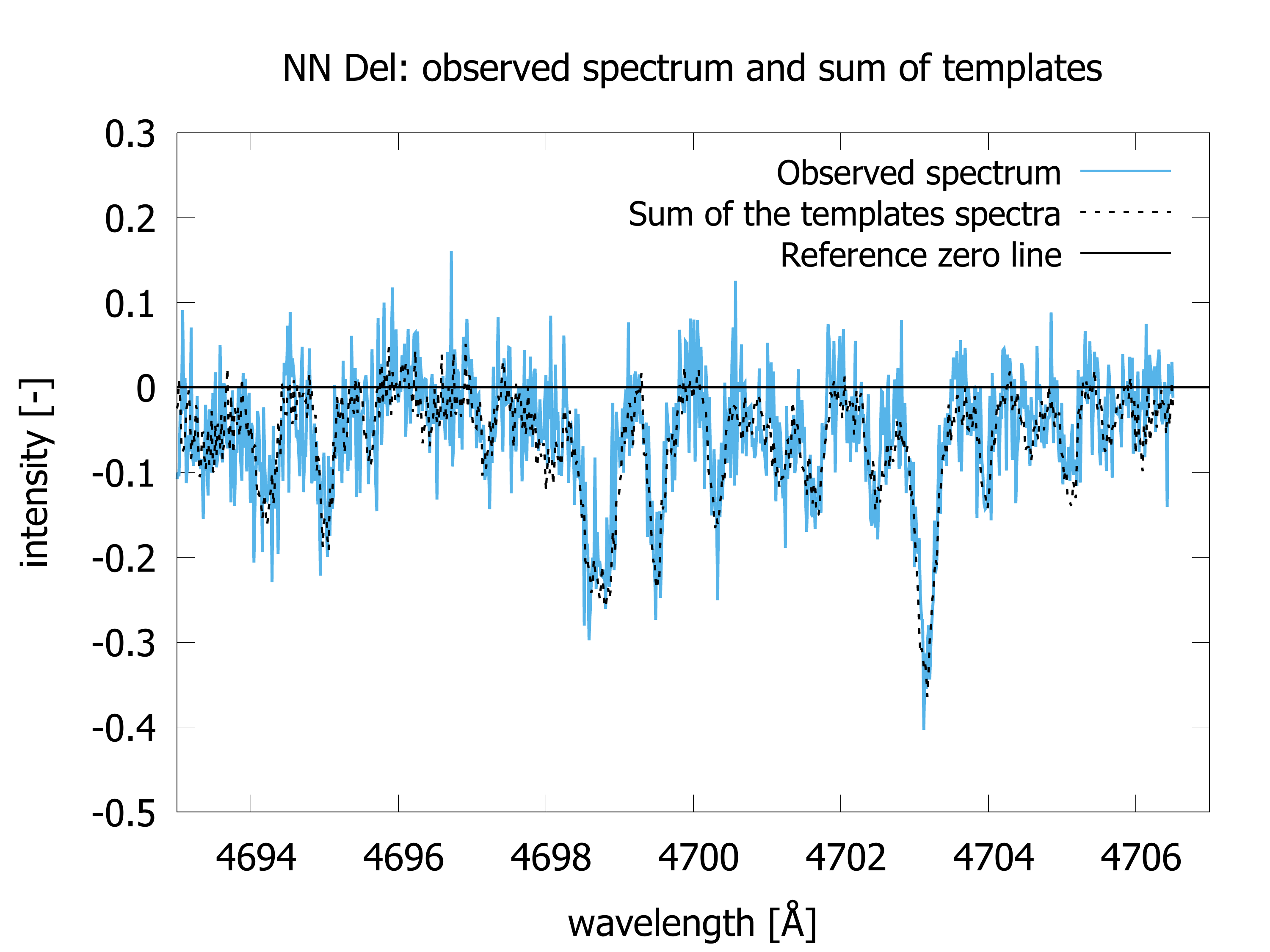}
		\caption{}\label{sfig:nndel-bfGrid-d}
	\end{subfigure}
	
	\begin{subfigure}[b]{.45\textwidth}
		\includegraphics[width=\textwidth]{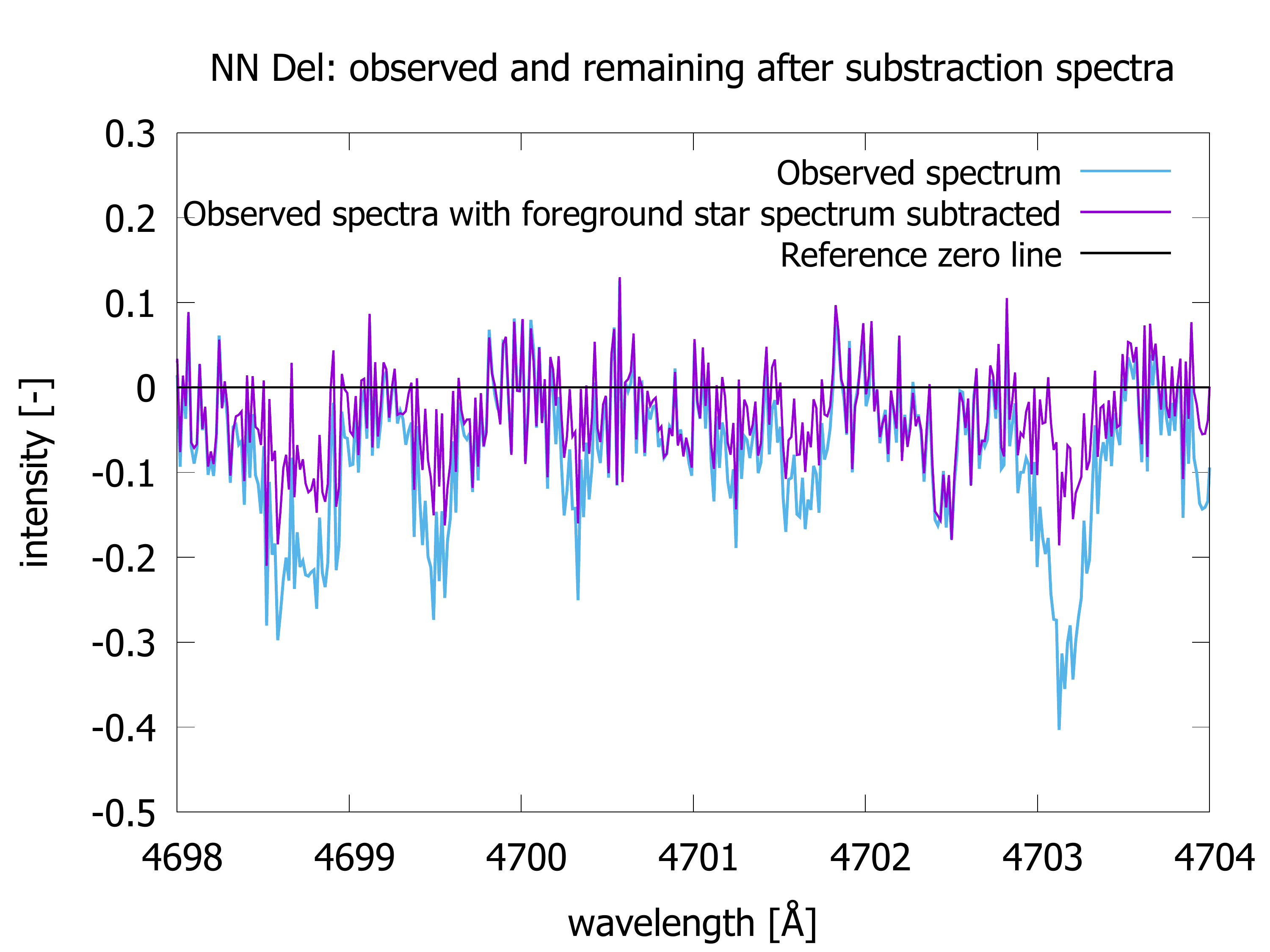}
		\caption{}\label{sfig:nndel-bfGrid-e}
	\end{subfigure}
	\begin{subfigure}[b]{.45\textwidth}
		\includegraphics[width=\textwidth]{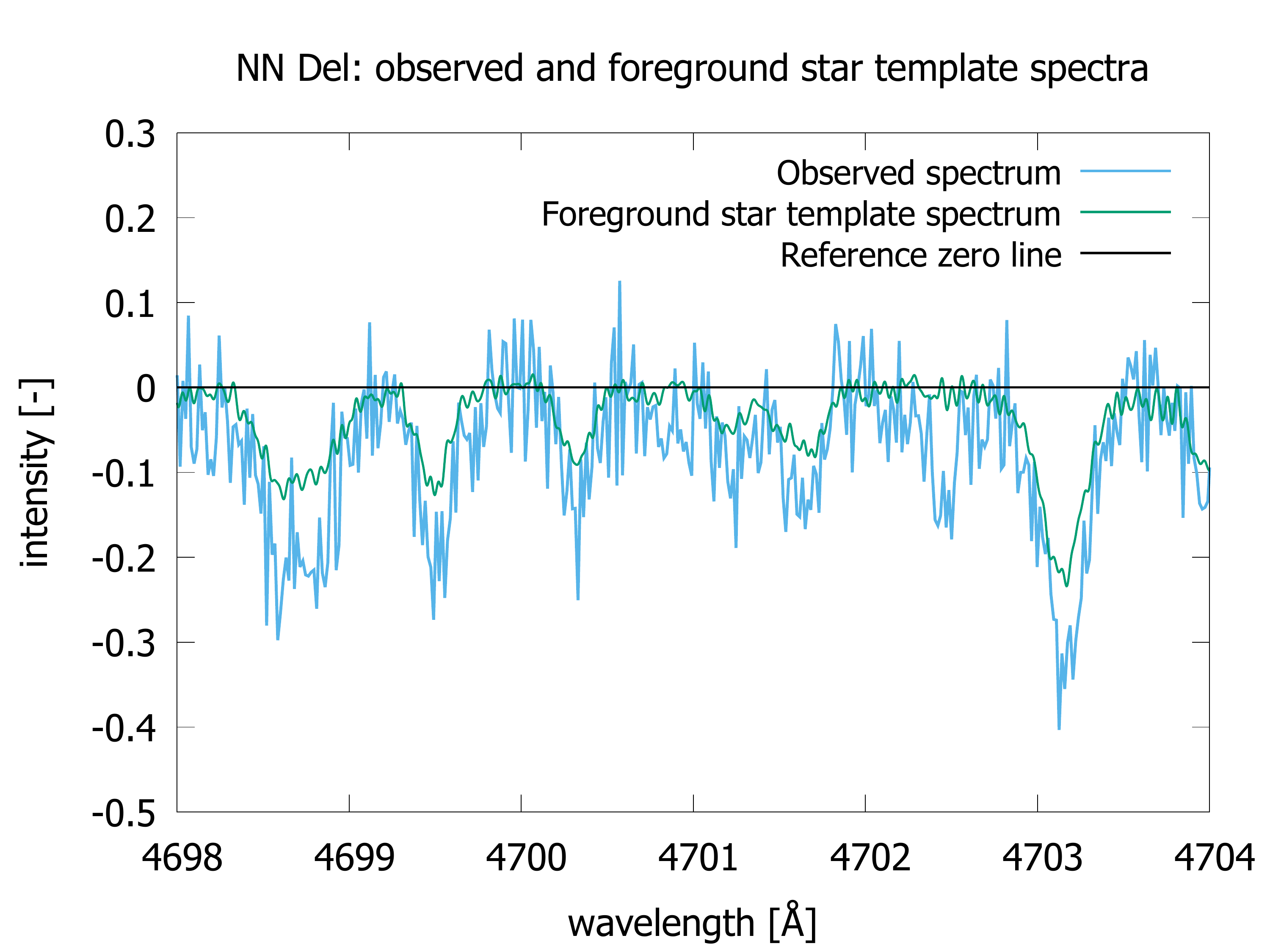}
		\caption{}\label{sfig:nndel-bfGrid-f}
	\end{subfigure}
	\caption{As in Figure \protect\ref{fig:fmleo-bfGrid1} but for primary component of NN Del. }
	\label{fig:nndel-bfGrid}
\end{figure*}

%V963 Cen
\begin{figure*}
	\centering
	\begin{subfigure}[b]{.45\textwidth}
		\includegraphics[width=\linewidth]{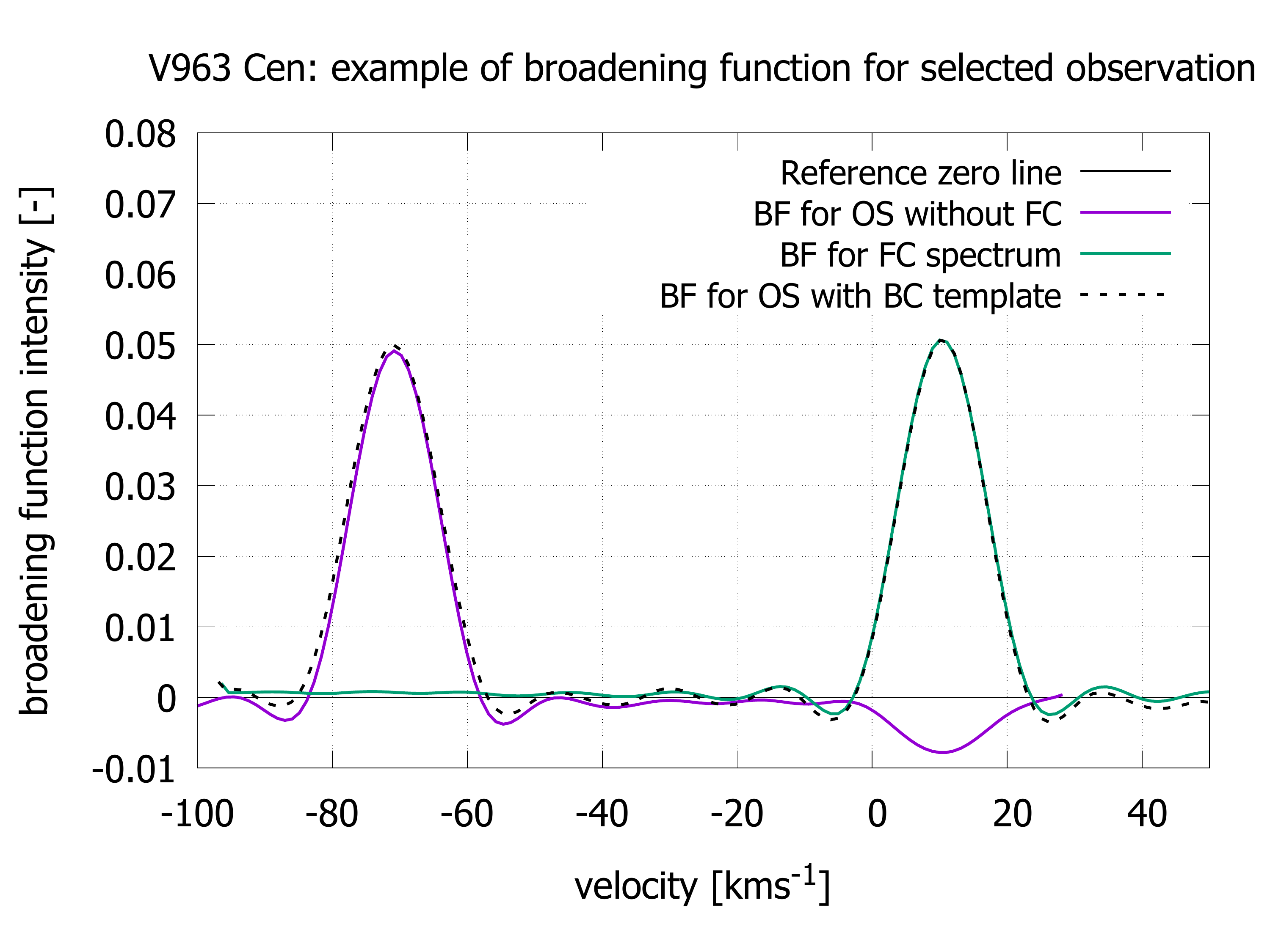}
		\caption{}\label{sfig:v963cen-bfGrid-a}
	\end{subfigure}
	\begin{subfigure}[b]{.45\textwidth}
		\includegraphics[width=\linewidth]{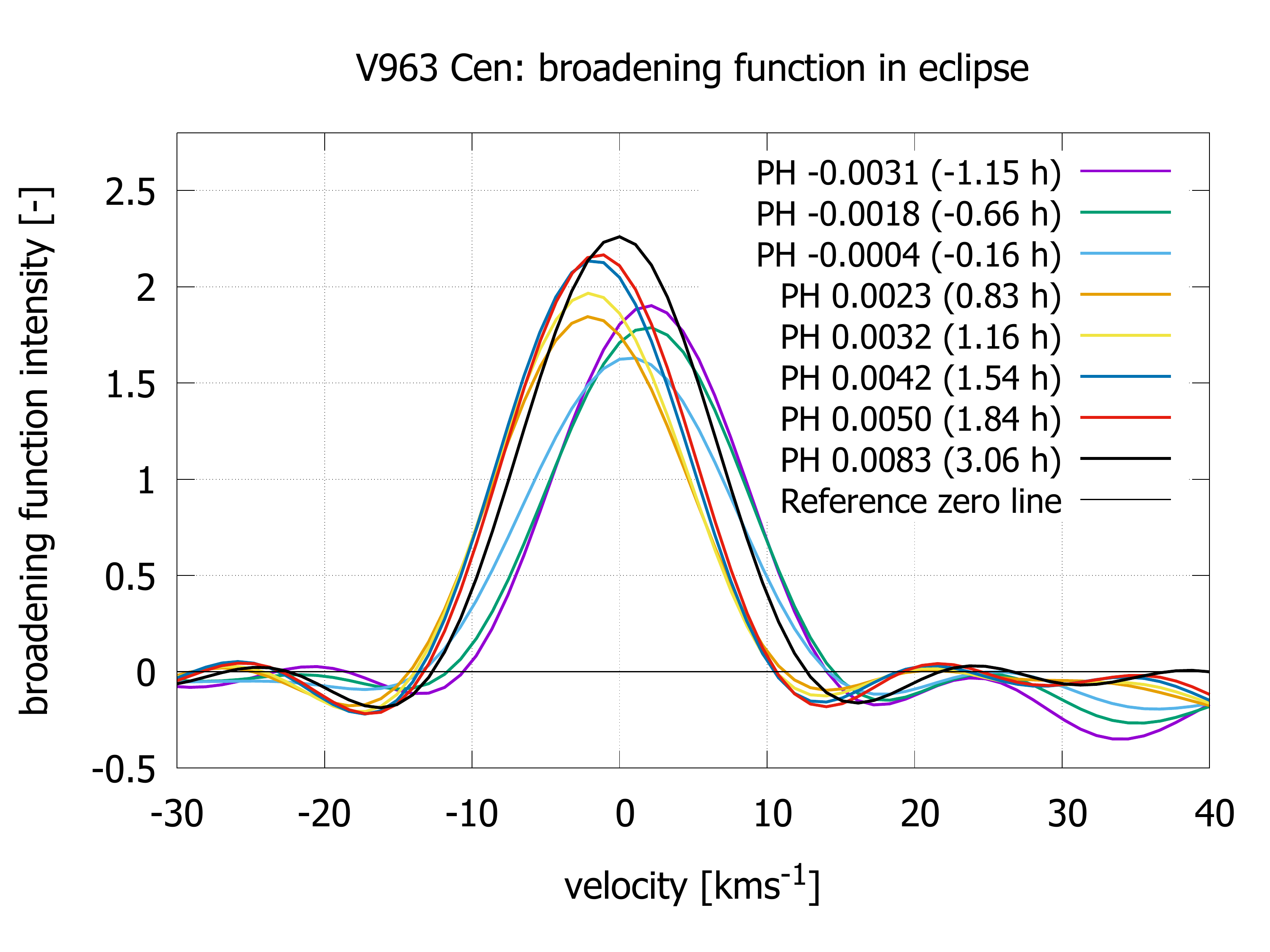}
		\caption{}\label{sfig:v963cen-bfGrid-b}
	\end{subfigure}
	
	\begin{subfigure}[b]{.45\textwidth}
		\includegraphics[width=\textwidth]{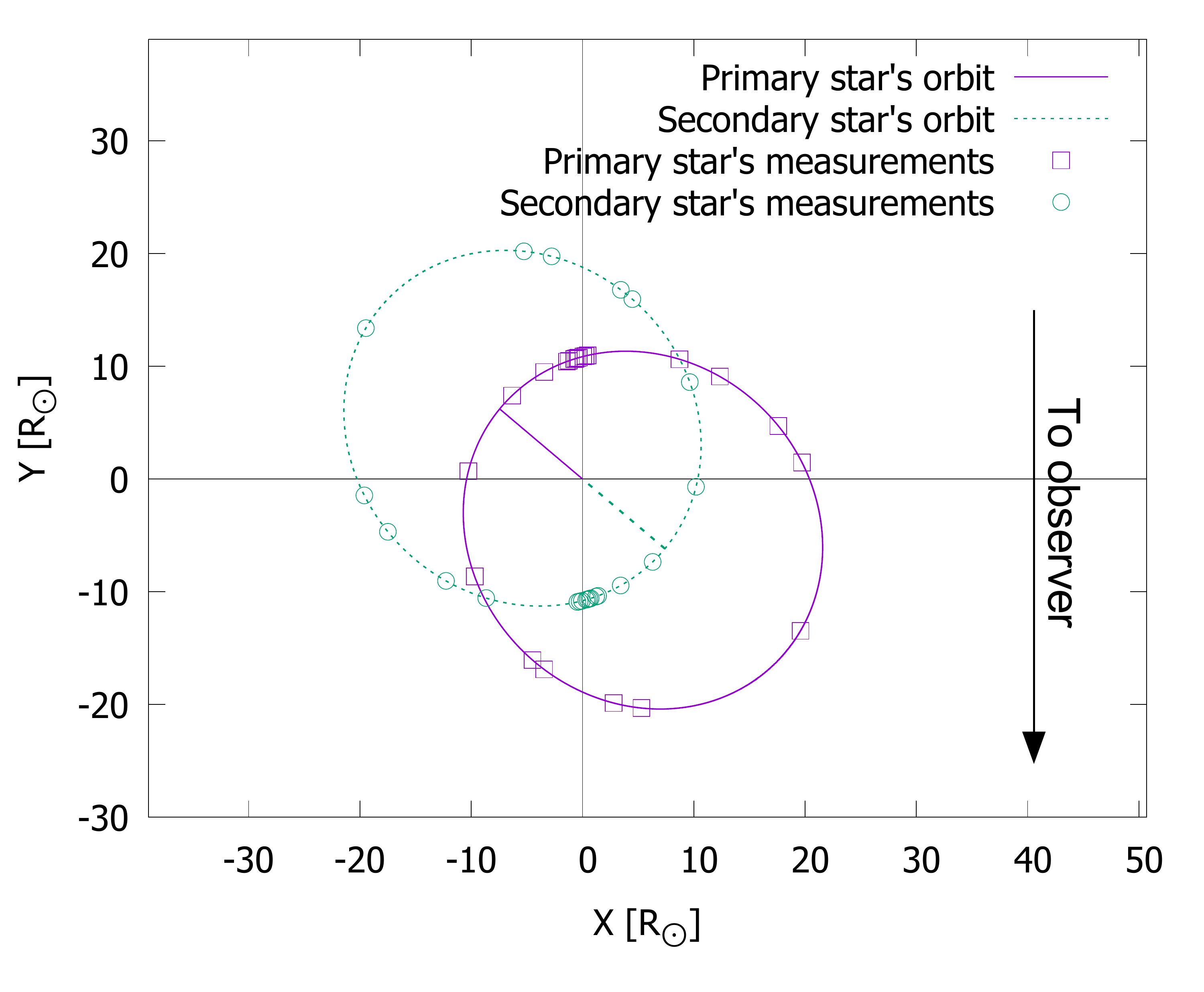}
		\caption{}\label{sfig:v963cen-bfGrid-c}
	\end{subfigure}
	\begin{subfigure}[b]{.45\textwidth}
		\includegraphics[width=\textwidth]{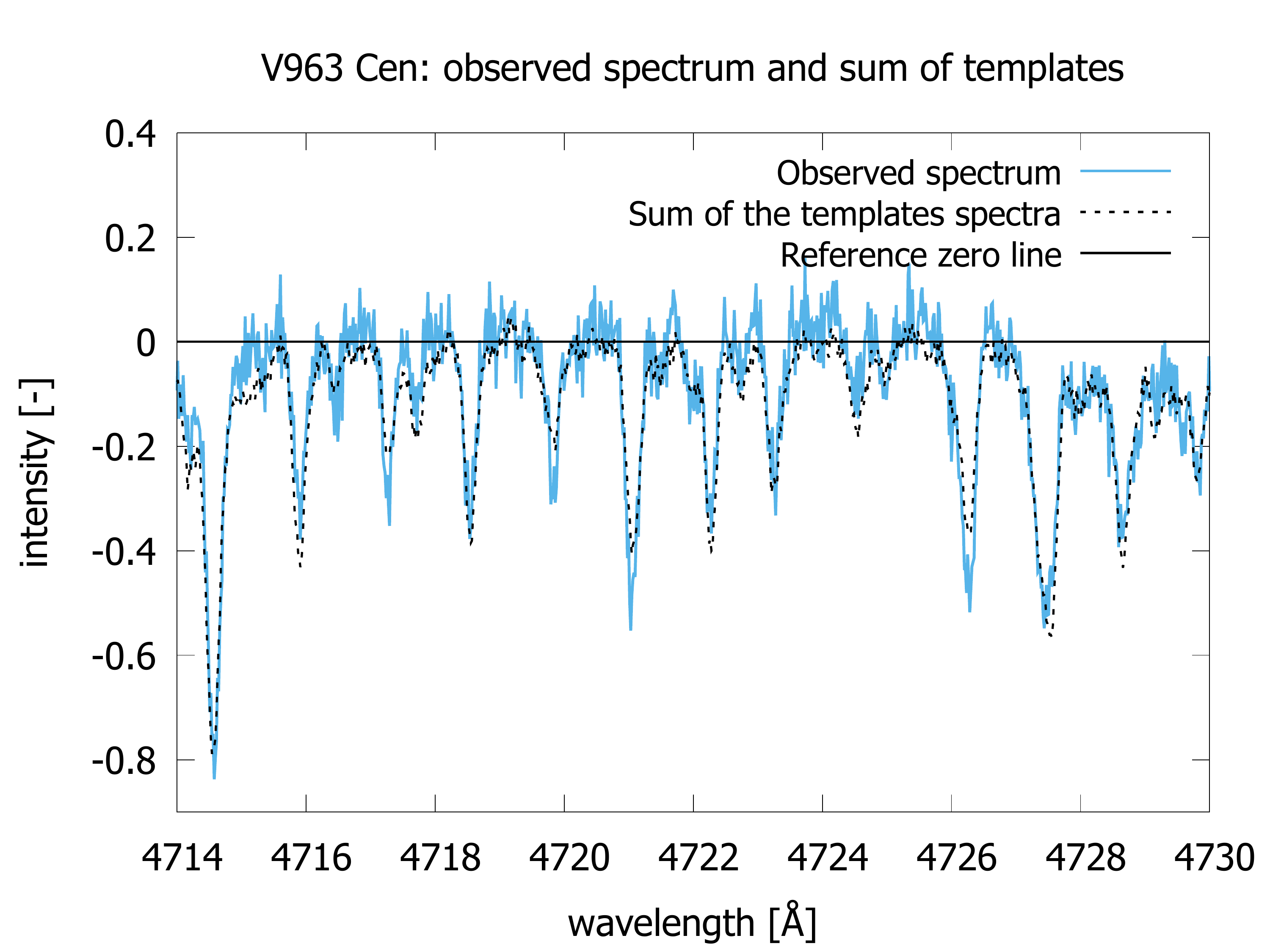}
		\caption{}\label{sfig:v963cen-bfGrid-d}
	\end{subfigure}
	
	\begin{subfigure}[b]{.45\textwidth}
		\includegraphics[width=\textwidth]{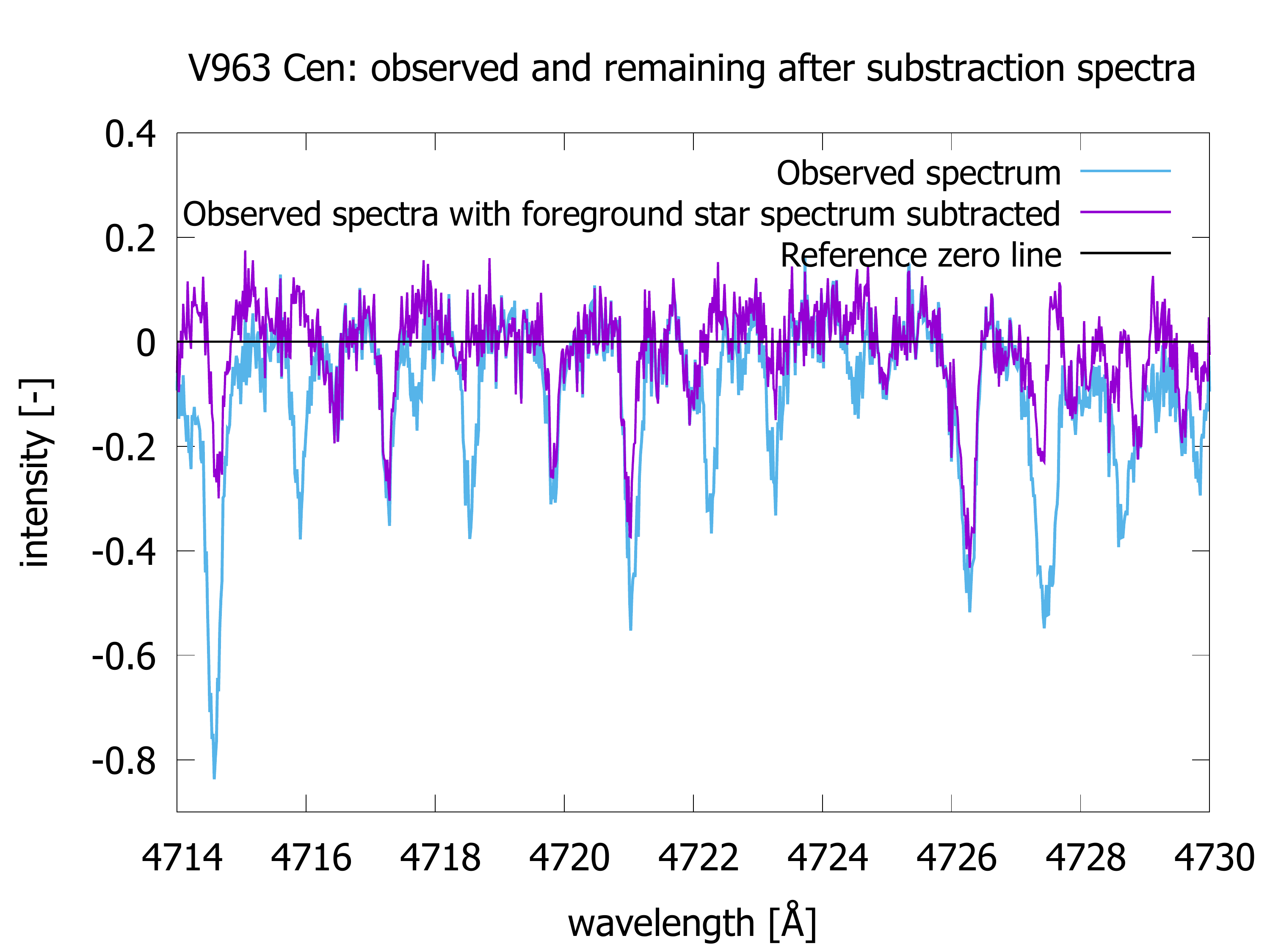}
		\caption{}\label{sfig:v963cen-bfGrid-e}
	\end{subfigure}
	\begin{subfigure}[b]{.45\textwidth}
		\includegraphics[width=\textwidth]{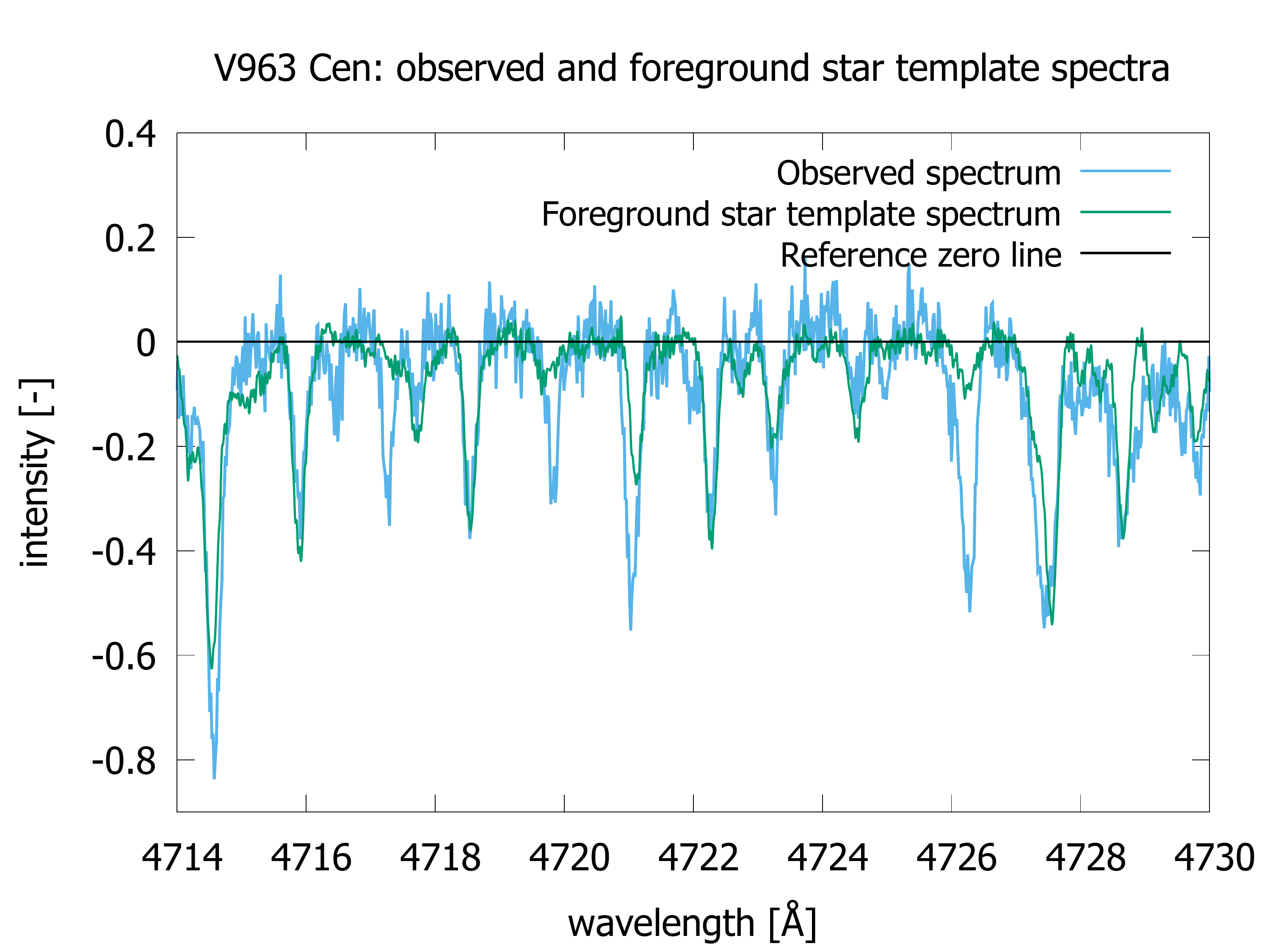}
		\caption{}\label{sfig:v963cen-bfGrid-f}
	\end{subfigure}
	\caption{As in Figure \protect\ref{fig:fmleo-bfGrid1} but for primary component of V963 Cen. }
	\label{fig:v963cen-bfGrid}
\end{figure*}

%AI Phe
\begin{figure*}
	\centering
	\begin{subfigure}[b]{.45\textwidth}
		\includegraphics[width=\linewidth]{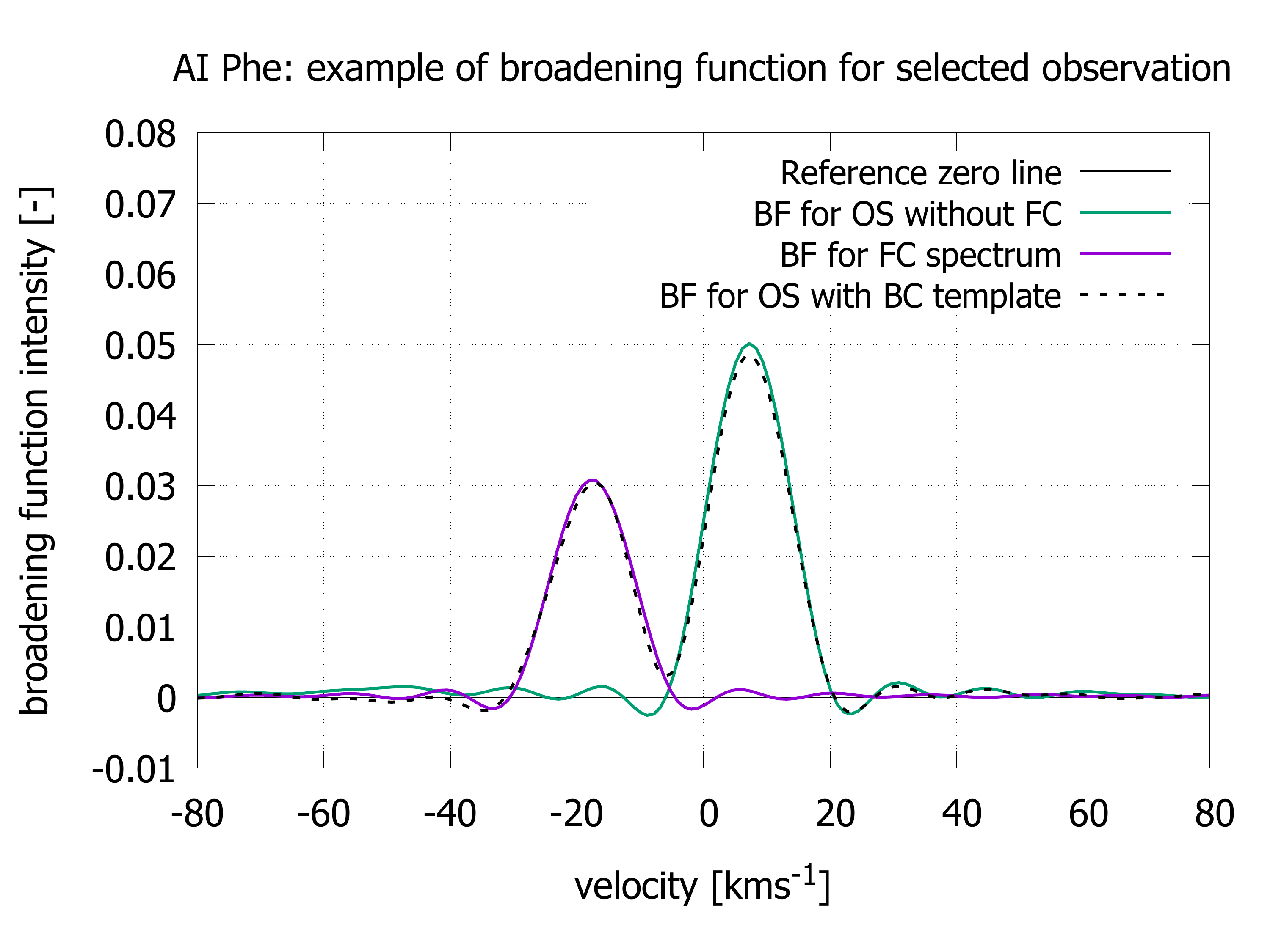}
		\caption{}\label{sfig:aiphe-bfGrid-a}
	\end{subfigure}
	\begin{subfigure}[b]{.45\textwidth}
		\includegraphics[width=\linewidth]{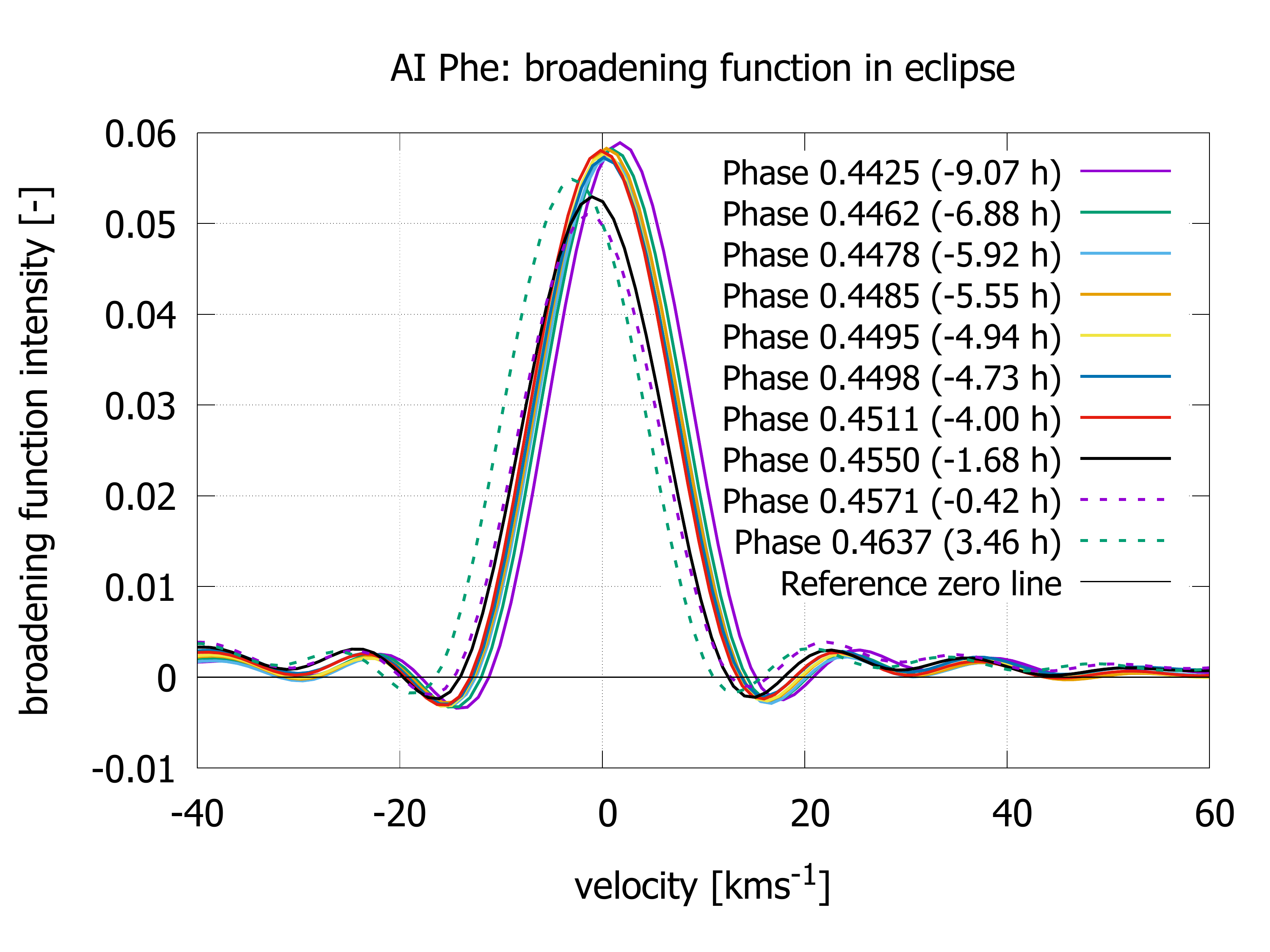}
		\caption{}\label{sfig:aiphe-bfGrid-b}
	\end{subfigure}
	
	\begin{subfigure}[b]{.45\textwidth}
		\includegraphics[width=\textwidth]{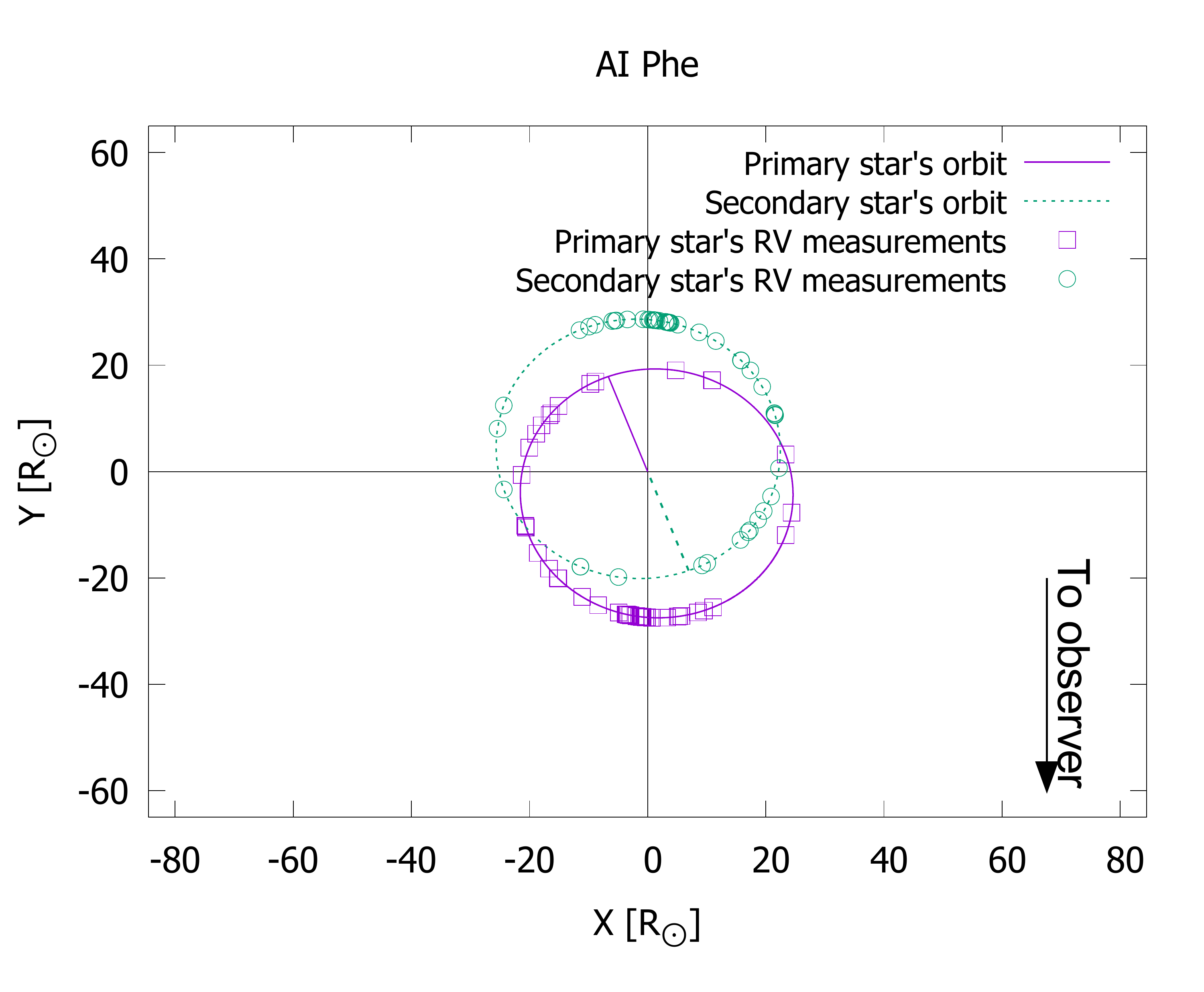}
		\caption{}\label{sfig:aiphe-bfGrid-c}
	\end{subfigure}
	\begin{subfigure}[b]{.45\textwidth}
		\includegraphics[width=\textwidth]{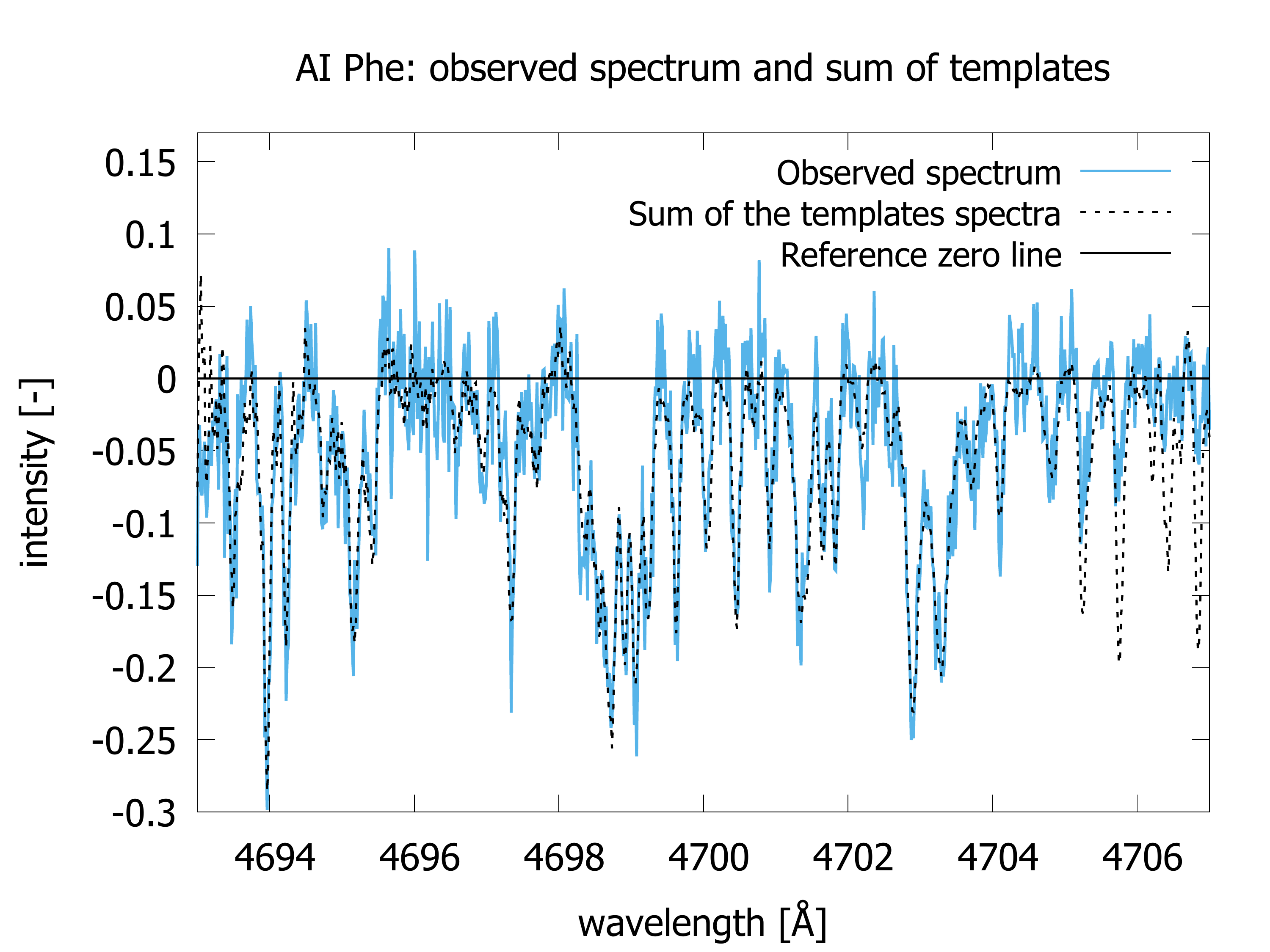}
		\caption{}\label{sfig:aiphe-bfGrid-d}
	\end{subfigure}
	
	\begin{subfigure}[b]{.45\textwidth}
		\includegraphics[width=\textwidth]{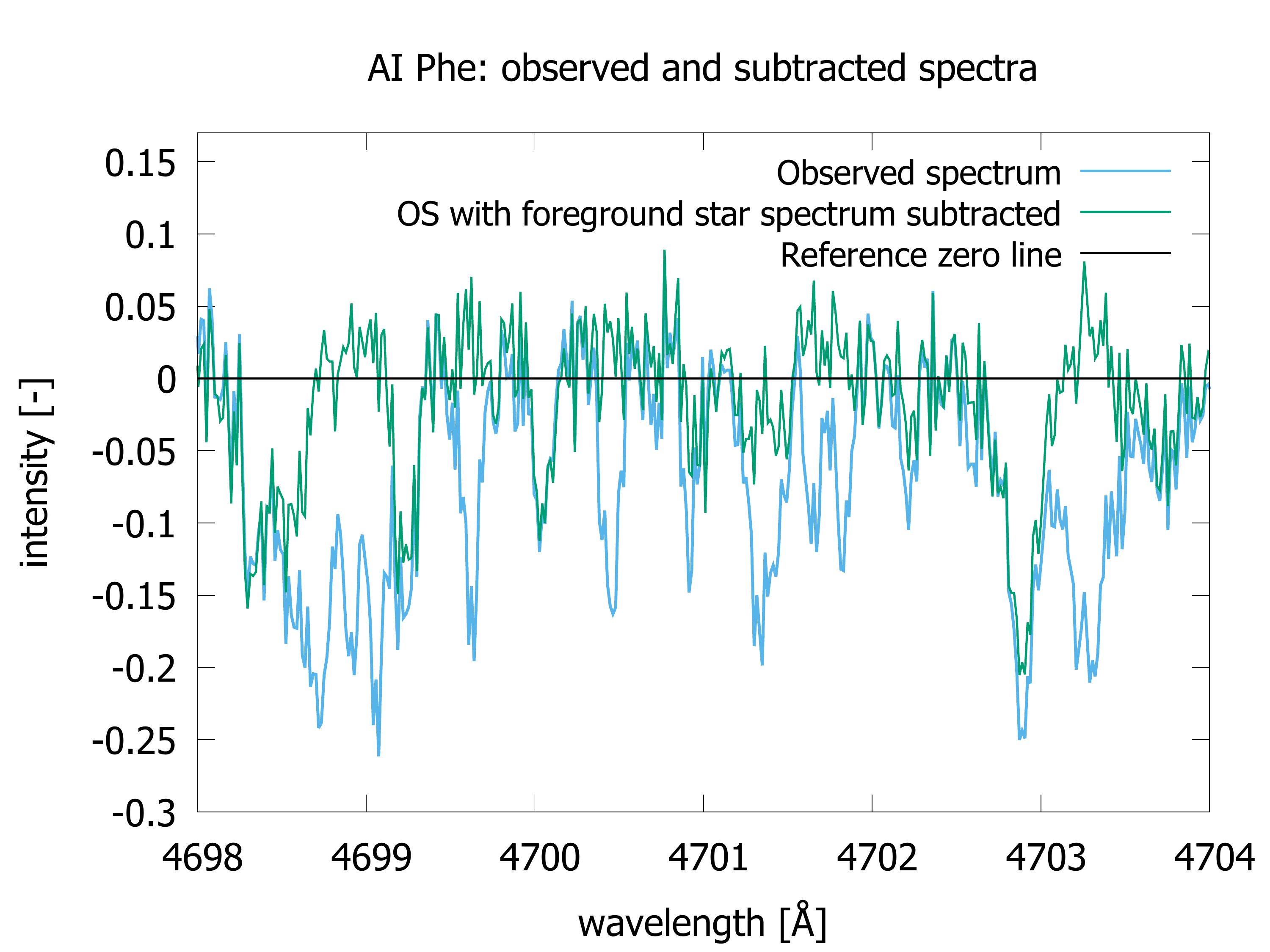}
		\caption{}\label{sfig:aiphe-bfGrid-e}
	\end{subfigure}
	\begin{subfigure}[b]{.45\textwidth}
		\includegraphics[width=\textwidth]{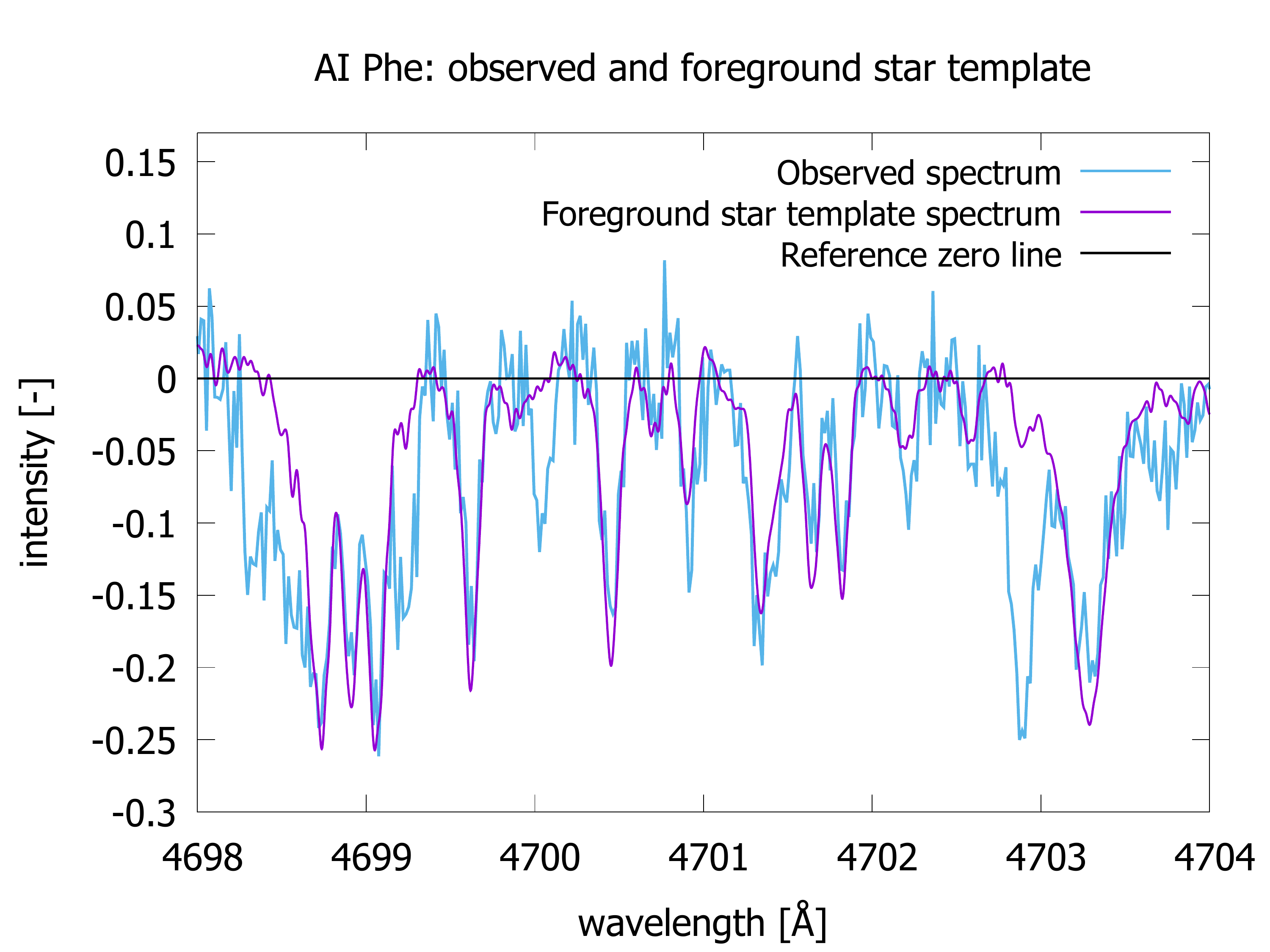}
		\caption{}\label{sfig:aiphe-bfGrid-f}
	\end{subfigure}
	\caption{As in Figure \protect\ref{fig:fmleo-bfGrid2} but for secondary component of AI Phe. }
	\label{fig:aiphe-bfGrid}
\end{figure*}

\begin{table*}
	\caption{Best fit parameters with errors of the studied systems obtained with JKTEBOP and MCMC code. The error of the time of secondary minimum has been estimated as a sum of primary minimum and period errors. In the case of the V963\,Cen with fixed period the error is based on the worst case of significant digit from literature provided period.  }
	\label{tab:JKTEBOP_Parameters}
\begin{centering}
	\begin{tabular}{r r r r r}
		\hline
		Parameter name														& FM\,Leo			&	NN\,Del			&	V963\,Cen		& AI\,Phe			\\
		\hline
		Period, $P$ [days]													&	6.7286134(36)	&	99.26849(15)	&	15.26932385(fixed)	&	24.592349(15)	\\
		Time of primary minimum, $T_0p$ [HJD-2450000]						&	2499.1810(25)	&	227.6023(11)	&	1248.33519(37)	    &	3247.6284(21)	\\
		Time of secondary minimum, $T_0s$ [HJD-2450000]						&	2502.5453(41)	&	246.2821(26)	&	1252.7658(38)	    &	3258.8882(22)	\\
		Major semi-axis, $a$ [$R_{\sun}$]									&	20.5609(77)		&	127.09(34)		&	33.4573(89)		    &	47.865(12)		\\
		Eccentricity, $e$													&	0.0 (fixed)		&	0.51944(55)		&	0.4217(fixed)    	&	0.18669(16)		\\
		Argument of periastron, $\omega$ [deg]								&	0.0 (fixed)		&	169.75(45)		&	140.101(29)		    &	110.450(50)		\\
		Orbital inclination, $i$ [deg]										&	89.07(63)		&	89.6342(76)		&	87.38(19)			&	88.60(31)		\\
		Velocity semi-amplitude (primary), $K_p$ [km $\textrm{s}^{-1}$]		&	75.992(29)		&	39.62(15)		&	60.924(18)			&	51.158(15)		\\
		Velocity semi-amplitude (secondary), $K_s$ [km $\textrm{s}^{-1}$]	&	78.653(36)		&	36.22(11)		&	61.259(19)			&	49.087(11)		\\
		Velocity offset (primary), $\gamma_p$ [km $\textrm{s}^{-1}$]		&	12.376(22)		&	-9.311(75)		&	-30.382(12)			&	-4.917(11)		\\
		Velocity offset (secondary), $\gamma_s$ [km $\textrm{s}^{-1}$]		&	12.378(27)		&	-9.500(43)		&	-30.406(12)			&	-4.8725(78)		\\
		Light scale factor, $L$ [mag]										&	8.4533(21)		&	-0.50053(25)	&	8.6351(14)			&	8.5930(13)		\\
		Surface brightness ratio, $J$										&	1.022(49)		&	0.8883(40)		&	1.012(85)			&	0.379(25)		\\
		Light ratio, $L_s$/$L_p$											&	0.49(11)		&	1.6564(76)		&	0.98(29)			&	1.098(66)		\\
		Primary mass, $M_p$ [$M_{\sun}$]									&	1.3119(16)		&	1.337(11)		&	1.08203(90)			&	1.19286(91)		\\
		Secondary mass, $M_s$ [$M_{\sun}$]									&	1.2675(14)		&	1.462(13)		&	1.07612(87)			&	1.2432(11)		\\
		Primary radius, $R_p$ [$R_{\sun}$]									&	1.764(78)		&	1.604(14)		&	1.41(13)			&	1.70(12)		\\
		Secondary radius, $R_s$ [$R_{\sun}$]								&	1.22(11)		&	2.188(15)		&	1.38(12)			&	2.901(86)		\\
		Logarithm of surface gravity (primary), $log g_p$ [cgs]				&	4.062(40)		&	4.1532(51)		&	4.175(78)			&	4.051(61)		\\
		Logarithm of surface gravity (secondary), $log g_s$ [cgs]			&	4.365(72)		&	3.9226(35)		&	4.187(70)			&	3.607(25)		\\
		Gravity darkening coefficient (primary), $g_p$						&	0.32 (fixed)	&	0.32 (fixed)	&	0.32 (fixed)		&	0.32 (fixed)	\\
		Gravity darkening coefficient (secondary), $g_s$					&	0.32 (fixed)	&	0.32 (fixed)	&	0.32 (fixed)		&	0.32 (fixed)	\\
		Limb darkening parameters, fixed (primary)							&	logarithmic		&	linear			&	logarithmic			&	logarithmic		\\
		&	$u_{p1}=$0.629, $u_{p2}=$0.18	&	$u_p=$0.531, $u_{p2}=$0.32	&	$u_{p1}=$0.629, $u_{p2}=$0.18	&	$u_{p1}=$0.629, $u_{p2}=$0.18	\\
		Limb darkening parameters, fixed (secondary)						&	logarithmic		&	linear	&	logarithmic	&	logarithmic	\\
		&	$u_{p1}=$0.629, $u_{p2}=$0.18	&	$u_p=$0.525, $u_{p2}=$0.32	&	$u_{p1}=$0.629, $u_{p2}=$0.18	&	$u_{p1}=$0.629, $u_{p2}=$0.18	\\
		\hline
	\end{tabular}
\end{centering}
\end{table*}

\begin{table}
	\caption{Geometric estimates of the synchronous rotation speed at the equator for aligned components based on the studied systems' parameters.}
	\label{rotation-parameters-table}
	\centering
	\begin{tabular}{r r r}
		\hline 
		Component			& Geometric synchronous $V_{rot}$	& Synchronous $V_{rot}$		\\
							& [km $\textrm{s}^{-1}$]			& [km $\textrm{s}^{-1}$]	\\
		\hline 
		FM\,Leo Primary		&	13.5							& 13.5(1.4)					\\
		FM\,Leo Secondary	&	9.3								& 9.3(0.5)					\\
		NN\,Del Primary		&	1.1								& 0.833(6)					\\
		NN\,Del Secondary	&	2.1								& 1.141(8)					\\
		V963\,Cen Primary	&	4.7								& 4.9(1.3)					\\
		V963\,Cen Secondary	&	4.6								& 4.5(1.2)					\\
		AI\,Phe Primary		&	8.3								& 13.3(6)					\\
		AI\,Phe Secondary	&	7.4								& 9.2(8)					\\
		\hline 
	\end{tabular} 
\end{table}

\begin{table*}
	\caption{Calculated with JKTABSDIM synchronous rotation velocity, timescale for circularisation, synchronization and spin-orbit angle change for studied binaries.}
	\label{absdim-table}
\begin{centering}
	\begin{tabular}{c r r r r}
		\hline 
									& synchronisation timescale		& circularisation timescale 	&	$a/R$	&	spin-orbit timescale	\\
									& [Gyr] 						& [Gyr]							&	[-]		&	[Myr]					\\
		\hline 
		FM\,Leo Primary	    		&	\multirow{2}{*}{7.3266(9)}	&	\multirow{2}{*}{9.9043(6)}	&	11.7	&	0.2$^a$					\\
		FM\,Leo Secondary			&								&								&	11.6	&	0.66$^a$				\\
		NN\,Del Primary				&	\multirow{2}{*}{11.949(6)}	&	\multirow{2}{*}{16.174(4)}	&	79.2	&	15000$^a$				\\
		NN\,Del Secondary			&								&								&	58.1	&	590$^b$					\\
		V963\,Cen Primary			&	\multirow{2}{*}{8.738(4)}	&	\multirow{2}{*}{11.810(3)}	&	23.7	&	3.2$^a$					\\
		V963\,Cen Secondary			&								&								&	24.2	&	1.6$^b$					\\
		AI\,Phe Primary				&	\multirow{2}{*}{7.326(3)}	&	\multirow{2}{*}{9.904(2)}	&	28.2	&	37$^b$					\\
		AI\,Phe Secondary 			&								&								&	16.5	&	5$^b$					\\
		\hline 
	\end{tabular} 
\end{centering}\\
\vspace{2mm} %5mm vertical space
Notes:
$^a$ Rotation period estimation based on $V_{rot}\sin{i_{rot}}$ with $\sin{i_{rot}}$ equal to one.
$^b$ Rotation period estimation based on synchronous $V_{rot}$.
\end{table*}

\begin{table*}
\caption{Calculated angles between stellar rotation axes and the angular momentum vectors for the sample presented in the paper. Two possible alignments for AI\,Phe Secondary component are in agreement with the Table~\ref{tab:JKTEBOP_Parameters} where by convention only the inclination $i$ lower than 90 degrees is provided, even if the second one, slightly higher than that would be fitting the \textsc{JKTEBOP} model as well. }
\label{rm-table}
\centering
\begin{tabular}{c r r r r r r r r}
\hline 
							& $\beta$		& $i$ 			& $V_{rot}\sin{i_{rot}}$	& $\omega$ 				& $T_0p$ 				&	$u$			&	$e$			& RMS					 \\ 
							& [deg]			& [deg] 		& [km $\textrm{s}^{-1}$]	& [deg] 				& [HJD-2450000] 		&	[-]			&	[-]			& [km $\textrm{s}^{-1}$] \\
\hline 
FM\,Leo Primary	    		&	0.0(1.1)	&	90(4)		&	22(9)					&	0 (fixed)			&	2499.1810(8)		&	0.21(0.22)	&	0.0(fixed)	&	0.2			 		 \\
FM\,Leo Secondary			&	-0.3(27.4)	&	90(3)		&	13.0(7.3)				&	0 (fixed)			&	2499.181(3)			&	0.27(0.33)	&	0.0(fixed)	&	0.04				 \\
NN\,Del Primary				&	0.0(2.6)	&	90.0(0.7)	&	18.0(8.0)				&	169.95(0.11)		&	227.60234(1)		&	0.6(0.3)	&	0.51943(43)	&	0.2					 \\
V963\,Cen Primary			&	0.0(2.4)	&	90.0(3.5)	&	15(11)					&	140.1010(5)			&	1248.33518(1)		&	0.8(0.3)	&	0.4216(9)	&	0.06				 \\
AI\,Phe Secondary Model 1	&	-87.5(16.2)	&	92.2(0.8)	&	1.3(4.3)				&	110.83(0.34)		&	3247.6248 (13)		&	0.06(0.21)	&	0.18698(11)	&	0.031				 \\
AI\,Phe Secondary Model 2	&	87(17)		&	87.8(0.9)	&	2.4(4.7)				&	110.80(0.35)		&	3247.6249 (15)		&	0.08(0.22)	&	0.18697(11)	&	0.031				 \\
\hline 
\end{tabular} 
\end{table*}

\clearpage

\section*{Acknowledgments}
PS would like to thank Simon Albrecht for fruitful discussions on the different possibilities of data reduction and analysis approaches related to Rossiter-McLaughlin effect. This work is supported by the Polish National Science Center grant 2011/03/N/ST9/03192 (PS) and grant 2016/21/B/ST9/01613 (KGH). This research has made use of the following web resources: http://simbad.u-strasbg.fr, http://adsabs.harvard.edu, http://arxiv.org.

\bibliographystyle{mn2e}
\bibliography{biblio}

\appendix
\section{Tabulated radial velocity measurements}
In Tables A1-A8 we provide for each observed system radial velocities obtained with {\sc todcor} (Tables \ref{appendix1-FMLeo}, \ref{appendix1-NNDel}, \ref{appendix1-V963Cen}, \ref{appendix1-AIPhe}) as well as radial velocities obtained for the RM effect of the first and/or secondary component (Tables \ref{appendix1-FMLeoRmP}, \ref{appendix1-NNDelRmP}, \ref{appendix1-V963CenRmP}, \ref{appendix1-AIPheRmP}).

\begin{table*}
	\caption{Radial velocities obtained with {\sc todcor} for FM\,Leo.}
	\label{appendix1-FMLeo}
	\centering
	\begin{tabular}{c r r r r}
\hline 
Time of the photometric middle  &	Primary star's	&	Primary star's	&	Secondary star's &	Secondary star's \\
of the observation $T_{pm}$		&	RV measurement 	& RV measurement error &	RV measurement 	& RV measurement error	\\
{[HJD-2450000]}	& [km $\textrm{s}^{-1}$] & [km $\textrm{s}^{-1}$] & [km $\textrm{s}^{-1}$] & [km $\textrm{s}^{-1}$]	\\
\hline
7430.79959	&	43.624	&	0.078	&	-19.977	&	0.094	\\
7431.74784	&	-21.359	&	0.077	&	47.234	&	0.089	\\
7433.74765	&	-42.881	&	0.077	&	69.430	&	0.090	\\
7435.77651	&	79.481	&	0.074	&	-57.067	&	0.087	\\
7436.81735	&	79.783	&	0.076	&	-57.421	&	0.090	\\
7446.76618	&	-59.000	&	0.072	&	86.466	&	0.086	\\
7448.76520	&	57.895	&	0.073	&	-34.756	&	0.085	\\
7450.70380	&	60.639	&	0.078	&	-37.479	&	0.090	\\
7458.67651	&	-22.273	&	0.074	&	48.259	&	0.085	\\
7462.65212	&	78.126	&	0.075	&	-55.748	&	0.087	\\
7472.65382	&	-49.936	&	0.081	&	76.792	&	0.096	\\
7475.64823	&	56.115	&	0.077	&	-32.953	&	0.090	\\
7476.60051	&	88.169	&	0.075	&	-65.938	&	0.092	\\
		\hline 
	\end{tabular} 
\end{table*}

\begin{table*}
	\caption{Radial velocities obtained for the RM effect of the primary component of FM\,Leo. Error for the measurement is calculated as a standard deviation of the average value from the list of RV measurement for each accepted order from the echelle spectrum. The order is accepted when at least three line profiles' peeks are higher in intensity amplitude than the calculated standard deviation of average from the difference of sum of disentangled spectra and observed spectra.}
	\label{appendix1-FMLeoRmP}
	\centering
	\begin{tabular}{c r r}
		\hline 
		Time of the photometric middle  &	Primary star's	&	Error of the   \\
		of the observation $T_{pm}$		&	RM measurement 	& RM measurement	\\
		{[HJD-2450000]}	& [km $\textrm{s}^{-1}$] & [km $\textrm{s}^{-1}$]	\\
		\hline 
7444.720157	&	-0.743	&	0.482	\\
7444.732192	&	-2.484	&	0.416	\\
7444.744771	&	-3.653	&	0.747	\\
7444.757937	&	-3.647	&	0.324	\\
7444.771247	&	-3.397	&	0.204	\\
7444.785491	&	-2.851	&	0.14	\\
7444.799568	&	-2.027	&	0.406	\\
7444.811342	&	-1.693	&	0.197	\\
7444.820337	&	-1.166	&	0.399	\\
7444.610834	&	1.37	&	0.431	\\
7444.66347	&	3.753	&	0.251	\\
7444.6796	&	3.353	&	0.842	\\
7444.693189	&	2.725	&	0.752	\\
7444.707722	&	0.769	&	0.574	\\
		\hline 
	\end{tabular} 
\end{table*}

\begin{table*}
	\caption{Radial velocities obtained with {\sc todcor} for NN\,Del.}
	\label{appendix1-NNDel}
	\centering
	\begin{tabular}{c r r r r}
	\hline 
	Time of the photometric middle  &	Primary star's	&	Primary star's	&	Secondary star's &	Secondary star's \\
	of the observation $T_{pm}$		&	RV measurement 	& RV measurement error &	RV measurement 	& RV measurement error	\\
	{[HJD-2450000]}	& [km $\textrm{s}^{-1}$] & [km $\textrm{s}^{-1}$] & [km $\textrm{s}^{-1}$] & [km $\textrm{s}^{-1}$]	\\
	\hline
7276.60144	&	-34.435	&	0.145	&	13.294	&	0.071	\\
7286.62434	&	-62.870	&	0.155	&	39.596	&	0.075	\\
7291.59619	&	-39.782	&	0.142	&	18.457	&	0.068	\\
7292.58543	&	-35.884	&	0.144	&	14.803	&	0.071	\\
7293.61946	&	-32.233	&	0.151	&	11.290	&	0.074	\\
7324.56059	&	7.368	&	0.143	&	-24.695	&	0.070	\\
7333.52344	&	9.537	&	0.148	&	-26.684	&	0.072	\\
		\hline 
	\end{tabular} 
\end{table*}

\begin{table*}
	\caption{Radial velocities obtained for RM effect of the primary component of NN\,Del.}
	\label{appendix1-NNDelRmP}
	\centering
	\begin{tabular}{c r r}
		\hline 
		Time of the photometric middle  &	Primary star's	&	Error of the   \\
		of the observation $T_{pm}$		&	RM measurement 	& RM measurement	\\
		{[HJD-2450000]}	& [km $\textrm{s}^{-1}$] & [km $\textrm{s}^{-1}$]	\\
		\hline 
6878.586268	&	0.901	&	1.615	\\
6878.592251	&	0.397	&	1.816	\\
6878.599639	&	-0.424	&	1.855	\\
6878.610656	&	-1.147	&	2.146	\\
6878.622342	&	-2.689	&	2.091	\\
6878.631481	&	-3.327	&	1.659	\\
6878.643283	&	-3.358	&	1.796	\\
6878.653281	&	-3.982	&	1.909	\\
6878.662593	&	-5.115	&	0.96	\\
6878.672044	&	-4.773	&	1.169	\\
6878.680813	&	-5.065	&	0.93	\\
6878.692221	&	-5.136	&	0.536	\\
6878.701269	&	-5.017	&	0.354	\\
6878.710819	&	-4.587	&	0.673	\\
6878.719885	&	-4.588	&	0.5	\\
6878.729648	&	-4.425	&	0.421	\\
6878.738523	&	-4.325	&	0.29	\\
6878.747768	&	-4.105	&	0.302	\\
6878.757992	&	-3.775	&	0.308	\\
6878.766888	&	-3.668	&	0.246	\\
6878.776209	&	-3.404	&	0.221	\\
7276.601442	&	0.101	&	0.212	\\
		\hline 
	\end{tabular} 
\end{table*}

\begin{table*}
	\caption{Radial velocities obtained with {\sc todcor} for V963\,Cen.}
	\label{appendix1-V963Cen}
	\centering
	\begin{tabular}{c r r r r}
	\hline 
	Time of the photometric middle  &	Primary star's	&	Primary star's	&	Secondary star's &	Secondary star's \\
	of the observation $T_{pm}$		&	RV measurement 	& RV measurement error &	RV measurement 	& RV measurement error	\\
	{[HJD-2450000]}	& [km $\textrm{s}^{-1}$] & [km $\textrm{s}^{-1}$] & [km $\textrm{s}^{-1}$] & [km $\textrm{s}^{-1}$]	\\
	\hline
6378.65055	&	-40.915	&	0.085	&	-19.904	&	0.086	\\
6379.66038	&	-96.260	&	0.087	&	35.876	&	0.092	\\
6381.61673	&	-86.249	&	0.078	&	25.755	&	0.083	\\
6746.71652	&	-111.003	&	0.074	&	-58.724	&	0.068	\\
6753.74153	&	-3.622	&	0.065	&	50.479	&	0.077	\\
6754.56431	&	1.777	&	0.072	&	-57.261	&	0.068	\\
6754.66568	&	2.338	&	0.075	&	-62.774	&	0.077	\\
6755.58955	&	7.054	&	0.086	&	-63.432	&	0.080	\\
6755.71260	&	7.607	&	0.090	&	-68.127	&	0.092	\\
6756.62214	&	10.300	&	0.068	&	-68.679	&	0.095	\\
6756.75798	&	10.529	&	0.066	&	-71.417	&	0.072	\\
6757.63765	&	10.382	&	0.069	&	-71.632	&	0.070	\\
6757.79114	&	10.086	&	0.068	&	-71.439	&	0.073	\\
6760.68846	&	-57.240	&	0.067	&	-70.967	&	0.071	\\
6760.70651	&	-58.121	&	0.067	&	-3.425	&	0.070	\\
6760.78383	&	-62.596	&	0.065	&	-2.412	&	0.070	\\
6760.82063	&	-64.801	&	0.068	&	2.038	&	0.069	\\
6765.60026	&	-40.076	&	0.070	&	4.276	&	0.072	\\
6765.62708	&	-39.640	&	0.073	&	-20.842	&	0.073	\\
6765.73756	&	-37.972	&	0.079	&	-21.210	&	0.075	\\
6765.75630	&	-37.751	&	0.081	&	-22.853	&	0.082	\\
6766.70221	&	-25.064	&	0.102	&	-23.210	&	0.083	\\
6692.87042	&	-2.153	&	0.065	&	-35.559	&	0.101	\\
6767.57073	&	-15.795	&	0.068	&	-45.078	&	0.070	\\
6778.62696	&	-86.019	&	0.074	&	25.594	&	0.078	\\
6797.58447	&	-21.238	&	0.115	&	-39.518	&	0.121	\\
6798.50377	&	-12.155	&	0.096	&	-48.773	&	0.101	\\
6802.73116	&	10.808	&	0.072	&	-71.771	&	0.077	\\
6803.63290	&	9.995	&	0.096	&	-70.883	&	0.100	\\
6804.60168	&	3.076	&	0.073	&	-64.005	&	0.076	\\
6824.52987	&	-83.654	&	0.071	&	23.173	&	0.074	\\
7013.85824	&	0.015	&	0.079	&	-60.907	&	0.083	\\
7020.83718	&	-89.624	&	0.082	&	29.220	&	0.085	\\
7031.84158	&	10.786	&	0.079	&	-71.811	&	0.083	\\
7034.80524	&	-22.560	&	0.092	&	-37.983	&	0.096	\\
7062.83796	&	10.681	&	0.068	&	-71.705	&	0.072	\\
7079.74965	&	-1.119	&	0.067	&	-59.818	&	0.070	\\
7083.64404	&	-95.581	&	0.069	&	35.171	&	0.071	\\
7084.82288	&	-66.531	&	0.069	&	6.003	&	0.070	\\
7116.69009	&	-41.654	&	0.069	&	-19.127	&	0.071	\\
7124.56009	&	8.798	&	0.068	&	-69.855	&	0.072	\\
7142.65610	&	-70.897	&	0.071	&	10.369	&	0.072	\\
7156.59672	&	-11.461	&	0.069	&	-49.498	&	0.071	\\
7176.63293	&	-62.261	&	0.081	&	1.631	&	0.084	\\
7189.53542	&	-110.953	&	0.075	&	50.558	&	0.075	\\
6766.69198	&	-25.119	&	0.104	&	-35.386	&	0.102	\\
6766.73756	&	-24.636	&	0.097	&	-35.954	&	0.096	\\
		\hline 
	\end{tabular} 
\end{table*}

\begin{table*}
	\caption{Radial velocities obtained for RM effect of the primary component of V936 Cen.}
	\label{appendix1-V963CenRmP}
	\centering
	\begin{tabular}{c r r}
		\hline 
		Time of the photometric middle  &	Primary star's	&	Error of the   \\
		of the observation $T_{pm}$		&	RM measurement 	& RM measurement	\\
		{[HJD-2450000]}	& [km $\textrm{s}^{-1}$] & [km $\textrm{s}^{-1}$]	\\
		\hline 
6760.595769	&	-1.753	&	0.159	\\
6760.609491	&	-1.868	&	0.102	\\
6760.624984	&	-1.62	&	0.137	\\
6760.637513	&	-1.339	&	0.081	\\
6760.688448	&	-0.02	&	0.088	\\
6760.706574	&	0.093	&	0.083	\\
7142.656212	&	0.064	&	0.078	\\
7020.837172	&	0.106	&	0.229	\\
7189.535418	&	0.022	&	0.086	\\
7079.749621	&	0.08	&	0.139	\\
7156.596762	&	0.048	&	0.075	\\
6760.512935	&	1.906	&	0.118	\\
6760.53366	&	1.797	&	0.124	\\
6760.554448	&	0.542	&	0.244	\\
		\hline 
	\end{tabular} 
\end{table*}

\begin{table*}
	\caption{Radial velocities obtained with {\sc todcor} for AI\,Phe.}
	\label{appendix1-AIPhe}
	\centering
	\begin{tabular}{c r r r r}
	\hline 
	Time of the photometric middle  &	Primary star's	&	Primary star's	&	Secondary star's &	Secondary star's \\
	of the observation $T_{pm}$		&	RV measurement 	& RV measurement error &	RV measurement 	& RV measurement error	\\
	{[HJD-2450000]}	& [km $\textrm{s}^{-1}$] & [km $\textrm{s}^{-1}$] & [km $\textrm{s}^{-1}$] & [km $\textrm{s}^{-1}$]	\\
	\hline
7269.71808	&	12.078	&	0.07	&	-21.303	&	0.022	\\
7293.7606	&	7.561	&	0.064	&	-16.792	&	0.021	\\
7298.82019	&	40.506	&	0.034	&	-48.536	&	0.018	\\
7303.76621	&	19.422	&	0.063	&	-28.104	&	0.03	\\
7308.64853	&	-56.018	&	0.06	&	44.252	&	0.023	\\
7318.56462	&	9.349	&	0.074	&	-18.538	&	0.023	\\
7322.60931	&	37.391	&	0.051	&	-45.498	&	0.034	\\
7333.6615	&	-58.057	&	0.038	&	46.246	&	0.037	\\
7362.53035	&	-39.08	&	0.058	&	27.989	&	0.038	\\
7362.54193	&	-38.95	&	0.053	&	27.888	&	0.038	\\
7374.54536	&	42.512	&	0.066	&	-50.283	&	0.023	\\
7377.54809	&	19.362	&	0.051	&	-27.971	&	0.022	\\
7260.67722	&	-59.463	&	0.039	&	47.316	&	0.044	\\
7263.74581	&	-42.796	&	0.102	&	31.427	&	0.038	\\
7306.64059	&	-31.987	&	0.054	&	21.085	&	0.016	\\
7380.55051	&	-34.131	&	0.051	&	23.249	&	0.026	\\
7258.80305	&	-50.759	&	0.05	&	39.035	&	0.02	\\
7332.64318	&	-51.311	&	0.038	&	39.72	&	0.016	\\
7283.79444	&	-54.243	&	0.048	&	42.485	&	0.024	\\
7360.57558	&	-54.007	&	0.065	&	42.368	&	0.031	\\
7360.58795	&	-53.907	&	0.051	&	42.338	&	0.021	\\
7360.62412	&	-53.722	&	0.044	&	42.089	&	0.02	\\
7360.63378	&	-53.645	&	0.049	&	42.033	&	0.021	\\
7286.7925	&	-54.078	&	0.06	&	42.385	&	0.024	\\
7287.72707	&	-47.734	&	0.054	&	36.233	&	0.031	\\
7363.52079	&	-30.06	&	0.053	&	19.315	&	0.016	\\
7265.74279	&	-24.509	&	0.078	&	13.788	&	0.04	\\
		\hline 
	\end{tabular} 
\end{table*}

\begin{table*}
	\caption{Radial velocities obtained for RM effect of the secondary component of AI\,Phe.}
	\label{appendix1-AIPheRmP}
	\centering
	\begin{tabular}{c r r}
		\hline 
		Time of the photometric middle  &	Primary star's	&	Error of the   \\
		of the observation $T_{pm}$		&	RM measurement 	& RM measurement	\\
		{[HJD-2450000]}	& [km $\textrm{s}^{-1}$] & [km $\textrm{s}^{-1}$]	\\
		\hline 
7315.646084	&	0.019	&	0.034	\\
7266.696172	&	-0.008	&	0.042	\\
7266.723055	&	0.01	&	0.043	\\
7266.750258	&	-0.02	&	0.048	\\
7266.771829	&	-0.002	&	0.058	\\
7266.79907	&	-0.008	&	0.06	\\
7266.841123	&	0.031	&	0.07	\\
7266.868665	&	0	&	0.076	\\
7291.655631	&	0.042	&	0.124	\\
7365.523879	&	0.13	&	0.147	\\
7365.563771	&	0.091	&	0.157	\\
7365.579298	&	0.06	&	0.202	\\
7365.604907	&	0.087	&	0.182	\\
7365.613354	&	0.136	&	0.181	\\
7365.643891	&	0.157	&	0.185	\\
7316.555926	&	0.147	&	0.15	\\
7316.608468	&	0.22	&	0.093	\\
7390.547102	&	0.144	&	0.115	\\
7292.690709	&	-0.054	&	0.069	\\
7317.646829	&	0.002	&	0.089	\\
7317.687913	&	0.015	&	0.134	\\
7317.776205	&	0	&	0.094	\\
		\hline 
	\end{tabular} 
\end{table*}

\label{lastpage}
\end{document}